\documentstyle[aps]{revtex}

\begin{document}
\title{Closed expression of the interaction kernel in the Bethe-Salpeter equation
for quark-antiquark bound states}
\author{Jun-Chen Su}
\address{. Center for Theoretical Physics, Physics College, Jilin University,\\
Changchun 130023, People's Republic of China}
\maketitle

\begin{abstract}
The interaction kernel in the Bethe-Salpeter equation for quark-antiquark
bound states is derived from QCD in the general case where the quark and the
antiquark can be of different flavors or the same flavor. The kernel is
derived with the aid of the Bethe-Salpeter equations satisfied by the
quark-antiquark four-point Green function. The latter equations are
established based on the equations of motion obeyed by the quark and
antiquark propagators, the quark-antiquark four-point Green function and
some other kinds of Green functions in which the gluon field is involved.
The interaction kernel derived is given a closed and explicit expression.
Since the kernel is represented in terms of only a few-types of Green
functions, it is not only convenient for perturbative calculations, but also
suitable for nonperturbative investigations.


\end{abstract}

\section{Introduction}

The Bethe-Salpeter (B-S) equation which was established early in Refs. [1,
2] is recognized as a rigorous approach to the relativistic bound state
problem and has extensively been investigated in the period of more than
half century [3-21]. The distinctive features of the equation are: (1) The
equation is derived from quantum field theory and hence set up on the firm
dynamical basis; (2) The interaction kernel in the equation contains all the
interactions taking place in the bound states and therefore the equation
provides a possibility of exactly solving the problem of relativistic bound
states; (3) The equation is elegantly formulated in a manifestly
Lorentz-covariant form in the Minkowski space which allows us to discuss the
equation in any coordinate frame. However, there are tremendous difficulties
in practical applications of the equation. One of the difficulties arises
from the fact that the kernel in the equation was not given a closed form in
the past. Usually, the kernel is defined as a sum of all B-S irreducible (or
say, two-particle-irreducible) graphs. According to this definition, the
kernel can only be calculated by the perturbation method. In practical
applications, the perturbation series of the kernel has to be truncated in a
ladder approximation. The ladder approximation has been proved to be
successful in QED for studying the bound states formed by the
electromagnetic interaction [22-26]. Nevertheless, it is not feasible in QCD
for exploring multi-quark bound states because the confining force which
must be taken into account in this case could not follow from a perturbative
calculation. It was remarked in Ref.[27] that '' The approach using the
Bethe-Salpeter equation has not led to a real breakthrough in our
understanding of quark-quark force''. The reason is mainly due to that'' we
have no method for computing the kernel of the B-S equation'' because we
have not known the closed expression of the kernel which can be used to
evaluate the confining force. In the previous application of the B-S
equation to investigate the hadron structure, a phenomenological confining
potential is necessarily introduced and added to the one-gluon exchange
kernel so as to obtain reasonable theoretical results [27-29]. The confining
potential was originally proposed from a nonperturbative computation, for
example, the lattice gauge calculation for a special Wilson loop [30-33].

Since the B-S equation is an exact formalism for bound states and the B-S
kernel contains all the interactions responsible for the formation of the
bound states, the B-S kernel is supposed to be the most suitable starting
point of examining the quark confinement. The question now is concentrated
on whether a closed expression of the B-S kernel exists and can be derived?
The aim of this paper is to give a definite answer to this question. It will
be shown that opposite to the conventional concept that '' The kernel $K$
can not be given in a closed form expression'' [34], a closed expression of
the B-S kernel for quark-antiquark bound states will perfectly be derived in
this paper. The method of deriving the expression of the kernel\ is usage of
various equations of motion satisfied by the $q\overline{q}$ four-point
Green functions some of which involve the gluon field in them [19-21]. The
kernel derived\ is compactly represented \ in terms of only a few types of
Green functions. Therefore, the kernel is not only easily calculated by the
perturbation method by employing the familiar perturbative expansions of the
Green functions, but also suitable for nonperturbative investigations by the
lattice approach by making use of the path-integral representations of the
Green functions.

The remainder of this paper is arranged as follows. In Sect.2, we
recapitulate the derivation of the B-S equation and show the definition of
the B-S kernel. Section 3 serves to describe the procedure of deriving the
B-S kernel. Section 4 is used to derive the final expression of the kernel.
In the last section, a summary and discussions are presented. In Appendix,
the equations of motion satisfied by some Green functions which are
necessary to use in the derivation of the B-S kernel are derived.

\section{Definition of the B-S kernel}

To derive the B-S interaction kernel appearing in the B-S equation for $q%
\overline{q}$ bound states, we first show how the kernel is defined through
Green functions. Before doing this, it is necessary to sketch the derivation
of the B-S equation satisfied by the $q\overline{q}$ four-point Green
function. For the $q\overline{q}$ bound states in which the quark and the
antiquark may have same flavor, the four-point Green function is
appropriately defined in the Heisenberg picture as follows [35] 
\begin{equation}
{\cal G}(x_{1,}x_2;y_1,y_2)_{\alpha \beta \rho \sigma }=\left\langle
0^{+}\left| T\{N[{\bf \psi }_\alpha (x_1){\bf \psi }_\beta ^c(x_2)]N[%
\overline{{\bf \psi }}_\rho (y_1)\overline{{\bf \psi }}_\sigma
^c(y_2)]\}\right| 0^{-}\right\rangle  \eqnum{2.1}
\end{equation}
where ${\bf \psi }(x)$ and ${\bf \psi }^c(x)$ are the quark and antiquark
field operators respectively, $\overline{{\bf \psi }}(x)$ and $\overline{%
\text{ }{\bf \psi }}^c(x)$ are their corresponding Dirac conjugates [5] 
\begin{equation}
{\bf \psi }^c(x)=C\overline{{\bf \psi }}^T(x),\overline{{\bf \psi }}^c(x)=-%
{\bf \psi }^T(x)C^{-1}  \eqnum{2.2}
\end{equation}
here $C=i\gamma ^2\gamma ^0$ is the charge conjugation operator, $\mid
0^{\pm }\rangle $ denote the physical vacuum states, $T$ symbolizes the
time-ordering product and $N$ designates the normal product which is
defined, for example, by 
\begin{equation}
N[{\bf \psi }_\alpha (x_1){\bf \psi }_\beta ^c(x_2)]=T[{\bf \psi }_\alpha
(x_1){\bf \psi }_\beta ^c(x_2)]-\left\langle 0^{+}\left| T[{\bf \psi }%
_\alpha (x_1){\bf \psi }_\beta ^c(x_2)]\right| 0^{-}\right\rangle . 
\eqnum{2.3}
\end{equation}
It is emphasized here that the above normal product can only \ be viewed as
a definition in the Heisenberg picture. With the definition shown in Eq.
(2.3), the Green function in Eq. (2.1) can be represented as 
\begin{equation}
{\cal G}(x_{1,}x_2;y_1,y_2)_{\alpha \beta \rho \sigma
}=G(x_{1,}x_2;y_1,y_2)_{\alpha \beta \rho \sigma }+S_F^{*}(x_1-x_2)_{\alpha
\beta }\overline{S}_F^{*}(y_1-y_2)_{\rho \sigma }  \eqnum{2.4}
\end{equation}
where 
\begin{equation}
G(x_{1,}x_2;y_1,y_2)_{\alpha \beta \rho \sigma }=\left\langle 0^{+}\left| T\{%
{\bf \psi }_\alpha (x_1){\bf \psi }_\beta ^c(x_2)\overline{{\bf \psi }}_\rho
(y_1)\overline{{\bf \psi }}_\sigma ^c(y_2)\}\right| 0^{-}\right\rangle 
\eqnum{2.5}
\end{equation}
is the conventional $q\overline{q}$ four-point Green function, 
\begin{equation}
\begin{tabular}{l}
$S_F^{*}(x_1-x_2)_{\alpha \beta }=\frac 1i\left\langle 0^{+}\left| T[{\bf %
\psi }_\alpha (x_1){\bf \psi }_\beta ^c(x_2)]\right| 0^{-}\right\rangle $ \\ 
$=S_F(x_1-x_2)_{\alpha \gamma }(C^{-1})_{\gamma \beta
}=S_F^c(x_2-x_1)_{\beta \lambda }C_{\lambda \alpha }$%
\end{tabular}
\eqnum{2.6}
\end{equation}
and 
\begin{equation}
\begin{tabular}{l}
$\overline{S_F^{*}}(y_1-y_2)_{\rho \sigma }=\frac 1i\left\langle 0^{+}\left|
T\{\overline{{\bf \psi }}_\rho (y_1)\overline{{\bf \psi }}_\sigma
^c(y_2)\}\right| 0^{-}\right\rangle $ \\ 
$=C_{\sigma \tau }S_F(y_2-y_1)_{\tau \rho }=(C^{-1})_{\rho \delta
}S_F^c(y_1-y_2)_{\delta \sigma }$%
\end{tabular}
\eqnum{2.7}
\end{equation}
in which 
\begin{equation}
S_F(x_1-x_2)_{\alpha \gamma }=\frac 1i\left\langle 0^{+}\left| T[{\bf \psi }%
_\alpha (x_1)\overline{{\bf \psi }}_\gamma (x_2)]\right| 0^{-}\right\rangle 
\eqnum{2.8}
\end{equation}
and 
\begin{equation}
S_F^c(y_1-y_2)_{\delta \sigma }=\frac 1i\left\langle 0^{+}\left| T\{{\bf %
\psi }_\delta ^c(y_1)\overline{{\bf \psi }}_\sigma ^c(y_2)\}\right|
0^{-}\right\rangle  \eqnum{2.9}
\end{equation}
are the ordinary quark and antiquark propagators respectively [5]. It is
clear that the propagators defined in Eqs. (2.6) and (2.7) are nonzero only
for such a quark and an antiquark that they are of the same flavor. For the
quark and antiquark of different flavors, the Green function defined in Eq.
(2.1) is reduced to the conventional form shown in Eq. (2.5) since the
second term on the right hand side (RHS) of Eq. (2.4) vanishes. In the case
of the quark and antiquark of the same flavor, the normal product in Eq.
(2.1) plays a role of excluding the contraction between the quark field and
the antiquark one from the Green function. Physically, this avoids the $q%
\overline{q}$ annihilation to break stability of the bound state. It is
pointed out that use of $\psi ^c(x)$ other than $\overline{\psi }(x)$ to
represent the antiquark field in this paper has an advantage that the
antiquark field would behave as a quark one in the B-S equation so that the
quark-antiquark equation formally is the same as the corresponding two-quark
equation in the case that the quark and antiquark have different flavors.

Now let us proceed to the derivation of the B-S equation for the $q\overline{%
q}$ bound system. This equation can be set up by operating on the Green's
function ${\cal G}(x_{1,}x_2;y_1,y_2)$ successively with the inverses of
quark and antiquark propagators, i.e. the operator $(i{\bf \partial }_{x_1}%
{\bf -}m_1{\bf +}$ $\Sigma )$ defined at the coordinate $x_1$ and the
operator $(i{\bf \partial }_{x_2}{\bf -}m_2{\bf +}$ $\Sigma ^c)$ \ defined
at the coordinate $x_2$ where ${\bf \partial }_x{\bf =}\gamma ^\mu \partial
_\mu ^x$, $m_{1\text{ }}$ and $m_2$ are the quark and antiquark masses, and $%
\Sigma $ and $\Sigma ^c$ stand for the quark and antiquark proper
self-energies which will be defined below. In Appendix, various equations of
motion satisfied by several Green functions have been derived from the QCD
generating functional. These equations of motion are necessary for deriving
the B-S equation. First, we show the equations of motion for the quark and
antiquark two-point Green functions (the propagators). From Eq. (A11) in the
appendix A, we get the equation of motion satisfied by the quark propagator
[5] 
\begin{equation}
\lbrack (i{\bf \partial }_{x_1}-m_1+\Sigma )S_F]_{\alpha \rho
}(x_1,y_1)=\delta _{\alpha \rho }\delta ^4(x_1-y_1)  \eqnum{2.10}
\end{equation}
where 
\begin{equation}
\begin{tabular}{l}
$(\Sigma S_F)_{\alpha \rho }(x_1,y_1)\equiv \int d^4z_1\Sigma
(x_1,z_1)_{\alpha \gamma }S_F(z_1-y_1)_{\gamma \rho }$ \\ 
$=(\Gamma ^{a\mu })_{\alpha \gamma }\Lambda _\mu ^a(x_1\mid x_1,y_1)_{\gamma
\beta }$%
\end{tabular}
\eqnum{2.11}
\end{equation}
here 
\begin{equation}
(\Gamma ^{a\mu })_{\alpha \gamma }=g(\gamma ^\mu T^a)_{\alpha \gamma } 
\eqnum{2.12}
\end{equation}
in which $T^a=\lambda ^a/2$ are the quark color matrices and 
\begin{equation}
\Lambda _\mu ^a(x_1\mid x_1,y_1)_{\gamma \rho }=\frac 1i\left\langle
0^{+}\left| T[{\bf A}_\mu ^a(x_1){\bf \psi }_\gamma (x_1)\overline{{\bf \psi 
}}_\rho (y_1)]\right| 0^{-}\right\rangle  \eqnum{2.13}
\end{equation}
in which ${\bf A}_\mu ^a(x_1)$ are the gluon field operators. Similarly,
from Eq. (B2) given in the appendix B, we can write the equation of motion
for the antiquark propagator 
\begin{equation}
\lbrack (i{\bf \partial }_{x_2}-m_2+\Sigma ^c)S_F^c]_{\beta \sigma
}(x_2,y_2)=\delta _{\beta \sigma }\delta ^4(x_2-y_2)  \eqnum{2.14}
\end{equation}
where 
\begin{equation}
\begin{tabular}{l}
$(\Sigma ^cS_F^c)_{\beta \sigma }(x_2,y_2)\equiv \int d^4z_2\Sigma
^c(x_2,z_2)_{\beta \lambda }S_F^c(z_2-y_2)_{\lambda \sigma }$ \\ 
$=(\overline{\Gamma }^{b\nu })_{\beta \lambda }\Lambda _\nu ^{{\bf c}%
b}(x_2\mid x_2,y_2)_{\lambda \sigma }$%
\end{tabular}
\eqnum{2.15}
\end{equation}
here 
\begin{equation}
\overline{\Gamma }^{b\nu }=g\gamma ^\nu \overline{T}^b  \eqnum{2.16}
\end{equation}
with $\overline{T}^b=-\lambda ^{b*}/2$ being the antiquark color matrices
and 
\begin{equation}
\Lambda _\nu ^{{\bf c}b}(x_2\mid x_2,y_2)_{\lambda \sigma }=\frac 1i%
\left\langle 0^{+}\left| T[{\bf A}_\nu ^b(x_2){\bf \psi }_\lambda ^c(x_2)%
\overline{{\bf \psi }}_\sigma ^c(y_2)]\right| 0^{-}\right\rangle . 
\eqnum{2.17}
\end{equation}
Equations (2.11) and (2.15) just give the definition of the proper
self-energies. The explicit representations of the self-energies can be
found from the one-particle-irreducible decomposition of the Green functions 
$\Lambda _\mu ^a(x_1\mid x_1,y_1)$ and $\Lambda _\nu ^{{\bf c}b}(x_2\mid
x_2,y_2)$ [5, 36].

Here, we have a place to give general definitions of the Green functions
involved in this paper for simplifying later statements. These Green's
functions are 
\begin{equation}
\begin{tabular}{l}
$\Lambda _{\mu ...\nu \kappa ...\theta
}^{a...bc...d}(x_i,...,x_j;y_k,...,y_l\mid x_1,y_1)_{\alpha \rho }$ \\ 
$=\frac 1i\langle 0^{+}\mid T[{\bf A}_\mu ^a(x_i)...{\bf A}_\nu ^b(x_j){\bf A%
}_\kappa ^c(y_k)...{\bf A}_\theta ^d(y_l){\bf \psi }_\alpha (x_1)\overline{%
{\bf \psi }}_\rho (y_1)]\mid 0^{-}\rangle $%
\end{tabular}
\eqnum{2.18}
\end{equation}

\begin{equation}
\begin{tabular}{l}
$\Lambda _{\mu ...\nu \kappa ...\theta }^{{\bf c}%
a...bc...d}(x_i,...,x_j;y_k,...,y_l\mid x_2,y_2)_{\beta \sigma }$ \\ 
$=\frac 1i\langle 0^{+}\mid T[{\bf A}_\mu ^a(x_i)...{\bf A}_\nu ^b(x_j){\bf A%
}_\kappa ^c(y_k)...{\bf A}_\theta ^d(y_l){\bf \psi }_\beta ^c(x_2)\overline{%
{\bf \psi }}_\sigma ^c(y_2)]\mid 0^{-}\rangle $%
\end{tabular}
\eqnum{2.19}
\end{equation}

\begin{equation}
\begin{tabular}{l}
$\Lambda _{\mu ...\nu \kappa ...\theta
}^{*a...bc...d}(x_i,...,x_j;y_k,...,y_l\mid x_1,x_2)_{\alpha \beta }$ \\ 
$=\frac 1i\langle 0^{+}\mid T[{\bf A}_\mu ^a(x_i)...{\bf A}_\nu ^b(x_j){\bf A%
}_\kappa ^c(y_k)...{\bf A}_\theta ^d(y_l){\bf \psi }_\alpha (x_1)\psi _\beta
^c(x_2)]\mid 0^{-}\rangle $%
\end{tabular}
\eqnum{2.20}
\end{equation}

\begin{equation}
\begin{tabular}{l}
$\overline{\Lambda }_{\mu ...\nu \kappa ...\theta
}^{*a...bc...d}(x_i,...,x_j;y_k,...,y_l\mid y_1,y_2)_{\rho \sigma }$ \\ 
$=\frac 1i\langle 0^{+}\mid T[{\bf A}_\mu ^a(x_i)...{\bf A}_\nu ^b(x_j){\bf A%
}_\kappa ^c(y_k)...{\bf A}_\theta ^d(y_l)\overline{\psi }_\rho (y_1)%
\overline{{\bf \psi }}_\sigma ^c(y_2)]\mid 0^{-}\rangle $%
\end{tabular}
\eqnum{2.21}
\end{equation}
\begin{equation}
\begin{tabular}{l}
$G_{\mu ...\nu \kappa ...\theta }^{a...bc...d}(x_i,...,x_j;y_k,...,y_l\mid
x_1,x_2;y_1,y_2)_{\alpha \beta \rho \sigma }$ \\ 
$=\langle 0^{+}\mid T[{\bf A}_\mu ^a(x_i)...{\bf A}_\nu ^b(x_j){\bf A}%
_\kappa ^c(y_k)...{\bf A}_\theta ^d(y_l){\bf \psi }_\alpha (x_1){\bf \psi }%
_\beta ^c(x_2)\overline{{\bf \psi }}_\rho (y_1)\overline{{\bf \psi }}_\sigma
^c(y_2)]\mid 0^{-}\rangle $%
\end{tabular}
\eqnum{2. 22}
\end{equation}
and 
\begin{equation}
\begin{tabular}{l}
${\cal G}_{\mu ...\nu \kappa ...\theta
}^{a...bc...d}(x_i,...,x_j;y_k,...,y_l\mid x_1,x_2;y_1,y_2)_{\alpha \beta
\rho \sigma }$ \\ 
$=\langle 0^{+}\mid T\{N[{\bf A}_\mu ^a(x_i)...{\bf A}_\nu ^b(x_j){\bf \psi }%
_\alpha (x_1){\bf \psi }_\beta ^c(x_2)]N[\overline{{\bf \psi }}_\rho (y_1)%
\overline{{\bf \psi }}_\sigma ^c(y_2){\bf A}_\kappa ^c(y_k)...{\bf A}_\theta
^d(y_l)]\}\mid 0^{-}\rangle $%
\end{tabular}
\eqnum{2.23}
\end{equation}
where $i,j.k,l=1,2$. The normal products in Eq. (2.23) are defined in the
same way as that in Eq. (2.3). Hereafter, we no longer write individual
representations of the Green functions encountered in later derivations .
They can be read off from the above expressions.

According to the derivations given in\ the appendices, we can write from
Eqs. (A24) and (B4) the equations of motion obeyed by the Green function
defined in Eq. (2.5) [5] 
\begin{equation}
\begin{tabular}{l}
$\lbrack (i{\bf \partial }_{x_1}-m_1+\Sigma )G]_{\alpha \beta \rho \sigma
}(x_{1,}x_2;y_1,y_2)=\delta _{\alpha \rho }\delta
^4(x_1-y_1)S_F^c(x_2-y_2)_{\beta \sigma }$ \\ 
$+C_{\alpha \beta }\delta ^4(x_1-x_2)\overline{S}_F^{*}(y_1-y_2)_{\rho
\sigma }-(\Gamma ^{a\mu })_{\alpha \gamma }G_\mu ^a(x_1\mid
x_1,x_2;y_1,y_2)_{\gamma \beta \rho \sigma }$ \\ 
$+\int d^4z_1\Sigma (x_1,z_1)_{\alpha \gamma }G(z_1,x_2;y_1,y_2)_{\gamma
\beta \rho \sigma }$%
\end{tabular}
\eqnum{2.24}
\end{equation}
and 
\begin{equation}
\begin{tabular}{l}
$\lbrack (i{\bf \partial }_{x_2}-m_2+\Sigma ^c)G]_{\alpha \beta \rho \sigma
}(x_{1,}x_2;y_1,y_2)=\delta _{\beta \sigma }\delta
^4(x_2-y_2)S_F(x_1-y_1)_{\alpha \rho }$ \\ 
$+C_{\alpha \beta }\delta ^4(x_1-x_2)\overline{S}_F^{*}(y_1-y_2)_{\rho
\sigma }-(\overline{\Gamma }^{b\nu })_{\beta \lambda }G_\nu ^b(x_2\mid
x_1,x_2;y_1,y_2)_{\alpha \lambda \rho \sigma }$ \\ 
$+\int d^4z_2\Sigma (x_2,z_2)_{\beta \lambda }G(x_1,z_2;y_1,y_2)_{\alpha
\lambda \rho \sigma }$%
\end{tabular}
\eqnum{2.25}
\end{equation}
where the self-energy-related terms have been added on the both sides of the
equations and $G_\mu ^a(x_i\mid x_1,x_2;y_1,y_2)_{\alpha \beta \rho \sigma }$
was defined in Eq. (2.22) with one gluon field operator ${\bf A}_\mu ^a(x_i)$
in it. It is noted that the second terms on the RHS of Eqs. (2.24) and
(2.25) will disappear when the quark and the antiquark have different
flavors.

The equations of motion satisfied by the Green function defined in Eq. (2.1)
can be written out by virtue of Eqs. (2.24) and (2.25). To do this, we need
to introduce new Green functions which follows immediately from Eq. (2.23)
as follows 
\begin{equation}
\begin{tabular}{l}
${\cal G}_\mu ^a(x_i\mid x_1,x_2;y_1,y_2)_{\alpha \beta \rho \sigma }=G_\mu
^a(x_i\mid x_1,x_2;y_1,y_2)_{\alpha \beta \rho \sigma }$ \\ 
$+\Lambda _\mu ^{*a}(x_i\mid x_1,x_2)_{\alpha \beta }\overline{S}%
_F^{*}(y_1-y_2)_{\rho \sigma }$%
\end{tabular}
\eqnum{2.26}
\end{equation}
where $i=1,2$. On substituting the definitions in Eqs. (2.4) and (2.26) into
Eqs.(2.24) and (2.25), employing the equations in Eqs. (2.10) and (2.14) and
the relations in Eqs. (2.6), (2.7), (2.11) and (2.15), it is not difficult
to derive the following equations 
\begin{equation}
\begin{tabular}{l}
$\lbrack (i{\bf \partial }_{x_1}-m_1+\Sigma ){\cal G}]_{\alpha \beta \rho
\sigma }(x_{1,}x_2;y_1,y_2)=\delta _{\alpha \rho }\delta
^4(x_1-y_1)S_F^c(x_2-y_2)_{\beta \sigma }$ \\ 
$-(\Gamma ^{a\mu })_{\alpha \gamma }{\cal G}_\mu ^a(x_1\mid
x_1,x_2;y_1,y_2)_{\gamma \beta \rho \sigma }+\int d^4z_1\Sigma
(x_1,z_1)_{\alpha \gamma }{\cal G}(z_1,x_2;y_1,y_2)_{\gamma \beta \rho
\sigma }$%
\end{tabular}
\eqnum{2.27}
\end{equation}
and 
\begin{equation}
\begin{tabular}{l}
$\lbrack (i{\bf \partial }_{x_2}-m_2+\Sigma ^c){\cal G}]_{\alpha \beta \rho
\sigma }(x_{1,}x_2;y_1,y_2)=\delta _{\beta \sigma }\delta
^4(x_2-y_2)S_F(x_1-y_1)_{\alpha \rho }$ \\ 
$-(\overline{\Gamma }^{b\nu })_{\beta \lambda }{\cal G}_\nu ^b(x_2\mid
x_1,x_2;y_1,y_2)_{\alpha \lambda \rho \sigma }+\int d^4z_2\Sigma
^c(x_2,z_2)_{\beta \lambda }{\cal G}(x_1,z_2;y_1,y_2)_{\alpha \lambda \rho
\sigma }.$%
\end{tabular}
\eqnum{2.28}
\end{equation}
In comparison of Eqs. (2.24) and (2.25) with Eqs. (2.27) and (2.28), we see
that the second terms on the RHS of Eqs. (2.24) and (2.25), which are
related to the annihilation between the quark and antiquark with the same
flavor, are absent in Eqs. (2.27) and (2.28). That is to say, although the
equations in Eqs. (2.27) and (2.28) suit to the both cases that the flavors
of quark and antiquark are either different or the same, they formally are
the same as those equations written in Eqs. (2.24) and (2.25) for the case
that the quark and antiquark have different flavors.

To derive the B-S equation, we need to operate on Eq. (2.28) with the
operator( $i{\bf \partial }_{x_1}{\bf -}m_1{\bf +}$ $\Sigma $ ) ( or
equivalently, operate on Eq. (2.27) with ($i{\bf \partial }_{x_2}{\bf -}m_2%
{\bf +}$ $\Sigma ^c$)) 
\begin{equation}
\begin{tabular}{l}
$\ [(i{\bf \partial }_{x_1}-m_1+\Sigma )(i{\bf \partial }_{x_2}-m_2+\Sigma
^c){\cal G}]_{\alpha \beta \rho \sigma }(x_{1,}x_2;y_1,y_2)$ \\ 
$=\delta _{\beta \sigma }\delta ^4(x_2-y_2)[(i{\bf \partial }%
_{x_1}-m_1+\Sigma )S_F]_{\alpha \rho }(x_1-y_1)$ \\ 
$-(\overline{\Gamma }^{b\nu })_{\beta \lambda }[(i{\bf \partial }%
_{x_1}-m_1+\Sigma ){\cal G}_\nu ^b]_{\alpha \lambda \rho \sigma }(x_2\mid
x_1,x_2;y_1,y_2)$ \\ 
$+\int d^4z_2\Sigma ^c(x_2,z_2)_{\beta \lambda }[(i{\bf \partial }%
_{x_1}-m_1+\Sigma ){\cal G]}_{\alpha \lambda \rho \sigma }(x_1,z_2;y_1,y_2).$%
\end{tabular}
\eqnum{2.29}
\end{equation}
The first and third terms in the above can directly be evaluated by
employing Eqs. (2.10) and (2.27). To compute the second term, we first write
down the equation of motion for the Green function $G_\nu ^b(x_2\mid
x_1,x_2;y_1,y_2)$ which is given by supplementing the self- energy-related
term to the both sides of Eq. (A25) 
\begin{equation}
\begin{tabular}{l}
$\lbrack (i{\bf \partial }_{x_1}-m_1+\Sigma )G_\nu ^b]_{\alpha \lambda \rho
\sigma }(x_2\mid x_{1,}x_2;y_1,y_2)=\delta _{\alpha \rho }\delta
^4(x_1-y_1)\Lambda _\nu ^{{\bf c}b}(x_2\mid x_2,y_2)_{\lambda \sigma }$ \\ 
$+C_{\alpha \lambda }\delta ^4(x_1-x_2)\overline{\Lambda }_\nu ^{*b}(x_2\mid
y_1,y_2)_{\rho \sigma }-(\Gamma ^{a\mu })_{\alpha \gamma }G_{\mu \nu
}^{ab}(x_1,x_2\mid x_1,x_2;y_1,y_2)_{\gamma \lambda \rho \sigma }$ \\ 
$+\int d^4z_1\Sigma (x_1,z_1)_{\alpha \gamma }G_\nu ^b(x_2\mid
z_1,x_2;y_1,y_2)_{\gamma \lambda \rho \sigma }$%
\end{tabular}
\eqnum{2.30}
\end{equation}
where $\Lambda _\nu ^{{\bf c}b}(x_2\mid x_2,y_2)_{\lambda \sigma },$ $%
\overline{\Lambda }_\nu ^{*b}(x_2\mid y_1,y_2)_{\rho \sigma }$ and $G_{\mu
\nu }^{ab}(x_1,x_2\mid x_1,x_2;y_1,y_2)_{\gamma \lambda \rho \sigma }$ are
the Green functions which can be read off from Eqs. (2.19), (2.21) and
(2.22) respectively. According to the relation shown in Eq. (2.26) and the
following relation which follows from Eq. (2.23) 
\begin{equation}
\begin{tabular}{l}
${\cal G}_{\mu \nu }^{ab}(x_1,x_2\mid x_1,x_2;y_1,y_2)_{\alpha \beta \rho
\sigma }=G_{\mu \nu }^{ab}(x_1,x_2\mid x_1,x_2;y_1,y_2)_{\alpha \beta \rho
\sigma }$ \\ 
$+\Lambda _{\mu \nu }^{*ab}(x_1,x_2\mid x_1,x_2)_{\alpha \beta }\overline{S}%
_F^{*}(y_1-y_2)_{\rho \sigma }$%
\end{tabular}
\eqnum{2.31}
\end{equation}
where $\Lambda _{\mu \nu }^{*ab}(x_1,x_2\mid x_1,x_2)_{\alpha \beta }$ was
defined in Eq. (2.20), and employing the equation of motion for the Green
function $\Lambda _\nu ^{*b}(x_2\mid x_1,x_2)_{\alpha \beta }$ which is
obtained from Eq. (A27) by introducing the self-energy-related term to the
equation 
\begin{equation}
\begin{tabular}{l}
$\lbrack (i{\bf \partial }_{x_1}-m_1+\Sigma )\Lambda _\nu ^{*b}]_{\alpha
\beta }(x_2\mid x_1,x_2)=-(\Gamma ^{a\mu })_{\alpha \gamma }\Lambda _{\mu
\nu }^{*ab}(x_1,x_2\mid x_1,x_2)_{\gamma \beta }$ \\ 
$+\int d^4z_1\Sigma (x_1,z_1)_{\alpha \gamma }\Lambda _\nu ^{*b}(x_2\mid
z_1,x_2)_{\gamma \beta }$%
\end{tabular}
\eqnum{2.32}
\end{equation}
it is easy to find from Eq. (2.30) an equation of motion for the Green
function ${\cal G}_\nu ^b(x_2\mid x_1,x_2;y_1,y_2)$ such that 
\begin{equation}
\begin{tabular}{l}
$\lbrack (i{\bf \partial }_{x_1}-m_1+\Sigma ){\cal G}_\nu ^b]_{\alpha
\lambda \rho \sigma }(x_2\mid x_1,x_2;y_1,y_2)=\delta _{\alpha \rho }\delta
^4(x_1-y_1)\Lambda _\nu ^{{\bf c}b}(x_2\mid x_2,y_2)_{\lambda \sigma }$ \\ 
$+C_{\alpha \lambda }\delta ^4(x_1-x_2)\overline{\Lambda }_\nu ^{*b}(x_2\mid
y_1,y_2)_{\rho \sigma }-(\Gamma ^{a\mu })_{\alpha \gamma }{\cal G}_{\mu \nu
}^{ab}(x_1,x_2\mid x_1,x_2;y_1,y_2)_{\gamma \lambda \rho \sigma }$ \\ 
$+\int d^4z_1\Sigma (x_1,z_1)_{\alpha \gamma }{\cal G}_\nu ^b(x_2\mid
z_1,x_2;y_1,y_2)_{\gamma \lambda \rho \sigma }.$%
\end{tabular}
\eqnum{2.33}
\end{equation}
Upon inserting Eqs. (2.10), (2.27) and (2.33) into Eq. (2.29), we arrive at 
\begin{equation}
\begin{tabular}{l}
$\lbrack (i{\bf \partial }_{x_1}-m_1+\Sigma )(i{\bf \partial }%
_{x_2}-m_2+\Sigma ^c){\cal G}]_{\alpha \beta \rho \sigma
}(x_{1,}x_2;y_1,y_2) $ \\ 
$=\delta _{\alpha \rho }\delta _{\beta \sigma }\delta ^4(x_1-y_1)\delta
^4(x_2-y_2)+{\cal H}_1(x_1,x_2;y_1,y_2)_{\alpha \beta \rho \sigma }$%
\end{tabular}
\eqnum{2.34}
\end{equation}
where 
\begin{equation}
{\cal H}_1(x_1,x_2;y_1,y_2)_{\alpha \beta \rho \sigma }=\sum_{i=1}^5{\cal H}%
_1^{(i)}(x_1,x_2;y_1,y_2)_{\alpha \beta \rho \sigma }  \eqnum{2.35}
\end{equation}
in which 
\begin{equation}
{\cal H}_1^{(1)}(x_1,x_2;y_1,y_2)_{\alpha \beta \rho \sigma }=-(\Gamma
^{a\mu })_{\alpha \gamma }\int d^4z_2\Sigma ^c(x_2,z_2)_{\beta \lambda }%
{\cal G}_\mu ^a(x_1\mid x_1,z_2;y_1,y_2)_{\gamma \lambda \rho \sigma }, 
\eqnum{2.36}
\end{equation}

\begin{equation}
{\cal H}_1^{(2)}(x_1,x_2;y_1,y_2)_{\alpha \beta \rho \sigma }=-(\overline{%
\Gamma }^{b\nu })_{\beta \lambda }\int d^4z_1\Sigma (x_1,z_1)_{\alpha \gamma
}{\cal G}_\nu ^b(x_2\mid z_1,x_2;y_1,y_2)_{\gamma \lambda \rho \sigma }, 
\eqnum{2.37}
\end{equation}
\begin{equation}
{\cal H}_1^{(3)}(x_1,x_2;y_1,y_2)_{\alpha \beta \rho \sigma }=(\Gamma ^{a\mu
})_{\alpha \gamma }(\overline{\Gamma }^{b\nu })_{\beta \lambda }{\cal G}%
_{\mu \nu }^{ab}(x_1,x_2\mid x_1,x_2;y_1,y_2)_{\gamma \lambda \rho \sigma },
\eqnum{2.38}
\end{equation}
\begin{equation}
{\cal H}_1^{(4)}(x_1,x_2;y_1,y_2)_{\alpha \beta \rho \sigma }=\delta
^4(x_1-x_2)(\overline{\Gamma }^{b\nu })_{\beta \lambda }C_{\lambda \alpha }%
\overline{\Lambda }_\nu ^{*b}(x_2\mid y_1,y_2)_{\rho \sigma }  \eqnum{2.39}
\end{equation}
and 
\begin{equation}
{\cal H}_1^{(5)}(x_1,x_2;y_1,y_2)_{\alpha \beta \rho \sigma }=\int
d^4z_1d^4z_2\Sigma (x_1,z_1)_{\alpha \gamma }\Sigma ^c(x_2,z_2)_{\beta
\lambda }{\cal G}(z_1,z_2;y_1,y_2)_{\gamma \lambda \rho \sigma }. 
\eqnum{2.40}
\end{equation}
From perturbative calculations or irreducible decompositions of the Green
functions [1-5, 21], it may be found that the function ${\cal H}%
_1(x_1,x_2;y_1,y_2)$ is B-S reducible and can be written in the form 
\begin{equation}
{\cal H}_1(x_1,x_2;y_1,y_2)_{\alpha \beta \rho \sigma }=\int
d^4z_1d^4z_2K(x_1,x_2;z_1,z_2)_{\alpha \beta \gamma \lambda }{\cal G}%
(z_1,z_2;y_1,y_2)_{\gamma \lambda \rho \sigma }  \eqnum{2.41}
\end{equation}
where $K(x_1,x_2;z_1,z_2)$ is precisely the B-S irreducible kernel. With the
above expression for the function ${\cal H}_1(x_1,x_2;y_1,y_2),$ the
equation in Eq. (2.34) can be expressed in a closed form 
\begin{equation}
\begin{tabular}{l}
$\lbrack (i{\bf \partial }_{x_1}-m_1+\Sigma )(i{\bf \partial }%
_{x_2}-m_2+\Sigma ^c){\cal G}]_{\alpha \beta \rho \sigma }(x_{1,}x_2;y_1,y_2)
$ \\ 
$=\delta _{\alpha \rho }\delta _{\beta \sigma }\delta ^4(x_1-y_1)\delta
^4(x_2-y_2)+\int d^4z_1d^4z_2K(x_1,x_2;z_1,z_2)_{\alpha \beta \gamma \lambda
}{\cal G}(z_1,z_2;y_1,y_2)_{\gamma \lambda \rho \sigma }.$%
\end{tabular}
\eqnum{2.42}
\end{equation}
This just is the B-S equation satisfied by the Green function ${\cal G}%
(x_1,x_2;y_1,y_2).$ By making use of the Lehmann representation of the Green
function ${\cal G}(x_1,x_2;y_1,y_2)$ or the well-known procedure proposed by
Gell-Mann and Low [2], one may readily derive from Eq. (2.42) the B-S
equation satisfied by B-S amplitudes describing the $q\overline{q\text{ }}$
bound states [1-5] 
\begin{equation}
\lbrack (i{\bf \partial }_{x_1}-m_1+\Sigma )(i{\bf \partial }%
_{x_2}-m_2+\Sigma ^c){\cal \chi }_{P\varsigma }](x_{1,}x_2)=\int
d^4y_1d^4y_2K(x_1,x_2;y_1,y_2)\chi _{P\varsigma }(y_1,y_2)  \eqnum{2.43}
\end{equation}
where 
\begin{equation}
\chi _{P\varsigma }(x_{1,}x_2)=\left\langle 0^{+}\left| N[{\bf \psi }(x_1)%
{\bf \psi }^c(x_2)]\right| P\varsigma \right\rangle   \eqnum{2.44}
\end{equation}
represents the B-S amplitude in which $P$ denotes the total momentum of a $q%
\overline{q\text{ }}$ bound state and$\ \varsigma $ marks the other quantum
numbers of the state. The above equation can be written in the form of an
integral equation when we operate on the both sides of the above equation
with $(i{\bf \partial }_{x_1}-m_1+\Sigma )^{-1}(i{\bf \partial }%
_{x_2}-m_2+\Sigma ^c)^{-1}$[1-5] 
\begin{equation}
{\cal \chi }_{P\varsigma }(x_{1,}x_2)=\int
d^4z_1d^4z_2d^4y_1d^4y_2S_F(x_1-z_1)S_F^c(x_2-z_2)K(z_1,z_2;y_1,y_2)\chi
_{P\varsigma }(y_1,y_2)  \eqnum{2.45}
\end{equation}

\section{Derivation of the B-S kernel}

Beyond the perturbation method, the B-S kernel may be derived by starting
from its definition shown in Eq. (2.41). One method of the derivation is
usage of the technique of irreducible decomposition of Green functions. By
this method, once the one and two-particle-irreducible decompositions of the
Green functions contained in the function ${\cal H}_1(x_1,x_2;y_1,y_2)$ are
completed, it can be found that the function ${\cal H}_1(x_1,x_2;y_1,y_2)$
can really be represented in the form as written in Eq. (2.41) and, at the
same time, the kernel will be explicitly worked out. Nevertheless, the
expression of the kernel given in this way is much complicated ( We leave
the detailed discussions in the future). Another method of deriving the
kernel is to use B-S equations which describe variations of the Green
functions involved in the function ${\cal H}_1(x_1,x_2;y_1,y_2)$ with the
coordinates $y_1$ and $y_2$. In this paper, we adopt the latter method and
describe the procedure of the derivation in this section. Let us operate on
the both sides of Eq. (2.41) from the right with the operator $(i%
\overleftarrow{{\bf \partial }}_{y_1}+m_1-\Sigma )(i\overleftarrow{{\bf %
\partial }}_{y_2}+m_2-\Sigma ^c)$ such that 
\begin{equation}
\begin{tabular}{l}
$\int d^4z_1d^4z_2K(x_1,x_2;z_1,z_2)_{\alpha \beta \gamma \lambda }[{\cal G}%
(i\overleftarrow{{\bf \partial }}_{y_1}+m_1-\Sigma )(i\overleftarrow{{\bf %
\partial }}_{y_2}+m_2-\Sigma ^c)]_{\gamma \lambda \rho \sigma
}(z_1,z_2;y_1,y_2)$ \\ 
$={\cal Q}(x_1,x_2;y_1,y_2)_{\alpha \beta \rho \sigma }$%
\end{tabular}
\eqnum{3.1}
\end{equation}
where 
\begin{equation}
{\cal Q}(x_1,x_2;y_1,y_2)_{\alpha \beta \rho \sigma }=[{\cal H}_1(i%
\overleftarrow{{\bf \partial }}_{y_1}+m_1-\Sigma )(i\overleftarrow{{\bf %
\partial }}_{y_2}+m_2-\Sigma ^c)]_{\alpha \beta \rho \sigma
}(x_1,x_2;y_1,y_2).  \eqnum{3.2}
\end{equation}

As seen from Eqs. (3.1) and (3.2), to derive the B-S kernel, it is necessary
to use the B-S equations with respect to $y_1$ and $y_2$ for the Green
function ${\cal G}(x_1,x_2;y_1,y_2)$ and the other Green functions appearing
in the function ${\cal H}_1(x_1,x_2;y_1,y_2).$ Analogous to the procedure
stated in the preceding section, the above-mentioned B-S equations can be
derived with the aid of equations of motion with respect to $y_1$ and $y_2$
for those Green functions. First, we show the equations of motion obeyed by
the quark and antiquark propagators 
\begin{equation}
\lbrack S_F(i\overleftarrow{{\bf \partial }}_{y_1}+m_1-\Sigma )]_{\alpha
\rho }(x_1,y_1)=-\delta _{\alpha \rho }\delta ^4(x_1-y_1)  \eqnum{3.3}
\end{equation}
and 
\begin{equation}
\lbrack S_F^c(i\overleftarrow{{\bf \partial }}_{y_2}+m_2-\Sigma ^c)]_{\beta
\sigma }(x_2,y_2)=-\delta _{\beta \sigma }\delta ^4(x_2-y_2)  \eqnum{3.4}
\end{equation}
which are given by Eqs. (C2) and (D2) respectively with the source $J$ being
set to vanish. We note here that the self-energies, as easily proved, are
the same as those defined in Eqs. (2.11) and (2.15).

Next, we derive the B-S equation satisfied by the Green function ${\cal G}%
(x_1,x_2;y_1,y_2).$ For this purpose, we first write down the equation of
motion describing variation of the conventional Green function $%
G(x_1,x_2;y_1,y_2)$ with $y_1$ which is obtained from Eq. (C5) by setting $%
J=0$ and adding the term related to the self-energy 
\begin{equation}
\begin{tabular}{l}
$\lbrack G(i\overleftarrow{{\bf \partial }}_{y_1}+m_1-\Sigma )]_{\alpha
\beta \rho \sigma }(x_1,x_2;y_1,y_2)=-\delta _{\alpha \rho }\delta
^4(x_1-y_1)S_F^c(x_2-y_2)_{\beta \sigma }$ \\ 
$-C_{\rho \sigma }\delta ^4(y_1-y_2)S_F^{*}(x_1-x_2)_{\alpha \beta }+G_\mu
^a(y_1\mid x_1,x_2;y_1,y_2)_{\alpha \beta \tau \sigma }(\Gamma ^{a\mu
})_{\tau \rho }$ \\ 
$-\int d^4z_1G(x_1,x_2;z_1,y_2)_{\alpha \beta \tau \sigma }\Sigma
(z_1,y_1)_{\tau \rho }$%
\end{tabular}
\eqnum{3.5}
\end{equation}
where $G_\mu ^a(y_1\mid x_1,x_2;y_1,y_2)$ is the Green function with a gluon
field at $y_1$ in it whose expression can be read from Eq. (2.22).
Substituting in Eq. (3.5) the relation given in Eq. (2.4) and the following
relation which follows from Eq. (2.23) 
\begin{equation}
{\cal G}_\mu ^a(y_i\mid x_1,x_2;y_1,y_2)_{\alpha \beta \rho \sigma }=G_\mu
^a(y_i\mid x_1,x_2;y_1,y_2)_{\alpha \beta \rho \sigma
}+S_F^{*}(x_1-x_2)_{\alpha \beta }\overline{\Lambda }_\mu ^{*a}(y_i\mid
y_1,y_2)_{\rho \sigma }  \eqnum{3.6}
\end{equation}
and employing the equation 
\begin{equation}
\lbrack \overline{S}_F^{*}(i\overleftarrow{{\bf \partial }}_{y_1}+m_1-\Sigma
)]_{\rho \sigma }(y_1,y_2)=C_{\rho \sigma }\delta ^4(y_1-y_2)  \eqnum{3.7}
\end{equation}
which is given by Eqs. (C8)-(C10), it can be found 
\begin{equation}
\begin{tabular}{l}
$\lbrack {\cal G}(i\overleftarrow{{\bf \partial }}_{y_1}+m_1-\Sigma
)]_{\alpha \beta \rho \sigma }(x_1,x_2;y_1,y_2)=-\delta _{\alpha \rho
}\delta ^4(x_1-y_1)S_F^c(x_2-y_2)_{\beta \sigma }$ \\ 
$+{\cal G}_\mu ^a(y_1\mid x_1,x_2;y_1,y_2)_{\alpha \beta \tau \sigma
}(\Gamma ^{a\mu })_{\tau \rho }-\int d^4z_1{\cal G}(x_1,x_2;z_1,y_2)_{\alpha
\beta \tau \sigma }\Sigma (z_1,y_1)_{\sigma \rho }.$%
\end{tabular}
\eqnum{3.8}
\end{equation}

In order to derive the B-S equation, we need to operate on the both sides of
the above equation with the operator $(i\overleftarrow{{\bf \partial }}%
_{y_2}+m_2-\Sigma ^c)$, 
\begin{equation}
\begin{tabular}{l}
$\lbrack {\cal G}(i\overleftarrow{{\bf \partial }}_{y_1}+m_1-\Sigma )(i%
\overleftarrow{{\bf \partial }}_{y_2}+m_2-\Sigma ^c)]_{\alpha \beta \rho
\sigma }(x_1,x_2;y_1,y_2)=\delta _{\alpha \rho }\delta _{\beta \sigma
}\delta ^4(x_1-y_1)\delta ^4(x_2-y_2)$ \\ 
$+[{\cal G}_\mu ^a(i\overleftarrow{{\bf \partial }}_{y_2}+m_2-\Sigma
^c)]_{\alpha \beta \tau \sigma }(y_1\mid x_1,x_2;y_1,y_2)(\Gamma ^{a\mu
})_{\tau \rho }$ \\ 
$-\int d^4z_1[{\cal G(}i\overleftarrow{{\bf \partial }}_{y_2}+m_2-\Sigma
^c)]_{\alpha \beta \tau \sigma }(x_1,x_2;z_1,y_2)\Sigma (z_1,y_1)_{\tau \rho
}$%
\end{tabular}
\eqnum{3.9}
\end{equation}
where Eq. (3.4) has been used. For computing the third term in the equation,
we start from the equation of motion which is obtained from Eq. (D8) with
setting $J=0$ and supplementing the self-energy-related term 
\begin{equation}
\begin{tabular}{l}
$\lbrack G(i\overleftarrow{{\bf \partial }}_{y_2}+m_2-\Sigma ^c)]_{\alpha
\beta \rho \sigma }(x_1,x_2;y_1,y_2)=-\delta _{\beta \sigma }\delta
^4(x_2-y_2)S_F(x_1-y_1)_{\alpha \rho }$ \\ 
$-C_{\rho \sigma }\delta ^4(y_1-y_2)S_F^{*}(x_1-x_2)_{\alpha \beta }+G_\nu
^b(y_2\mid x_1,x_2;y_1,y_2)_{\alpha \beta \rho \delta }(\overline{\Gamma }%
^{b\nu })_{\delta \sigma }$ \\ 
$-\int d^4z_2G(x_1,x_2;y_1,z_2)_{\alpha \beta \rho \delta }\Sigma
^c(z_2,y_2)_{\delta \sigma }.$%
\end{tabular}
\eqnum{3.10}
\end{equation}
By applying the relations shown in Eqs. (2.4) and (3.6) to Eq. (3.10) and
considering the following equation 
\begin{equation}
\lbrack \overline{S}_F^{*}(i\overleftarrow{{\bf \partial }}_{y_2}+m_2-\Sigma
^c)]_{\rho \sigma }(y_1,y_2)=C_{\rho \sigma }\delta ^4(y_1-y_2)  \eqnum{3.11}
\end{equation}
which is given by Eqs. (D14)-(D16) and the relations represented in Eq.
(2.15), one can get such an equation that 
\begin{equation}
\begin{tabular}{l}
$\lbrack {\cal G}(i\overleftarrow{{\bf \partial }}_{y_2}+m_2-\Sigma
^c)]_{\alpha \beta \rho \sigma }(x_1,x_2;y_1,y_2)=-\delta _{\beta \sigma
}\delta ^4(x_2-y_2)S_F(x_1-y_1)_{\alpha \rho }$ \\ 
$+{\cal G}_\nu ^b(y_2\mid x_1,x_2;y_1,y_2)_{\alpha \beta \rho \delta }(%
\overline{\Gamma }^{b\nu })_{\delta \sigma }-\int d^4z_2{\cal G}%
(x_1,x_2;y_1,z_2)_{\alpha \beta \rho \delta }\Sigma ^c(z_2,y_2)_{\delta
\sigma }.$%
\end{tabular}
\eqnum{3.12}
\end{equation}

To calculate the second term in Eq. (3.9), let us operate on the both sides
of Eq. (3.6) with $i=1$ by the operator $(i\overleftarrow{{\bf \partial }}%
_{y_2}+m_2-\Sigma ^c)$ 
\begin{equation}
\begin{tabular}{l}
$\lbrack {\cal G}_\mu ^a(i\overleftarrow{{\bf \partial }}_{y_2}+m_2-\Sigma
^c)]_{\alpha \beta \rho \sigma }(y_1\mid x_1,x_2;y_1,y_2)$ \\ 
$=[G_\mu ^a(i\overleftarrow{{\bf \partial }}_{y_2}+m_2-\Sigma ^c)]_{\alpha
\beta \rho \sigma }(y_1\mid x_1,x_2;y_1,y_2)$ \\ 
$+S_F^{*}(x_1-x_2)_{\alpha \beta }[\overline{\Lambda }_\mu ^{*a}(i%
\overleftarrow{{\bf \partial }}_{y_2}+m_2-\Sigma ^c)]_{\rho \sigma }(y_1\mid
y_1,y_2).$%
\end{tabular}
\eqnum{3.13}
\end{equation}
The first term can be calculated by virtue of the equation of motion derived
in Eq. (D9). With addition of the self-energy-relevant term to Eq. (D9), we
have 
\begin{equation}
\begin{tabular}{l}
$\lbrack G_\mu ^a(i\overleftarrow{{\bf \partial }}_{y_2}+m_2-\Sigma
^c)]_{\alpha \beta \rho \sigma }(y_1\mid x_1,x_2;y_1,y_2)=-\delta _{\beta
\sigma }\delta ^4(x_2-y_2)\Lambda _\mu ^a(y_1\mid x_1,y_1)_{\alpha \rho }$
\\ 
$-C_{\rho \sigma }\delta ^4(y_1-y_2)\Lambda _\mu ^{*a}(y_1\mid
x_1,x_2)_{\alpha \beta }+G_{\mu \nu }^{ab}(y_1,y_2\mid
x_1,x_2;y_1,y_2)_{\alpha \beta \rho \delta }(\overline{\Gamma }^{b\nu
})_{\delta \sigma }$ \\ 
$-\int d^4z_2G_\mu ^a(y_1\mid x_1,x_2;y_1,z_2)_{\alpha \beta \rho \delta
}\Sigma ^c(z_2,y_2)_{\delta \sigma }$%
\end{tabular}
\eqnum{3.14}
\end{equation}
where the Green functions have been defined in Eqs. (2.18), (2.20) and
(2.22). In accordance with Eq.(2.23), one can write 
\begin{equation}
\begin{tabular}{l}
${\cal G}_{\mu \nu }^{ab}(y_1,y_2\mid x_1,x_2;y_1,y_2)_{\alpha \beta \rho
\sigma }=G_{\mu \nu }^{ab}(y_1,y_2\mid x_1,x_2;y_1,y_2)_{\alpha \beta \rho
\sigma }$ \\ 
$+S_F^{*}(x_1-x_2)_{\alpha \beta }\overline{\Lambda }_{\mu \nu
}^{*ab}(y_1,y_2\mid y_1,y_2)_{\rho \sigma }$%
\end{tabular}
\eqnum{3.15}
\end{equation}
where $\overline{\Lambda }_{\mu \nu }^{*ab}(y_1,y_2\mid y_1,y_2)_{\rho
\sigma }$ was defined in Eq. (2.21). When the relations given in Eqs. (3.6)
and (3.15) are substituted into Eq. (3.14), we are led to 
\begin{equation}
\begin{tabular}{l}
$\lbrack G_\mu ^a(i\overleftarrow{{\bf \partial }}_{y_2}+m_2-\Sigma
^c)]_{\alpha \beta \rho \sigma }(y_1\mid x_1,x_2;y_1,y_2)=-\delta _{\beta
\sigma }\delta ^4(x_2-y_2)\Lambda _\mu ^a(y_1\mid x_1,y_1)_{\alpha \rho }$
\\ 
$-C_{\rho \sigma }\delta ^4(y_1-y_2)\Lambda _\mu ^{*a}(y_1\mid
x_1,x_2)_{\alpha \beta }+{\cal G}_{\mu \nu }^{ab}(y_1,y_2\mid
x_1,x_2;y_1,y_2)_{\alpha \beta \rho \delta }(\overline{\Gamma }^{b\nu
})_{\delta \sigma }$ \\ 
$-\int d^4z_2{\cal G}_\mu ^a(y_1\mid x_1,x_2;y_1,z_2)_{\alpha \beta \rho
\delta }\Sigma ^c(z_2,y_2)_{\delta \sigma }-S_F^{*}(x_1-x_2)_{\alpha \beta }%
\overline{\Lambda }_{\mu \nu }^{*ab}(y_1,y_2\mid y_1,y_2)_{\rho \delta }(%
\overline{\Gamma }^{b\nu })_{\delta \sigma }$ \\ 
$+S_F^{*}(x_1-x_2)_{\alpha \beta }\int d^4z_2\overline{\Lambda }_\mu
^{*a}(y_1\mid y_1,z_2)_{\rho \delta }\Sigma ^c(z_2,y_2)_{\delta \sigma }.$%
\end{tabular}
\eqnum{3.16}
\end{equation}
The second term in Eq. (3.13) is directly determined by the following
equation of motion 
\begin{equation}
\begin{tabular}{l}
$\lbrack \overline{\Lambda }_\mu ^{*a}(i\overleftarrow{{\bf \partial }}%
_{y_2}+m_2-\Sigma ^c)]_{\rho \sigma }(y_1\mid y_1,y_2)=\overline{\Lambda }%
_{\mu \nu }^{*ab}(y_1,y_2\mid y_1,y_2)_{\rho \delta }(\overline{\Gamma }%
^{b\nu })_{\delta \sigma }$ \\ 
$-\int d^4z_2\overline{\Lambda }_\mu ^{*a}(y_1\mid y_1,z_2)_{\rho \delta
}\Sigma ^c(z_2,y_2)_{\delta \sigma }$%
\end{tabular}
\eqnum{3.17}
\end{equation}
which is given in Eq. (D17) with setting $J=0$ and adding the term
associated with the self-energy. On inserting Eqs. (3.16) and (3.17) into
Eq. (3.13), it is easy to find 
\begin{equation}
\begin{tabular}{l}
$\lbrack {\cal G}_\mu ^a(i\overleftarrow{{\bf \partial }}_{y_2}+m_2-\Sigma
^c)]_{\alpha \beta \rho \sigma }(y_1\mid x_1,x_2;y_1,y_2)=-\delta _{\beta
\sigma }\delta ^4(x_2-y_2)\Lambda _\mu ^a(y_1\mid x_1,y_1)_{\alpha \rho }$
\\ 
$-C_{\rho \sigma }\delta ^4(y_1-y_2)\Lambda _\mu ^{*a}(y_1\mid
x_1,x_2)_{\alpha \beta }+{\cal G}_{\mu \nu }^{ab}(y_1,y_2\mid
x_1,x_2;y_1,y_2)_{\alpha \beta \rho \delta }(\overline{\Gamma }^{b\nu
})_{\delta \sigma }$ \\ 
$-\int d^4z_2{\cal G}_\mu ^a(y_1\mid x_1,x_2;y_1,z_2)_{\alpha \beta \rho
\delta }\Sigma ^c(z_2,y_2)_{\delta \sigma }.$%
\end{tabular}
\eqnum{3.18}
\end{equation}

Substitution of Eqs. (3.12) and (3.18) in Eq. (3.9) yields 
\begin{equation}
\begin{tabular}{l}
$\lbrack {\cal G}(i\overleftarrow{{\bf \partial }}_{y_1}+m_1-\Sigma )(i%
\overleftarrow{{\bf \partial }}_{y_2}+m_2-\Sigma ^c)]_{\alpha \beta \rho
\sigma }(x_1,x_2;y_1,y_2)$ \\ 
$=\delta _{\alpha \rho }\delta _{\beta \sigma }\delta ^4(x_1-y_1)\delta
^4(x_2-y_2)+{\cal H}_2(x_1,x_2;y_1,y_2)_{\alpha \beta \rho \sigma }$%
\end{tabular}
\eqnum{3.19}
\end{equation}
where 
\begin{equation}
{\cal H}_2(x_1,x_2;y_1,y_2)_{\alpha \beta \rho \sigma }=\sum_{j=1}^5{\cal H}%
_2^{(j)}(x_1,x_2;y_1,y_2)_{\alpha \beta \rho \sigma }  \eqnum{3.20}
\end{equation}
in which 
\begin{equation}
{\cal H}_2^{(1)}(x_1,x_2;y_1,y_2)_{\alpha \beta \rho \sigma }=-\int d^4z_2%
{\cal G}_\mu ^a(y_1\mid x_1,x_2;y_1,z_2)_{\alpha \beta \tau \delta }\Sigma
^c(z_2,y_2)_{\delta \sigma }(\Gamma ^{a\mu })_{\tau \rho },  \eqnum{3.21}
\end{equation}
\begin{equation}
{\cal H}_2^{(2)}(x_1,x_2;y_1,y_2)_{\alpha \beta \rho \sigma }=-\int d^4z_1%
{\cal G}_\nu ^b(y_2\mid x_1,x_2;z_1,y_2)_{\alpha \beta \tau \delta }\Sigma
(z_1,y_1)_{\tau \rho }(\overline{\Gamma }^{b\nu })_{\delta \sigma }, 
\eqnum{3.22}
\end{equation}
\begin{equation}
{\cal H}_2^{(3)}(x_1,x_2;y_1,y_2)_{\alpha \beta \rho \sigma }={\cal G}_{\mu
\nu }^{ab}(y_1,y_2\mid x_1,x_2;y_1,y_2)_{\alpha \beta \tau \delta }(\Gamma
^{a\mu })_{\tau \rho }(\overline{\Gamma }^{b\nu })_{\delta \sigma }, 
\eqnum{3.23}
\end{equation}
\begin{equation}
{\cal H}_2^{(4)}(x_1,x_2;y_1,y_2)_{\alpha \beta \rho \sigma }=\delta
^4(y_1-y_2)\Lambda _\mu ^{*a}(y_1\mid x_1,x_2)_{\alpha \beta }(C\Gamma
^{a\mu })_{\sigma \rho }  \eqnum{3.24}
\end{equation}
and 
\begin{equation}
{\cal H}_2^{(5)}(x_1,x_2;y_1,y_2)_{\alpha \beta \rho \sigma }=\int
d^4z_1d^4z_2{\cal G}(x_1,x_2;z_1,z_2)_{\alpha \beta \tau \delta }\Sigma
(z_1,y_1)_{\tau \rho }\Sigma ^c(z_2,y_2)_{\delta \sigma }.  \eqnum{3.25}
\end{equation}

By applying the B-S equation given in Eq. (3.19), Eq. (3.1) can be written
as 
\begin{equation}
\begin{tabular}{l}
$K(x_1,x_2;y_1,y_2)_{\alpha \beta \rho \sigma }={\cal Q}(x_1,x_2;y_1,y_2)_{%
\alpha \beta \rho \sigma }$ \\ 
$-\int d^4z_1d^4z_2K(x_1,x_2;z_1,z_2)_{\alpha \beta \tau \delta }{\cal H}%
_2(z_1,z_2;y_1,y_2)_{\tau \delta \rho \sigma }.$%
\end{tabular}
\eqnum{3.26}
\end{equation}
To reach a closed expression of the B-S kernel, it is necessary to eliminate
the kernel appearing in the second term on the RHS of Eq. (3.26). For this
purpose, we operate on the both sides of Eq. (2.41) with the inverse of the
Green function ${\cal G}(x_1,x_2;y_1,y_2)$. With this operation , noticing 
\begin{equation}
\int d^4z_1d^4z_2{\cal G}(x_1,x_2;z_1,z_2)_{\alpha \beta \gamma \lambda }%
{\cal G}^{-1}(z_1,z_2;y_1,y_2)_{\gamma \lambda \rho \sigma }=\delta _{\alpha
\rho }\delta _{\beta \sigma }\delta ^4(x_1-y_1)\delta ^4(x_2-y_2), 
\eqnum{3.27}
\end{equation}
one gets 
\begin{equation}
K(x_1,x_2;z_1,z_2)_{\alpha \beta \tau \delta }=\int d^4u_1d^4u_2{\cal H}%
_1(x_1,x_2;u_1,u_2)_{\alpha \beta \gamma \lambda }{\cal G}%
^{-1}(u_1,u_2;z_1,z_2)_{\gamma \lambda \tau \delta }.  \eqnum{3.28}
\end{equation}
Upon inserting the above expression into Eq. (3.26), we finally arrive at 
\begin{equation}
K(x_1,x_2;y_1,y_2)_{\alpha \beta \rho \sigma }={\cal Q}(x_1,x_2;y_1,y_2)_{%
\alpha \beta \rho \sigma }-{\cal S}(x_1,x_2;y_1,y_2)_{\alpha \beta \rho
\sigma }  \eqnum{3.29}
\end{equation}
where 
\begin{equation}
\begin{tabular}{l}
${\cal S}(x_1,x_2;y_1,y_2)_{\alpha \beta \rho \sigma }$ \\ 
$=\int d^4u_1d^4u_2d^4v_1d^4v_2{\cal H}_1(x_1,x_2;u_1,u_2)_{\alpha \beta
\gamma \lambda }{\cal G}^{-1}(u_1,u_2;v_1,v_2)_{\gamma \lambda \tau \delta }%
{\cal H}_2(v_1,v_2;y_1,y_2)_{\tau \delta \rho \sigma }.$%
\end{tabular}
\eqnum{3.30}
\end{equation}
As we see, the B-S kernel shown above contains two terms. The second term
has been explicitly written out. It contains a few types of the Green
functions as well as the self-energies appearing in the mutually conjugate
functions ${\cal H}_1(x_1,x_2;y_1,y_2)$ and ${\cal H}_2(x_1,x_2;y_1,y_2)$
shown in Eqs. (2.35)-(2.40) and (3. 20)-(3.25) respectively. Clearly, to
give a final expression of the B-S kernel, according to the definition in
Eq.(3.2), we need to calculate the function ${\cal Q}(x_1,x_2;y_1,y_2)$ by
means of the B-S equations with respect to $y_1$ and $y_2$ for the Green
functions involved in the function ${\cal H}_1(x_1,x_2;y_1,y_2).$ We leave
the calculations in the next section.

\section{Closed expression of the B-S kernel}

This section is used to calculate the function ${\cal Q}(x_1,x_2;y_1,y_2)$
so as to give the final expression of the B-S kernel. In accordance with the
expressions shown in Eqs. (3.2) and (2.35), we can write 
\begin{equation}
{\cal Q}(x_1,x_2;y_1,y_2)_{\alpha \beta \rho \sigma }=\sum\limits_{i=1}^5%
{\cal Q}_i(x_1,x_2;y_1,y_2)_{\alpha \beta \rho \sigma }  \eqnum{4.1}
\end{equation}
where 
\begin{equation}
{\cal Q}_i(x_1,x_2;y_1,y_2)_{\alpha \beta \rho \sigma }=[{\cal H}_1^{(i)}(i%
\overleftarrow{{\bf \partial }}_{y_1}+m_1-\Sigma )(i\overleftarrow{{\bf %
\partial }}_{y_2}+m_2-\Sigma ^c)]_{\alpha \beta \rho \sigma
}(x_1,x_2;y_1,y_2).  \eqnum{4.2}
\end{equation}
This expression indicates that to compute each ${\cal Q}_i(x_1,x_2;y_1,y_2)$%
, we need to derive the B-S equations with respect to $y_{1\text{ }}$and $%
y_2 $ for the Green functions contained in the function ${\cal H}%
_1(x_1,x_2;y_1,y_2)$. We will separately calculate each ${\cal Q}%
_i(x_1,x_2;y_1,y_2)$ below.

{\bf a. Calculation of }${\cal Q}_1(x_1,x_2;y_1,y_2)$

Based on Eqs. (4.2) and (2.36), we can write 
\begin{equation}
\begin{tabular}{l}
${\cal Q}_1(x_1,x_2;y_1,y_2)_{\alpha \beta \rho \sigma }$ \\ 
$=-(\Gamma ^{a\mu })_{\alpha \gamma }\int d^4u_2\Sigma ^c(x_2,u_2)_{\beta
\lambda }[{\cal G}_\mu ^a(i\overleftarrow{{\bf \partial }}_{y_1}+m_1-\Sigma
)(i\overleftarrow{{\bf \partial }}_{y_2}+m_2-\Sigma ^c)]_{\gamma \lambda
\rho \sigma }(x_1\mid x_1,u_2;y_1,y_2).$%
\end{tabular}
\eqnum{4.3}
\end{equation}
To calculate the above function, it is necessary to start from the following
equation of motion which is established from Eq. (C6) by setting the source $%
J=0$, 
\begin{equation}
\begin{tabular}{l}
$G_\mu ^a(x_1\mid x_1,u_2;y_1,y_2)_{\gamma \lambda \tau \sigma }(i%
\overleftarrow{{\bf \partial }}_{y_1}+m_1)_{\tau \rho }=-\delta _{\gamma
\rho }\delta ^4(x_1-y_1)\Lambda _\mu ^{{\bf c}a}(x_1\mid u_2,y_2)_{\lambda
\sigma }$ \\ 
$-C_{\rho \sigma }\delta ^4(y_1-y_2)\Lambda _\mu ^{*a}(x_1\mid
x_1,u_2)_{\gamma \lambda }+G_{\mu \nu }^{ab}(x_1;y_1\mid
x_1,u_2;y_1,y_2)_{\gamma \lambda \tau \sigma }(\Gamma ^{a\mu })_{\tau \rho }$%
\end{tabular}
\eqnum{4.4}
\end{equation}
where the Green functions have been represented respectively in Eqs. (2.19),
(2.20) and (2.22). Considering the definitions given in Eq. (2.26) and in
the following 
\begin{equation}
\begin{tabular}{l}
${\cal G}_{\mu \nu }^{ab}(x_i;y_j\mid x_1,u_2;y_1,y_2)_{\gamma \lambda \rho
\sigma }=G_{\mu \nu }^{ab}(x_i;y_j\mid x_1,u_2;y_1,y_2)_{\gamma \lambda \rho
\sigma }$ \\ 
$+\Lambda _\mu ^{*a}(x_i\mid x_1,u_2)_{\gamma \lambda }\overline{\Lambda }%
_\nu ^{*b}(y_j\mid y_1,y_2)_{\rho \sigma }$%
\end{tabular}
\eqnum{4.5}
\end{equation}
($i,j=1,2$) which immediately follows from Eq. (2.23) and employing Eqs.
(3.7) and (4.4), it is not difficult to derive 
\begin{equation}
\begin{tabular}{l}
$\lbrack {\cal G}_\mu ^a(i\overleftarrow{{\bf \partial }}_{y_1}+m_1-\Sigma
)]_{\gamma \lambda \rho \sigma }(x_1\mid x_1,u_2;y_1,y_2)=-\delta _{\gamma
\rho }\delta ^4(x_1-y_1)\Lambda _\mu ^{{\bf c}a}(x_1\mid u_2,y_2)_{\lambda
\sigma }$ \\ 
$+{\cal G}_{\mu \nu }^{ab}(x_1;y_1\mid x_1,u_2;y_1,y_2)_{\gamma \lambda \tau
\sigma }(\Gamma ^{a\mu })_{\tau \rho }-\int d^4v_1{\cal G}_\mu ^a(x_1\mid
x_1,u_2;v_1,y_2)_{\gamma \lambda \tau \sigma }\Sigma (v_1,y_1)_{\tau \rho }.$%
\end{tabular}
\eqnum{4.6}
\end{equation}

Now, let us operate on the above equation with $(i\overleftarrow{{\bf %
\partial }}_{y_2}+m_2-\Sigma ^c),$ giving 
\begin{equation}
\begin{tabular}{l}
$\lbrack {\cal G}_\mu ^a(i\overleftarrow{{\bf \partial }}_{y_1}+m_1-\Sigma
)(i\overleftarrow{{\bf \partial }}_{y_2}+m_2-\Sigma ^c)]_{\gamma \lambda
\rho \sigma }(x_1\mid x_1,u_2;y_1,y_2)$ \\ 
$=-\delta _{\gamma \rho }\delta ^4(x_1-y_1)[\Lambda _\mu ^{{\bf c}a}(i%
\overleftarrow{{\bf \partial }}_{y_2}+m_2-\Sigma ^c)]_{\lambda \sigma
}(x_1\mid u_2,y_2)$ \\ 
$+[{\cal G}_{\mu \nu }^{ab}(i\overleftarrow{{\bf \partial }}%
_{y_2}+m_2-\Sigma ^c)]_{\gamma \lambda \tau \sigma }(x_1;y_1\mid
x_1,u_2;y_1,y_2)(\Gamma ^{b\nu })_{\tau \rho }$ \\ 
$-\int d^4v_1[{\cal G}_\mu ^a(i\overleftarrow{{\bf \partial }}%
_{y_2}+m_2-\Sigma ^c)]_{\gamma \lambda \tau \sigma }(x_1\mid
x_1,u_2;v_1,y_2)\Sigma (v_1,y_1)_{\tau \rho }.$%
\end{tabular}
\eqnum{4.7}
\end{equation}
The first term can be calculated by using the equation derived in Eq. (D5), 
\begin{equation}
\begin{tabular}{l}
$\lbrack \Lambda _\mu ^{{\bf c}a}(i\overleftarrow{{\bf \partial }}%
_{y_2}+m_2-\Sigma ^c)]_{\lambda \sigma }(x_1\mid u_2,y_2)=\Lambda _{\mu \nu
}^{{\bf c}ab}(x_1;y_2\mid u_2,y_2)_{\lambda \delta }(\overline{\Gamma }%
^{b\nu })_{\delta \sigma }$ \\ 
$-\int d^4v_2\Lambda _\mu ^{ca}(x_1\mid u_2,v_2)_{\lambda \delta }\Sigma
^c(v_2,y_2)_{\delta \sigma }$%
\end{tabular}
\eqnum{4.8}
\end{equation}
where the two Green functions were represented in Eq. (2.19). The third term
in Eq. (4.7) can be calculated by virtue of the equation given in Eq. (D10) 
\begin{equation}
\begin{tabular}{l}
$G_\mu ^a(x_1\mid x_{1,}u_2;v_1,y_2)_{\gamma \lambda \tau \delta }(i%
\overleftarrow{{\bf \partial }}_{y_2}+m_2)_{\delta \sigma }=-\delta
_{\lambda \sigma }\delta ^4(u_2-y_2)\Lambda _\mu ^a(x_1\mid x_1,v_1)_{\gamma
\tau }$ \\ 
$-C_{\tau \sigma }\delta ^4(v_1-y_2)\Lambda _\mu ^{*a}(x_1\mid
x_1,u_2)_{\gamma \lambda }+G_{\mu \nu }^{ab}(x_1;y_2\mid
x_1,u_2;v_1,y_2)_{\gamma \lambda \tau \delta }(\overline{\Gamma }^{b\nu
})_{\delta \sigma }$%
\end{tabular}
\eqnum{4.9}
\end{equation}
in which the Green functions have been respectively defined in Eqs. (2.18),
(2.20) and (2.22). In view of the definitions in Eqs. (2.26) and (4.5) and
employing the equations in Eqs. (4.9) and (3.11), it can be found 
\begin{equation}
\begin{tabular}{l}
$\lbrack {\cal G}_\mu ^a(i\overleftarrow{{\bf \partial }}_{y_2}+m_2-\Sigma
^c)]_{\gamma \lambda \tau \sigma }(x_1\mid x_{1,}u_2;v_1,y_2)=-\delta
_{\lambda \sigma }\delta ^4(u_2-y_2)\Lambda _\mu ^a(x_1\mid x_1,v_1)_{\gamma
\tau }$ \\ 
$+{\cal G}_{\mu \nu }^{ab}(x_1;y_2\mid x_1,u_2;v_1,y_2)_{\gamma \lambda \tau
\delta }(\overline{\Gamma }^{b\nu })_{\delta \sigma }-\int d^4v_2{\cal G}%
_\mu ^a(x_1\mid x_{1,}u_2;v_1,v_2)_{\gamma \lambda \tau \delta }\Sigma
^c(v_2,y_2)_{\delta \sigma }.$%
\end{tabular}
\eqnum{4.10}
\end{equation}
For calculating the second term in Eq. (4.7), it is needed to use the
following equation obtained from Eq. (D11) by setting $J=0$ 
\begin{equation}
\begin{tabular}{l}
$\lbrack G_{\mu \nu }^{ab}(i\overleftarrow{{\bf \partial }}%
_{y_2}+m_2)]_{\gamma \lambda \tau \sigma }(x_1;y_1\mid
x_1,u_2;y_1,y_2)_{\gamma \lambda \tau \sigma }=-\delta _{\lambda \sigma
}\delta ^4(u_2-y_2)\Lambda _{\mu \nu }^{ab}(x_1;y_1\mid x_1,y_1)_{\gamma
\tau }$ \\ 
$-C_{\tau \sigma }\delta ^4(y_1-y_2)\Lambda _{\mu \nu }^{*ab}(x_1;y_1\mid
x_1,u_2)_{\gamma \lambda }+G_{\mu \nu \kappa }^{abc}(x_1;y_1,y_2\mid
x_1,u_2;y_1,y_2)_{\gamma \lambda \tau \delta }(\overline{\Gamma }^{c\kappa
})_{\delta \sigma }$%
\end{tabular}
\eqnum{4.11}
\end{equation}
where all the Green functions were defined respectively in Eqs. (2.18),
(2.20) and (2.22). By considering the definitions given in Eq. (4.5) and in
the following 
\begin{equation}
\begin{tabular}{l}
${\cal G}_{\mu \nu \kappa }^{abc}(x_1;y_1,y_2\mid x_1,u_2;y_1,y_2)_{\gamma
\lambda \tau \delta }=G_{\mu \nu \kappa }^{abc}(x_1;y_1,y_2\mid
x_1,u_2;y_1,y_2)_{\gamma \lambda \tau \delta }$ \\ 
$+\Lambda _\mu ^{*a}(x_1\mid x_1,u_2)_{\gamma \lambda }\overline{\Lambda }%
_{\nu \kappa }^{*bc}(y_1,y_2\mid y_1,y_2)_{\tau \sigma }$%
\end{tabular}
\eqnum{4.12}
\end{equation}
which is implied by Eq. (2.23) and utilizing Eqs. (4.11) and (3.17), an
equation of motion for the Green function ${\cal G}_{\mu \nu
}^{ab}(x_1;y_1\mid x_1,u_2;y_1,y_2)$ will be found to be 
\begin{equation}
\begin{tabular}{l}
$\lbrack {\cal G}_{\mu \nu }^{ab}(i\overleftarrow{{\bf \partial }}%
_{y_2}+m_2-\Sigma ^c)]_{\gamma \lambda \tau \sigma }(x_1;y_1\mid
x_1,u_2;y_1,y_2)_{\gamma \lambda \tau \sigma }=-\delta _{\lambda \sigma
}\delta ^4(u_2-y_2)\Lambda _{\mu \nu }^{ab}(x_1;y_1\mid x_1,y_1)_{\gamma
\tau }$ \\ 
$-C_{\tau \sigma }\delta ^4(y_1-y_2)\Lambda _{\mu \nu }^{*ab}(x_1;y_1\mid
x_1,u_2)_{\gamma \lambda }+{\cal G}_{\mu \nu \kappa }^{abc}(x_1;y_1,y_2\mid
x_1,u_2;y_1,y_2)_{\gamma \lambda \tau \delta }(\overline{\Gamma }^{c\kappa
})_{\delta \sigma }$ \\ 
$-\int d^4v_2{\cal G}_{\mu \nu }^{ab}(x_1;y_1\mid x_{1,}u_2;y_1,v_2)_{\gamma
\lambda \tau \delta }\Sigma ^c(v_2,y_2)_{\delta \sigma }.$%
\end{tabular}
\eqnum{4.13}
\end{equation}

When Eqs. (4.8), (4.10) and (4.13) are inserted into Eq. (4.7) and then Eq.
(4.7) is inserted into Eq. (4.3), an explicit expression of the function $%
{\cal Q}_1(x_1,x_2;y_1,y_2)$ will be derived as given in the following 
\begin{equation}
\begin{tabular}{l}
${\cal Q}_1(x_1,x_2;y_1,y_2)_{\alpha \beta \rho \sigma }={\cal R}%
_1(x_1,x_2;y_1,y_2)_{\alpha \beta \rho \sigma }+{\cal M}%
_1(x_1,x_2;y_1,y_2)_{\alpha \beta \rho \sigma }$ \\ 
$+\int d^4v_1d^4v_2{\cal H}_1^{(1)}(x_1,x_2;v_1,v_2)_{\alpha \beta \tau
\delta }\Sigma (v_1,y_1)_{\tau \rho }\Sigma ^c(v_2,y_2)_{\delta \sigma }$%
\end{tabular}
\eqnum{4.14}
\end{equation}
where 
\begin{equation}
\begin{tabular}{l}
${\cal R}_1(x_1,x_2;y_1,y_2)_{\alpha \beta \rho \sigma }=\delta
^4(x_1-y_1)(\Gamma ^{a\mu })_{\alpha \rho }\int d^4u_2\Sigma
^c(x_2,u_2)_{\beta \lambda }[\Lambda _{\mu \nu }^{{\bf c}ab}(x_1;y_2\mid
u_2,y_2)_{\lambda \delta }(\overline{\Gamma }^{b\nu })_{\delta \sigma }$ \\ 
$-\int d^4v_2\Lambda _\mu ^{{\bf c}a}(x_1\mid u_2,v_2)_{\lambda \delta
}\Sigma ^c(v_2,y_2)_{\delta \sigma }]+\Sigma ^c(x_2,y_2)_{\beta \sigma
}(\Gamma ^{a\mu })_{\alpha \gamma }[\Lambda _{\mu \nu }^{ab}(x_1;y_1\mid
x_1,y_1)_{\gamma \tau }(\Gamma ^{b\nu })_{\tau \rho }$ \\ 
$-\int d^4v_1\Lambda _\mu ^a(x_1\mid x_1,v_1)_{\gamma \tau }\Sigma
(v_1,y_1)_{\tau \rho }]$ \\ 
$-\delta ^4(y_1-y_2)(\Gamma ^{a\mu })_{\alpha \gamma }\int d^4u_2\Sigma
^c(x_2,u_2)_{\beta \lambda }\Lambda _{\mu \nu }^{*ab}(x_1;y_1\mid
x_1,u_2)_{\gamma \lambda }(C\Gamma ^{b\nu })_{\sigma \rho },$%
\end{tabular}
\eqnum{4.15}
\end{equation}
\begin{equation}
\begin{tabular}{l}
${\cal M}_1(x_1,x_2;y_1,y_2)_{\alpha \beta \rho \sigma }=-(\Gamma ^{a\mu
})_{\alpha \gamma }\int d^4u_2\Sigma ^c(x_2,u_2)_{\beta \lambda }[{\cal G}%
_{\mu \nu \kappa }^{abc}(x_1;y_1,y_2\mid x_1,u_2;y_1,y_2)_{\gamma \lambda
\tau \delta }(\Gamma ^{b\nu })_{\tau \rho }(\overline{\Gamma }^{c\kappa
})_{\delta \sigma }$ \\ 
$-\int d^4v_2{\cal G}_{\mu \nu }^{ab}(x_1;y_1\mid x_{1,}u_2;y_1,v_2)_{\gamma
\lambda \tau \delta }\Sigma ^c(v_2,y_2)_{\delta \sigma }(\Gamma ^{b\nu
})_{\tau \rho }$ \\ 
$-\int d^4v_1{\cal G}_{\mu \nu }^{ab}(x_1;y_2\mid x_{1,}u_2;v_1,y_2)_{\gamma
\lambda \tau \delta }\Sigma (v_1,y_1)_{\tau \rho }(\overline{\Gamma }^{b\nu
})_{\delta \sigma }]$%
\end{tabular}
\eqnum{4.16}
\end{equation}
and the function ${\cal H}_1^{(1)}(x_1,x_2;v_1,v_2)$ was defined in Eq.
(2.36).

{\bf b. Calculation of }${\cal Q}_2(x_1,x_2;y_1,y_2)$

From Eqs. (4.2) and (2.37), one can write 
\begin{equation}
\begin{tabular}{l}
${\cal Q}_2(x_1,x_2;y_1,y_2)_{\alpha \beta \rho \sigma }$ \\ 
$=-(\overline{\Gamma }^{a\mu })_{\beta \lambda }\int d^4u_1\Sigma
(x_1,u_1)_{\alpha \gamma }[{\cal G}_\mu ^a(i\overleftarrow{{\bf \partial }}%
_{y_1}+m_1-\Sigma )(i\overleftarrow{{\bf \partial }}_{y_2}+m_2-\Sigma
^c)]_{\gamma \lambda \rho \sigma }(x_2\mid u_1,x_2;y_1,y_2)$%
\end{tabular}
\eqnum{4.17}
\end{equation}
In comparison of the above expression with that denoted in Eq. (4.3), it is
clearly seen that the function ${\cal Q}_2(x_1,x_2;y_1,y_2)_{\alpha \beta
\rho \sigma }$ can explicitly be written out from the expression of ${\cal Q}%
_1(x_1,x_2;y_1,y_2)_{\alpha \beta \rho \sigma }$ shown in Eqs. (4.14)-(4.16)
by the following replacements: $(\Gamma ^{a\mu })_{\alpha \gamma }\int
d^4u_2\Sigma ^c(x_2,u_2)_{\beta \lambda }\rightarrow (\overline{\Gamma }%
^{a\mu })_{\beta \lambda }\int d^4u_1\Sigma (x_1,u_1)_{\alpha \gamma }$ and
in the $[{\cal G}_\mu ^a(i\overleftarrow{{\bf \partial }}_{y_1}+m_1-\Sigma
)(i\overleftarrow{{\bf \partial }}_{y_2}+m_2-\Sigma ^c)]_{\gamma \lambda
\rho \sigma }(x_1\mid x_1,u_2;y_1,y_2)$ whose expression is given in Eqs.
(4.7), (4.8), (4.10) and (4.13), the gluon field coordinate $x_1$ is changed
to $x_2$ and the quark and antiquark coordinates $x_1$ and $u_2$ are
respectively changed to $u_1$ and $x_2$. The result is 
\begin{equation}
\begin{tabular}{l}
${\cal Q}_2(x_1,x_2;y_1,y_2)_{\alpha \beta \rho \sigma }={\cal R}%
_2(x_1,x_2;y_1,y_2)_{\alpha \beta \rho \sigma }+{\cal M}%
_2(x_1,x_2;y_1,y_2)_{\alpha \beta \rho \sigma }$ \\ 
$+\int d^4v_1d^4v_2{\cal H}_1^{(2)}(x_1,x_2;v_1,v_2)_{\alpha \beta \tau
\delta }\Sigma (v_1,y_1)_{\tau \rho }\Sigma ^c(v_2,y_2)_{\delta \sigma }$%
\end{tabular}
\eqnum{4.18}
\end{equation}
where 
\begin{equation}
\begin{tabular}{l}
${\cal R}_2(x_1,x_2;y_1,y_2)_{\alpha \beta \rho \sigma }=\delta ^4(x_2-y_2)(%
\overline{\Gamma }^{a\mu })_{\beta \sigma }\int d^4u_1\Sigma
(x_1,u_1)_{\alpha \gamma }[\Lambda _{\mu \nu }^{ab}(x_2;y_1\mid
u_1,y_1)_{\gamma \tau }(\Gamma ^{b\nu })_{\tau \rho }$ \\ 
$-\int d^4v_1\Lambda _\mu ^a(x_2\mid u_1,v_1)_{\gamma \tau }\Sigma
(v_1,y_1)_{\tau \rho }]+\Sigma (x_1,y_1)_{\alpha \rho }(\overline{\Gamma }%
^{a\mu })_{\beta \lambda }[\Lambda _{\mu \nu }^{{\bf c}ab}(x_2;y_2\mid
x_2,y_2)_{\lambda \delta }(\overline{\Gamma }^{b\nu })_{\delta \sigma }$ \\ 
$-\int d^4v_2\Lambda _\mu ^{{\bf c}a}(x_2\mid x_2,v_2)_{\lambda \delta
}\Sigma ^c(v_2,y_2)_{\delta \sigma }]$ \\ 
$-\delta ^4(y_1-y_2)(\overline{\Gamma }^{a\mu })_{\beta \lambda }\int
d^4u_1\Sigma (x_1,u_1)_{\alpha \gamma }\Lambda _{\mu \nu }^{*ab}(x_2;y_1\mid
u_1,x_2)_{\gamma \lambda }(C\Gamma ^{b\nu })_{\sigma \rho },$%
\end{tabular}
\eqnum{4.19}
\end{equation}

\begin{equation}
\begin{tabular}{l}
${\cal M}_2(x_1,x_2;y_1,y_2)_{\alpha \beta \rho \sigma }$ \\ 
$=-(\overline{\Gamma }^{a\mu })_{\beta \lambda }\int d^4u_1\Sigma
(x_1,u_1)_{\alpha \gamma }[{\cal G}_{\mu \nu \kappa }^{abc}(x_2;y_1,y_2\mid
u_1,x_2;y_1,y_2)_{\gamma \lambda \tau \delta }(\Gamma ^{b\nu })_{\tau \rho }(%
\overline{\Gamma }^{c\kappa })_{\delta \sigma }$ \\ 
$-\int d^4v_2{\cal G}_{\mu \nu }^{ab}(x_2;y_1\mid u_{1,}x_2;y_1,v_2)_{\gamma
\lambda \tau \delta }\Sigma ^c(v_2,y_2)_{\delta \sigma }(\Gamma ^{b\nu
})_{\tau \rho }$ \\ 
$-\int d^4v_1{\cal G}_{\mu \nu }^{ab}(x_2;y_2\mid u_{1,}x_2;v_1,y_2)_{\gamma
\lambda \tau \delta }\Sigma (v_1,y_1)_{\tau \rho }(\overline{\Gamma }^{b\nu
})_{\delta \sigma }]$%
\end{tabular}
\eqnum{4.20}
\end{equation}
and the function ${\cal H}_1^{(2)}(x_1,x_2;v_1,v_2)$ was defined in Eq.
(2.37).

{\bf c. Calculation of }${\cal Q}_3(x_1,x_2;y_1,y_2)$

In accordance with Eqs. (4.2) and (2.38), the function ${\cal Q}%
_3(x_1,x_2;y_1,y_2)$ is known to be 
\begin{equation}
\begin{tabular}{l}
${\cal Q}_3(x_1,x_2;y_1,y_2)_{\alpha \beta \rho \sigma }$ \\ 
$=(\Gamma ^{a\mu })_{\alpha \gamma }(\overline{\Gamma }^{a\nu })_{\beta
\lambda }[{\cal G}_{\mu \nu }^{ab}(i\overleftarrow{{\bf \partial }}%
_{y_1}+m_1-\Sigma )(i\overleftarrow{{\bf \partial }}_{y_2}+m_2-\Sigma
^c)]_{\gamma \lambda \rho \sigma }(x_1,x_2\mid x_1,x_2;y_1,y_2)$%
\end{tabular}
\eqnum{4.21}
\end{equation}
For calculating this function, it is necessary to start from the equation of
motion given by Eq. (C7) 
\begin{equation}
\begin{tabular}{l}
$G_{\mu \nu }^{ab}(x_1,x_2\mid x_{1,}x_2;y_1,y_2)_{\gamma \lambda \tau
\sigma }(i\overleftarrow{{\bf \partial }}_{y_1}+m_1)_{\tau \rho }=-\delta
_{\gamma \rho }\delta ^4(x_1-y_1)\Lambda _{\mu \nu }^{{\bf c}ab}(x_1,x_2\mid
x_2,y_2)_{\lambda \sigma }$ \\ 
$-C_{\rho \sigma }\delta ^4(y_1-y_2)\Lambda _{\mu \nu }^{*ab}(x_1,x_2\mid
x_1,x_2)_{\gamma \lambda }+G_{\mu \nu \kappa }^{abc}(x_1,x_2;y_1\mid
x_1,x_2;y_1,y_2)_{\gamma \lambda \tau \sigma }(\Gamma ^{c\kappa })_{\tau
\rho }$%
\end{tabular}
\eqnum{4.22}
\end{equation}
where the Green functions were respectively represented in Eqs. (2.19),
(2.20) and (2. 22). Grounded on the definition given in Eqs. (2.31) and the
following one which can be written out from Eq. (2.23): 
\begin{equation}
\begin{tabular}{l}
${\cal G}_{\mu \nu \kappa }^{abc}(x_1,x_2,y_i\mid x_1,x_2;y_1,y_2)_{\gamma
\lambda \tau \sigma }=G_{\mu \nu \kappa }^{abc}(x_1,x_2,y_i\mid
x_1,x_2;y_1,y_2)_{\gamma \lambda \tau \sigma }$ \\ 
$+\Lambda _{\mu \nu }^{*ab}(x_1,x_2\mid x_1,x_2)_{\gamma \lambda }\overline{%
\Lambda }_\kappa ^{*c}(y_i\mid y_1,y_2)_{\tau \sigma }$%
\end{tabular}
\eqnum{4.23}
\end{equation}
here $i=1,2$ and the equations shown in Eqs. (4.22) and (3.7), it is easy to
derive 
\begin{equation}
\begin{tabular}{l}
$\lbrack {\cal G}_{\mu \nu }^{ab}(i\overleftarrow{{\bf \partial }}%
_{y_1}+m_1-\Sigma )]_{\gamma \lambda \rho \sigma }(x_1,x_2\mid
x_{1,}x_2;y_1,y_2)=-\delta _{\gamma \rho }\delta ^4(x_1-y_1)\Lambda _{\mu
\nu }^{{\bf c}ab}(x_1,x_2\mid x_2,y_2)_{\lambda \sigma }$ \\ 
$+{\cal G}_{\mu \nu \kappa }^{abc}(x_1,x_2;y_1\mid x_1,x_2;y_1,y_2)_{\gamma
\lambda \tau \sigma }(\Gamma ^{c\kappa })_{\tau \rho }-\int d^4v_1{\cal G}%
_{\mu \nu }^{ab}(x_1,x_2\mid x_1,x_2;v_1,y_2)_{\gamma \lambda \tau \sigma
}\Sigma (v_1,y_1)_{\tau \rho }.$%
\end{tabular}
\eqnum{4.24}
\end{equation}

Now, we operate on the above equation with $(i\overleftarrow{{\bf \partial }}%
_{y_2}+m_2-\Sigma ^c)$, giving 
\begin{equation}
\begin{tabular}{l}
$\lbrack {\cal G}_{\mu \nu }^{ab}(i\overleftarrow{{\bf \partial }}%
_{y_1}+m_1-\Sigma )(i\overleftarrow{{\bf \partial }}_{y_2}+m_2-\Sigma
^c)]_{\gamma \lambda \rho \sigma }(x_1,x_2\mid x_{1,}x_2;y_1,y_2)$ \\ 
$=-\delta _{\gamma \rho }\delta ^4(x_1-y_1)\Lambda _{\mu \nu }^{{\bf c}ab}(i%
\overleftarrow{{\bf \partial }}_{y_2}+m_2-\Sigma ^c)]_{\lambda \sigma
}(x_1,x_2\mid x_2,y_2)$ \\ 
$+[{\cal G}_{\mu \nu \kappa }^{abc}(i\overleftarrow{{\bf \partial }}%
_{y_2}+m_2-\Sigma ^c)]_{\gamma \lambda \tau \sigma }(x_1,x_2,y_1\mid
x_1,x_2;y_1,y_2)(\Gamma ^{c\kappa })_{\tau \rho }$ \\ 
$-\int d^4v_1[{\cal G}_{\mu \nu }^{ab}(i\overleftarrow{{\bf \partial }}%
_{y_2}+m_2-\Sigma ^c)]_{\gamma \lambda \tau \sigma }(x_1,x_2\mid
x_1,x_2;v_1,y_2)\Sigma (v_1,y_1)_{\tau \rho }.$%
\end{tabular}
\eqnum{4.25}
\end{equation}
The first term is easily computed by virtue of the equation given in Eq.
(D6) by adding the term associated to the self-energy 
\begin{equation}
\begin{tabular}{l}
$\Lambda _{\mu \nu }^{{\bf c}ab}(\overleftarrow{{\bf \partial }}%
_{y_2}+m_2-\Sigma ^c)_{\lambda \sigma }(x_1,x_2\mid x_2,y_2)=-\delta
_{\lambda \sigma }\delta ^4(x_2-y_2)i\Delta _{\mu \nu }^{ab}(x_1-x_2)$ \\ 
$+\Lambda _{\mu \nu \kappa }^{{\bf c}abc}(x_1,x_2,y_2\mid x_2,y_2)_{\lambda
\delta }(\overline{\Gamma }^{c\kappa })_{\delta \sigma }-\int d^4v_2\Lambda
_{\mu \nu }^{{\bf c}ab}(x_1,x_2\mid x_2,v_2)_{\lambda \delta }\Sigma
^c(v_2-y_2)_{\delta \sigma }$%
\end{tabular}
\eqnum{4.26}
\end{equation}
where 
\begin{equation}
i\Delta _{\mu \nu }^{ab}(x_1-x_2)=\left\langle 0^{+}\left| T[{\bf A}_\mu
^a(x_1){\bf A}_\nu ^b(x_2)]\right| 0^{-}\right\rangle  \eqnum{4.27}
\end{equation}
is the gluon propagator and the other Green functions were defined in Eq.
(2.19). The third term in Eq. (4.25) will be evaluated with the aid of the
equation given by Eq. (D12) 
\begin{equation}
\begin{tabular}{l}
$G_{\mu \nu }^{ab}(x_1,x_2\mid x_{1,}x_2;v_1,y_2)_{\gamma \lambda \tau
\delta }(i\overleftarrow{{\bf \partial }}_{y_2}+m_2)_{\delta \sigma
}=-\delta _{\lambda \sigma }\delta ^4(x_2-y_2)\Lambda _{\mu \nu
}^{ab}(x_1,x_2\mid x_1,v_1)_{\gamma \tau }$ \\ 
$-C_{\tau \sigma }\delta ^4(v_1-y_2)\Lambda _{\mu \nu }^{*ab}(x_1,x_2\mid
x_1,x_2)_{\gamma \lambda }+G_{\mu \nu \kappa }^{abc}(x_1,x_2;y_2\mid
x_1,x_2;v_1,y_2)_{\gamma \lambda \tau \delta }(\overline{\Gamma }^{c\kappa
})_{\delta \sigma }.$%
\end{tabular}
\eqnum{4.28}
\end{equation}
Noticing the definitions in Eqs. (2.31) and (4.23) and applying the
equations in Eqs. (3.11) and (4.28), we find 
\begin{equation}
\begin{tabular}{l}
$\lbrack {\cal G}_{\mu \nu }^{ab}(i\overleftarrow{{\bf \partial }}%
_{y_2}+m_2-\Sigma ^c)]_{\gamma \lambda \tau \sigma }(x_1,x_2\mid
x_{1,}x_2;v_1,y_2)=-\delta _{\lambda \sigma }\delta ^4(x_2-y_2)\Lambda _{\mu
\nu }^{ab}(x_1,x_2\mid x_1,v_1)_{\gamma \tau }$ \\ 
$+{\cal G}_{\mu \nu \kappa }^{abc}(x_1,x_2;y_2\mid x_1,x_2;v_1,y_2)_{\gamma
\lambda \tau \delta }(\overline{\Gamma }^{c\kappa })_{\delta \sigma }-\int
d^4v_2{\cal G}_{\mu \nu }^{ab}(x_1,x_2\mid x_1,x_2;v_1,v_2)_{\gamma \lambda
\tau \delta }\Sigma ^c(v_2,y_2)_{\delta \sigma }.$%
\end{tabular}
\eqnum{4.29}
\end{equation}
The second term in Eq. (4.25) can be calculated by means of the equation
derived in Eq. (D13) 
\begin{equation}
\begin{tabular}{l}
$G_{\mu \nu \kappa }^{abc}(x_1,x_2;y_1\mid x_{1,}x_2;y_1,y_2)_{\gamma
\lambda \tau \delta }(i\overleftarrow{{\bf \partial }}_{y_2}+m_2)_{\delta
\sigma }=-\delta _{\lambda \sigma }\delta ^4(x_2-y_2)\Lambda _{\mu \nu
\kappa }^{abc}(x_1,x_2;y_1\mid x_1,y_1)_{\gamma \tau }$ \\ 
$-C_{\tau \sigma }\delta ^4(y_1-y_2)\Lambda _{\mu \nu \kappa
}^{*abc}(x_1,x_2;y_1\mid x_1,x_2)_{\gamma \lambda }+G_{\mu \nu \kappa \theta
}^{abcd}(x_1,x_2;y_1,y_2\mid x_1,x_2;y_1,y_2)_{\gamma \lambda \tau \delta }(%
\overline{\Gamma }^{d\theta })_{\delta \sigma }$%
\end{tabular}
\eqnum{4.30}
\end{equation}
where all the Green functions can be read off from Eq. (2.18), (2.20) and
(2.22). Considering the definitions given in Eq. (4.23) and in the following 
\begin{equation}
\begin{tabular}{l}
${\cal G}_{\mu \nu \kappa \theta }^{abcd}(x_1,x_2;y_1,y_2\mid
x_1,x_2;y_1,y_2)_{\gamma \lambda \tau \sigma }=G_{\mu \nu \kappa \theta
}^{abcd}(x_1,x_2;y_1,y_2\mid x_1,x_2;y_1,y_2)_{\gamma \lambda \tau \sigma }$
\\ 
$+\Lambda _{\mu \nu }^{*ab}(x_1,x_2\mid x_1,x_2)_{\gamma \lambda }\overline{%
\Lambda }_{\kappa \theta }^{*cd}(y_1,y_2\mid y_1,y_2)_{\tau \sigma }$%
\end{tabular}
\eqnum{4.31}
\end{equation}
which follows from Eq. (2.23) and applying the equations in Eqs. (4.30) and
(3.17), it is easy to find 
\begin{equation}
\begin{tabular}{l}
$\lbrack {\cal G}_{\mu \nu \kappa }^{abc}(i\overleftarrow{{\bf \partial }}%
_{y_2}+m_2-\Sigma ^c)]_{\gamma \lambda \tau \sigma }(x_1,x_2;y_1\mid
x_{1,}x_2;y_1,y_2)=-\delta _{\lambda \sigma }\delta ^4(x_2-y_2)\Lambda _{\mu
\nu \kappa }^{abc}(x_1,x_2;y_1\mid x_1,y_1)_{\gamma \tau }$ \\ 
$-C_{\tau \sigma }\delta ^4(y_1-y_2)\Lambda _{\mu \nu \kappa
}^{*abc}(x_1,x_2;y_1\mid x_1,x_2)_{\gamma \lambda }+{\cal G}_{\mu \nu \kappa
\theta }^{abcd}(x_1,x_2;y_1,y_2\mid x_1,x_2;y_1,y_2)_{\gamma \lambda \tau
\delta }(\overline{\Gamma }^{d\theta })_{\delta \sigma }$ \\ 
$-\int d^4v_2{\cal G}_{\mu \nu \kappa }^{abc}(x_1,x_2;y_1\mid
x_1,x_2;y_1,v_2)_{\gamma \lambda \tau \delta }\Sigma ^c(v_2,y_2)_{\delta
\sigma }.$%
\end{tabular}
\eqnum{4.32}
\end{equation}

Upon substituting Eqs. (4.26), (4.29) and (4.32) in Eq. (4.25) and then
inserting Eq. (4.25) into Eq. (4.21), we eventually arrive at 
\begin{equation}
\begin{tabular}{l}
${\cal Q}_3(x_1,x_2;y_1,y_2)_{\alpha \beta \rho \sigma
}=K_t^{(0)}(x_1,x_2;y_1,y_2)_{\alpha \beta \rho \sigma }+\overline{K}%
(x_1,x_2;y_1,y_2)_{\alpha \beta \rho \sigma }+{\cal R}_3(x_1,x_2;y_1,y_2)_{%
\alpha \beta \rho \sigma }$ \\ 
$+{\cal M}_3(x_1,x_2;y_1,y_2)_{\alpha \beta \rho \sigma }+\int d^4v_1d^4v_2%
{\cal H}_1^{(3)}(x_1,x_2;v_1,v_2)_{\alpha \beta \tau \delta }\Sigma
(v_1,y_1)_{\tau \rho }\Sigma ^c(v_2,y_2)_{\delta \sigma }$%
\end{tabular}
\eqnum{4.33}
\end{equation}
where 
\begin{equation}
K_t^{(0)}(x_1,x_2;y_1,y_2)_{\alpha \beta \rho \sigma }=\delta
^4(x_1-y_1)\delta ^4(x_2-y_2)(\Gamma ^{a\mu })_{\alpha \rho }(\overline{%
\Gamma }^{a\nu })_{\beta \sigma }i\Delta _{\mu \nu }^{ab}(x_1-x_2) 
\eqnum{4.34}
\end{equation}
which is the t-channel one-gluon exchange kernel, 
\begin{equation}
\overline{K}(x_1,x_2;y_1,y_2)_{\alpha \beta \rho \sigma }=(\Gamma ^{a\mu
})_{\alpha \gamma }(\overline{\Gamma }^{a\nu })_{\beta \lambda }{\cal G}%
_{\mu \nu \kappa \theta }^{abcd}(x_1,x_2;y_1,y_2\mid
x_1,x_2;y_1,y_2)_{\gamma \lambda \tau \delta }(\Gamma ^{c\kappa })_{\tau
\rho }(\overline{\Gamma }^{d\theta })_{\delta \sigma },  \eqnum{4.35}
\end{equation}

\begin{equation}
\begin{tabular}{l}
${\cal R}_3(x_1,x_2;y_1,y_2)_{\alpha \beta \rho \sigma }=-\delta
^4(x_1-y_1)(\Gamma ^{a\mu })_{\alpha \rho }(\overline{\Gamma }^{a\nu
})_{\beta \lambda }[\Lambda _{\mu \nu \kappa }^{{\bf c}abc}(x_1,x_2;y_2\mid
x_2,y_2)_{\lambda \delta }(\overline{\Gamma }^{c\kappa })_{\delta \sigma }$
\\ 
$-\int d^4v_2\Lambda _{\mu \nu }^{{\bf c}ab}(x_1,x_2\mid x_2,v_2)_{\lambda
\delta }\Sigma ^c(v_2,y_2)_{\delta \sigma }]$ \\ 
$-\delta ^4(x_2-y_2)(\overline{\Gamma }^{b\nu })_{\beta \sigma }((\Gamma
^{a\mu })_{\alpha \gamma }[\Lambda _{\mu \nu \kappa }^{abc}(x_1,x_2;y_1\mid
x_1,y_1)_{\gamma \tau }(\Gamma ^{c\kappa })_{\tau \rho }$ \\ 
$-\int d^4v_1\Lambda _{\mu \nu }^{ab}(x_1,x_2\mid x_1,v_1)_{\gamma \tau
}\Sigma (v_1,y_1)_{\tau \rho }]$ \\ 
$+\delta ^4(y_1-y_2)(\Gamma ^{a\mu })_{\alpha \gamma }(\overline{\Gamma }%
^{a\nu })_{\beta \lambda }\Lambda _{\mu \nu \kappa }^{*abc}(x_1,x_2;y_1\mid
x_1,x_2)_{\gamma \lambda }(C\Gamma ^{c\kappa })_{\sigma \rho },$%
\end{tabular}
\eqnum{4.36}
\end{equation}
\begin{equation}
\begin{tabular}{l}
${\cal M}_3(x_1,x_2;y_1,y_2)_{\alpha \beta \rho \sigma }$ \\ 
$=-(\Gamma ^{a\mu })_{\alpha \gamma }(\overline{\Gamma }^{a\nu })_{\beta
\lambda }[\int d^4v_2{\cal G}_{\mu \nu \kappa }^{abc}(x_1,x_2;y_1\mid
x_1,x_2;y_1,v_2)_{\gamma \lambda \tau \delta }\Sigma ^c(v_2,y_2)_{\delta
\sigma }(\Gamma ^{c\kappa })_{\tau \rho }$ \\ 
$+\int d^4v_1{\cal G}_{\mu \nu \kappa }^{abc}(x_1,x_2;y_2\mid
x_1,x_2;v_1,y_2)_{\gamma \lambda \tau \delta }\Sigma (v_1,y_1)_{\tau \rho }(%
\overline{\Gamma }^{c\kappa })_{\delta \sigma }]$%
\end{tabular}
\eqnum{4.37}
\end{equation}
and the function ${\cal H}_1^{(3)}(x_1,x_2;v_1,v_2)$ was defined in Eq.
(2.38).

{\bf d. Calculation of }${\cal Q}_4(x_1,x_2;y_1,y_2)$

According to Eqs. (4.2) and (2.39), the function ${\cal Q}%
_4(x_1,x_2;y_1,y_2) $ is ought to be 
\begin{equation}
\begin{tabular}{l}
${\cal Q}_4(x_1,x_2;y_1,y_2)_{\alpha \beta \rho \sigma }$ \\ 
$=\delta ^4(x_1-x_2)(\overline{\Gamma }^{a\mu }C)_{\beta \alpha }[\overline{%
\Lambda }_\mu ^{*a}(i\overleftarrow{{\bf \partial }}_{y_1}+m_1-\Sigma )(i%
\overleftarrow{{\bf \partial }}_{y_2}+m_2-\Sigma ^c)]_{\rho \sigma }(x_2\mid
y_1,y_2)$%
\end{tabular}
\eqnum{4.38}
\end{equation}
Clearly, to evaluate the above function, we need at first the equation of
motion which is obtained from Eq. (C11) by setting the source $J$ to be zero
and adding the term associated with the self-energy 
\begin{equation}
\begin{tabular}{l}
$\lbrack \overline{\Lambda }_\mu ^{*a}(i\overleftarrow{{\bf \partial }}%
_{y_1}+m_1-\Sigma )]_{\rho \sigma }(x_2\mid y_1,y_2)=\overline{\Lambda }%
_{\mu \nu }^{*ab}(x_2;y_1\mid y_1,y_2)_{\tau \sigma }(\Gamma ^{b\nu })_{\tau
\rho }$ \\ 
$-\int d^4v_1\overline{\Lambda }_\mu ^{*a}(x_2\mid v_1,y_2)_{\tau \sigma
}\Sigma (v_1,y_1)_{\tau \rho }$%
\end{tabular}
\eqnum{4.39}
\end{equation}
where the Green functions was written in Eq. (2.21).

Now, let us operate on Eq. (4.39) with the operator $(i\overleftarrow{{\bf %
\partial }}_{y_2}+m_2-\Sigma ^c)$, 
\begin{equation}
\begin{tabular}{l}
$\lbrack \overline{\Lambda }_\mu ^{*a}(i\overleftarrow{{\bf \partial }}%
_{y_1}+m_1-\Sigma )(i\overleftarrow{{\bf \partial }}_{y_2}+m_2-\Sigma
^c)]_{\rho \sigma }(x_2\mid y_1,y_2)$ \\ 
$=[\overline{\Lambda }_{\mu \nu }^{*ab}(i\overleftarrow{{\bf \partial }}%
_{y_2}+m_2-\Sigma ^c)]_{\tau \sigma }(x_2;y_1\mid y_1,y_2)(\Gamma ^{b\nu
})_{\tau \rho }$ \\ 
$-\int d^4v_1[\overline{\Lambda }_\mu ^{*a}(i\overleftarrow{{\bf \partial }}%
_{y_2}+m_2-\Sigma ^c)]_{\tau \sigma }(x_2\mid v_1,y_2)\Sigma (v_1,y_1)_{\tau
\rho }.$%
\end{tabular}
\eqnum{4.40}
\end{equation}
Based on the equations shown in Eqs. (D18) and (D19), one can write
respectively the following equations: 
\begin{equation}
\begin{tabular}{l}
$\lbrack \overline{\Lambda }_\mu ^{*a}(i\overleftarrow{{\bf \partial }}%
_{y_2}+m_2-\Sigma ^c)]_{\tau \sigma }(x_2\mid v_1,y_2)=\overline{\Lambda }%
_{\mu \nu }^{*ab}(x_2;y_2\mid v_1,y_2)_{\tau \delta }(\overline{\Gamma }%
^{b\nu })_{\delta \sigma }$ \\ 
$-\int d^4v_2\overline{\Lambda }_\mu ^{*a}(x_2\mid v_1,v_2)_{\tau \delta
}\Sigma ^c(v_2,y_2)_{\delta \sigma }$%
\end{tabular}
\eqnum{4.41}
\end{equation}
and 
\begin{equation}
\begin{tabular}{l}
$\lbrack \overline{\Lambda }_{\mu \nu }^{*ab}(i\overleftarrow{{\bf \partial }%
}_{y_2}+m_2-\Sigma ^c)]_{\tau \sigma }(x_2,y_1\mid y_1,y_2)=C_{\tau \sigma
}\delta ^4(y_1-y_2)i\Delta _{\mu \nu }^{ab}(x_2-y_1)$ \\ 
$+\overline{\Lambda }_{\mu \nu \kappa }^{*abc}(x_2,y_1,y_2\mid
y_1,y_2)_{\tau \delta }(\overline{\Gamma }^{c\kappa })_{\delta \sigma }-\int
d^4v_2\overline{\Lambda }_{\mu \nu }^{*ab}(x_2,y_1\mid y_1,v_2)_{\tau \delta
}\Sigma ^c(v_2,y_2)_{\delta \sigma }$%
\end{tabular}
\eqnum{4.42}
\end{equation}
where $i\Delta _{\mu \nu }^{ab}(x_2-y_1)$ is the gluon propagator as defined
in Eq. (4.27) and the other Green functions can be read from Eq. (2.21).

Substituting at first Eqs. (4.41) and (4.42) into Eq. (4.40) and then Eq.
(4.40) into Eq. (4.38), the function ${\cal Q}_4(x_1,x_2;y_1,y_2)$ will be
explicitly written in the form 
\begin{equation}
\begin{tabular}{l}
${\cal Q}_4(x_1,x_2;y_1,y_2)_{\alpha \beta \rho \sigma
}=K_s^{(0)}(x_1,x_2;y_1,y_2)_{\alpha \beta \rho \sigma }+{\cal R}%
_4(x_1,x_2;y_1,y_2)_{\alpha \beta \rho \sigma }$ \\ 
$+\int d^4v_1d^4v_2{\cal H}_1^{(4)}(x_1,x_2;v_1,v_2)_{\alpha \beta \tau
\delta }\Sigma (v_1,y_1)_{\tau \rho }\Sigma ^c(v_2,y_2)$%
\end{tabular}
\eqnum{4.43}
\end{equation}
where 
\begin{equation}
K_s^{(0)}(x_1,x_2;y_1,y_2)_{\alpha \beta \rho \sigma }=\delta
^4(x_1-x_2)\delta ^4(y_1-y_2)(\overline{\Gamma }^{a\mu }C)_{\beta \alpha
}(C^{-1}\Gamma ^{b\nu })_{\sigma \rho }i\Delta _{\mu \nu }^{ab}(x_2-y_1) 
\eqnum{4.44}
\end{equation}
which is the s-channel one-gluon exchange kernel, 
\begin{equation}
\begin{tabular}{l}
${\cal R}_4(x_1,x_2;y_1,y_2)_{\alpha \beta \rho \sigma }=\delta ^4(x_1-x_2)(%
\overline{\Gamma }^{a\mu }C)_{\beta \alpha }\{[\overline{\Lambda }_{\mu \nu
\kappa }^{*abc}(x_2;y_1,y_2\mid y_1,y_2)_{\tau \delta }(\overline{\Gamma }%
^{c\kappa })_{\delta \sigma }$ \\ 
$-\int d^4v_2\overline{\Lambda }_{\mu \nu }^{*ab}(x_2;y_1\mid y_1,v_2)_{\tau
\delta }\Sigma ^c(v_2,y_2)_{\delta \sigma }](\Gamma ^{b\nu })_{\tau \rho }$
\\ 
$-\int d^4v_1\overline{\Lambda }_{\mu \nu }^{*ab}(x_2;y_2\mid v_1,y_2)_{\tau
\delta }\Sigma (v_1,y_1)_{\tau \rho }(\overline{\Gamma }^{b\nu })_{\delta
\sigma }\}$%
\end{tabular}
\eqnum{4.45}
\end{equation}
and the function ${\cal H}_1^{(4)}(x_1,x_2;v_1,v_2)$ was defined in Eq.
(2.39).. It is noted here that the s-channel one-gluon exchange kernel
describes the $q\overline{q}$ annihilation and creation process which takes
place between the quark (antiquark) in the initial state and the antiquark
(quark) in the final state as indicated by the gluon propagator in Eq.
(4.44), not between the quark and the antiquark both of which belong to the
initial state or the final state B-S amplitude for a bound state.

{\bf e. Calculation of} ${\cal Q}_5(x_1,x_2;y_1,y_2)$

From Eqs. (4.2) and (2.40), we have 
\begin{equation}
\begin{tabular}{l}
${\cal Q}_5(x_1,x_2;y_1,y_2)_{\alpha \beta \rho \sigma }$ \\ 
$=\int d^4u_1d^4u_2\Sigma (x_1,u_1)_{\alpha \gamma }\Sigma
^c(x_2,u_2)_{\beta \lambda }[{\cal G}(i\overleftarrow{{\bf \partial }}%
_{y_1}+m_1-\Sigma )(i\overleftarrow{{\bf \partial }}_{y_2}+m_2-\Sigma
^c)]_{\gamma \lambda \rho \sigma }(u_1,u_2;y_1,y_2).$%
\end{tabular}
\eqnum{4.46}
\end{equation}
By applying the equation given in Eq. (3.19), the above function can
explicitly be represented as 
\begin{equation}
\begin{tabular}{l}
${\cal Q}_5(x_1,x_2;y_1,y_2)_{\alpha \beta \rho \sigma }={\cal R}%
_0(x_1,x_2;y_1,y_2)_{\alpha \beta \rho \sigma }$ \\ 
$+\int d^4u_1d^4u_2\Sigma (x_1,u_1)_{\alpha \gamma }\Sigma
^c(x_2,u_2)_{\beta \lambda }{\cal H}_2(u_1,u_2;y_1,y_2)_{\gamma \lambda \rho
\sigma }$%
\end{tabular}
\eqnum{4.47}
\end{equation}
where 
\begin{equation}
{\cal R}_0(x_1,x_2;y_1,y_2)_{\alpha \beta \rho \sigma }=\Sigma
(x_1,y_1)_{\alpha \rho }\Sigma ^c(x_2,y_2)_{\beta \sigma }  \eqnum{4.48}
\end{equation}
and ${\cal H}_2(u_1,u_2;y_1,y_2)_{\gamma \lambda \rho \sigma }$ was shown in
Eqs. (3.20)-(3.25).

In summary, combining the expressions given in Eqs. (4.14), (4.18), (4.33),
(4.43) and (4.47), an explicit expression of the function ${\cal Q}%
(x_1,x_2;y_1,y_2)$ written in Eq. (4.1) will be eventually found as shown in
the following 
\begin{equation}
\begin{tabular}{l}
${\cal Q}(x_1,x_2;y_1,y_2)_{\alpha \beta \rho \sigma
}=K_t^{(0)}(x_1,x_2;y_1,y_2)_{\alpha \beta \rho \sigma
}+K_s^{(0)}(x_1,x_2;y_1,y_2)_{\alpha \beta \rho \sigma }$ \\ 
$+\overline{K}(x_1,x_2;y_1,y_2)_{\alpha \beta \rho \sigma }+{\cal R}%
(x_1,x_2;y_1,y_2)_{\alpha \beta \rho \sigma }+{\cal M}(x_1,x_2;y_1,y_2)_{%
\alpha \beta \rho \sigma }$ \\ 
$+\int d^4u_1d^4u_2\Sigma (x_1,u_1)_{\alpha \gamma }\Sigma
^c(x_2,u_2)_{\beta \lambda }{\cal H}_2(u_1,u_2;y_1,y_2)_{\gamma \lambda \rho
\sigma }$ \\ 
$+\int d^4v_1d^4v_2H_1(x_1,x_2;v_1,v_2)_{\alpha \beta \tau \delta }\Sigma
(v_1,y_1)_{\tau \rho }\Sigma ^c(v_2,y_2)_{\delta \sigma }$%
\end{tabular}
\eqnum{4.49}
\end{equation}
where we have defined 
\begin{equation}
{\cal R}(x_1,x_2;y_1,y_2)_{\alpha \beta \rho \sigma }=\sum\limits_{i=0}^4%
{\cal R}_i(x_1,x_2;y_1,y_2)_{\alpha \beta \rho \sigma },  \eqnum{4.50}
\end{equation}
\begin{equation}
{\cal M}(x_1,x_2;y_1,y_2)_{\alpha \beta \rho \sigma }=\sum\limits_{i=1}^3%
{\cal M}_i(x_1,x_2;y_1,y_2)_{\alpha \beta \rho \sigma }  \eqnum{4.51}
\end{equation}
and 
\begin{equation}
H_j(x_1,x_2;y_1,y_2)_{\alpha \beta \rho \sigma }=\sum\limits_{i=1}^4{\cal H}%
_j^{(i)}(x_1,x_2;y_1,y_2)_{\alpha \beta \rho \sigma }  \eqnum{4.52}
\end{equation}
here $j=1,2$ and all the functions on the RHS of Eqs.(4.50)-(4.52) have
already been explicitly given before.

Now let us turn to the function ${\cal S}(x_1,x_2;y_1,y_2)_{\alpha \beta
\rho \sigma }$ expressed in Eq. (3.30). According to the definitions denoted
in Eqs. (2.35), (3.20) and (4.52), we can write 
\begin{equation}
{\cal H}_j(x_1,x_2;y_1,y_2)_{\alpha \beta \rho \sigma
}=H_j(x_1,x_2;y_1,y_2)_{\alpha \beta \rho \sigma }+{\cal H}%
_j^{(5)}(x_1,x_2;y_1,y_2)_{\alpha \beta \rho \sigma }  \eqnum{4.53}
\end{equation}
.Thus, Eq. (3.30) can be rewritten in the form 
\begin{equation}
\begin{tabular}{l}
${\cal S}(x_1,x_2;y_1,y_2)_{\alpha \beta \rho \sigma }$ \\ 
$=\int d^4u_1d^4u_2d^4v_1d^4v_2H_1(x_1,x_2;u_1,u_2)_{\alpha \beta \gamma
\lambda }{\cal G}^{-1}(u_1,u_2;v_1,v_2)_{\gamma \lambda \tau \delta
}H_2(v_1,v_2;y_1,y_2)_{\tau \delta \rho \sigma }$ \\ 
$+\int d^4u_1d^4u_2d^4v_1d^4v_2{\cal H}_1^{(5)}(x_1,x_2;u_1,u_2)_{\alpha
\beta \gamma \lambda }{\cal G}^{-1}(u_1,u_2;v_1,v_2)_{\gamma \lambda \tau
\delta }{\cal H}_2(v_1,v_2;y_1,y_2)_{\tau \delta \rho \sigma }$ \\ 
$+\int d^4u_1d^4u_2d^4v_1d^4v_2H_1(x_1,x_2;u_1,u_2)_{\alpha \beta \gamma
\lambda }{\cal G}^{-1}(u_1,u_2;v_1,v_2)_{\gamma \lambda \tau \delta }{\cal H}%
_2^{(5)}(v_1,v_2;y_1,y_2)_{\tau \delta \rho \sigma }.$%
\end{tabular}
\eqnum{4.54}
\end{equation}
When the functions ${\cal H}_1^{(5)}(x_1,x_2;u_1,u_2)_{\alpha \beta \gamma
\lambda }$ and ${\cal H}_2^{(5)}(v_1,v_2;y_1,y_2)_{\tau \delta \rho \sigma }$
in the second and third terms are respectively replaced by the expressions
given in Eqs. (2.40) and (3.25) and utilizing the relation in Eq. (3.27), it
is easy to see that the last two terms in Eq. (4.54) exactly equal to the
last two terms in Eq. (4.49) respectively. Therefore, when we substitute
Eqs. (4.49) and (4.54) into Eq. (3.29), the above-mentioned terms are
cancelled with each other. Thus, the B-S kernel will be finally expressed by 
\begin{equation}
K(x_1,x_2;y_1,y_2)_{_{_{\alpha \beta \rho \sigma }}}={\bf Q}%
(x_1,x_2;y_1,y_2)_{\alpha \beta \rho \sigma }-{\bf S}(x_1,x_2;y_1,y_2)_{%
\alpha \beta \rho \sigma }  \eqnum{4.55}
\end{equation}
where 
\begin{equation}
\begin{tabular}{l}
${\bf Q}(x_1,x_2;y_1,y_2)_{\alpha \beta \rho \sigma
}=K_t^{(0)}(x_1,x_2;y_1,y_2)_{\alpha \beta \rho \sigma
}+K_s^{(0)}(x_1,x_2;y_1,y_2)_{\alpha \beta \rho \sigma }$ \\ 
$+\overline{K}(x_1,x_2;y_1,y_2)+{\cal R}(x_1,x_2;y_1,y_2)_{\alpha \beta \rho
\sigma }+{\cal M}(x_1,x_2;y_1,y_2)_{\alpha \beta \rho \sigma }$%
\end{tabular}
\eqnum{4.56}
\end{equation}
and 
\begin{equation}
\begin{tabular}{l}
${\bf S}(x_1,x_2;y_1,y_2)_{\alpha \beta \rho \sigma }$ \\ 
$=\int d^4u_1d^4u_2d^4v_1d^4v_2H_1(x_1,x_2;u_1,u_2)_{\alpha \beta \gamma
\lambda }{\cal G}^{-1}(u_1,u_2;v_1,v_2)_{\gamma \lambda \tau \delta
}H_2(v_1,v_2;y_1,y_2)_{\tau \delta \rho \sigma }.$%
\end{tabular}
\eqnum{4.57}
\end{equation}
In Eqs. (4.56) and (4.57), all the functions are represented through only a
few types of Green functions. Therefore, the kernel derived is not only
easily calculated by the perturbation method, but also provides a proper
basis for studying the QCD nonperturbative effect.

\section{Summary and discussions}

In the preceding sections, the B-S kernel has been derived by means of the
B-S equations satisfied by the $q\overline{q}$ Green functions and given a
closed and explicit expression shown in Eqs. (4.55)-(4.57). This expression
is manifestly Lorentz-covariant even in the space-like region of Minkowski
space where the bound states exist and it suits to not only the case that
the quark and antiquark have different flavors, but also the case where they
have the same flavor. In the latter case, the masses of the quark and
antiquark are the same, $m_1=m_2=m.$ Meanwhile, the B-S kernel contains the
part describing the $q\overline{q}$ annihilation and creation process such
as the s-channel one-gluon exchange kernel $K_s^{(0)}(x_1,x_2;y_1,y_2)$
shown in Eq (4.44) and the terms involving the Green functions $\Lambda
_{\mu ...\nu ...}^{*a...b...}(x_i,...,y_j...\mid x_1,x_2)$ and $\overline{%
\Lambda }_{\mu ...\nu ...}^{*a...b...}(x_i...,y_{j,...}\mid y_1,y_2)$ which
give the higher order perturbative corrections to the vertices of the
s-channel one-gluon exchange kernel. It is well known that the s-channel
one-gluon exchange kernel gives no contribution to the $q\overline{q}$ bound
state because the matrix element of the color operator in the kernel
vanishes in the $q\overline{q}$ color singlet. But, this does not imply that
the $q\overline{q}$ annihilation part of the kernel is meaningless. For
example, the s-channel one-gluon exchange kernel derived in this paper would
give a considerable contribution to the interaction of a
multi-quark-antiquark system as shown in our previous works [37]. For the
former case that the quark and antiquark are of different flavors,
certainly, the $q\overline{q}$ annihilation part of the kernel is absent. In
this case, the B-S kernel is of a simpler expression which can be written
out from the corresponding expression given in the preceding section by
deleting out the $q\overline{q}$ annihilation terms and replacing all the
Green functions by the conventional ones. For convenience of later
discussions, we summarize the expression of such a kernel in the following. 
\begin{equation}
K(x_1,x_2;y_1,y_2)_{\alpha \beta \rho \sigma }=Q(x_1,x_2;y_1,y_2)_{\alpha
\beta \rho \sigma }-S(x_1,x_2;y_1,y_2)_{\alpha \beta \rho \sigma } 
\eqnum{5.1}
\end{equation}
where $Q(x_1,x_2;y_1,y_2)$ and $S(x_1,x_2;y_1,y_2)$ are separately described
below.

\begin{equation}
\begin{tabular}{l}
$Q(x_1,x_2;y_1,y_2)_{\alpha \beta \rho \sigma
}=K_t^{(0)}(x_1,x_2;y_1,y_2)_{\alpha \beta \rho \sigma
}+R(x_1,x_2;y_1,y_2)_{\alpha \beta \rho \sigma }$ \\ 
$+\overline{K}(x_1,x_2;y_1,y_2)_{\alpha \beta \rho \sigma
}+M(x_1,x_2;y_1,y_2)_{\alpha \beta \rho \sigma }$%
\end{tabular}
\eqnum{5.2}
\end{equation}
where $K_t^{(0)}(x_1,x_2;y_1,y_2)_{\alpha \beta \rho \sigma }$ was given in
Eq. (4.34), 
\begin{equation}
\overline{K}(x_1,x_2;y_1,y_2)_{\alpha \beta \rho \sigma }=(\Gamma ^{a\mu
})_{\alpha \gamma }(\overline{\Gamma }^{a\nu })_{\beta \lambda }G_{\mu \nu
\kappa \theta }^{abcd}(x_1,x_2;y_1,y_2\mid x_1,x_2;y_1,y_2)_{\gamma \lambda
\tau \delta }(\Gamma ^{c\kappa })_{\tau \rho }(\overline{\Gamma }^{d\theta
})_{\delta \sigma }  \eqnum{5.3}
\end{equation}
which formally is the same as given in Eq. (4.35), but the Green function in
it becomes the conventional one as defined in Eq. (2.22), 
\begin{equation}
R(x_1,x_2;y_1,y_2)_{\alpha \beta \rho \sigma
}=\sum\limits_{i=0}^3R_i(x_1,x_2;y_1,y_2)_{\alpha \beta \rho \sigma } 
\eqnum{5.4}
\end{equation}
here the fact that ${\cal R}_4(x_1,x_2;y_1,y_2)$ defined in Eq.(4.45)
vanishes in this case has been considered, $R_0(x_1,x_2;y_1,y_2)_{\alpha
\beta \rho \sigma }$ is defined as the same as written in Eq. (4.48), while
the other $R_i(x_1,x_2;y_1,y_2)_{\alpha \beta \rho \sigma }$ are now defined
by 
\begin{equation}
\begin{tabular}{l}
$R_1(x_1,x_2;y_1,y_2)_{\alpha \beta \rho \sigma }=\delta ^4(x_1-y_1)(\Gamma
^{a\mu })_{\alpha \rho }\int d^4u_2\Sigma ^c(x_2,u_2)_{\beta \lambda
}[\Lambda _{\mu \nu }^{{\bf c}ab}(x_1;y_2\mid u_2,y_2)_{\lambda \delta }(%
\overline{\Gamma }^{b\nu })_{\delta \sigma }$ \\ 
$-\int d^4v_2\Lambda _\mu ^{{\bf c}a}(x_1\mid u_2,v_2)_{\lambda \delta
}\Sigma ^c(v_2,y_2)_{\delta \sigma }]+\Sigma ^c(x_2,y_2)_{\beta \sigma
}(\Gamma ^{a\mu })_{\alpha \gamma }[\Lambda _{\mu \nu }^{ab}(x_1;y_1\mid
x_1,y_1)_{\gamma \tau }(\Gamma ^{b\nu })_{\tau \rho }$ \\ 
$-\int d^4v_1\Lambda _\mu ^a(x_1\mid x_1,v_1)_{\gamma \tau }\Sigma
(v_1,y_1)_{\tau \rho }],$%
\end{tabular}
\eqnum{5.5}
\end{equation}
\begin{equation}
\begin{tabular}{l}
$R_2(x_1,x_2;y_1,y_2)_{\alpha \beta \rho \sigma }=\delta ^4(x_2-y_2)(%
\overline{\Gamma }^{a\mu })_{\beta \sigma }\int d^4u_1\Sigma
(x_1,u_1)_{\alpha \gamma }[\Lambda _{\mu \nu }^{ab}(x_2;y_1\mid
u_1,y_1)_{\gamma \tau }(\Gamma ^{b\nu })_{\tau \rho }$ \\ 
$-\int d^4v_1\Lambda _\mu ^a(x_2\mid u_1,v_1)_{\gamma \tau }\Sigma
(v_1,y_1)_{\tau \rho }]+\Sigma (x_1,y_1)_{\alpha \rho }(\overline{\Gamma }%
^{a\mu })_{\beta \lambda }[\Lambda _{\mu \nu }^{{\bf c}ab}(x_2;y_2\mid
x_2,y_2)_{\lambda \delta }(\overline{\Gamma }^{b\nu })_{\delta \sigma }$ \\ 
$-\int d^4v_2\Lambda _\mu ^{{\bf c}a}(x_2\mid x_2,v_2)_{\lambda \delta
}\Sigma ^c(v_2,y_2)_{\delta \sigma }]$%
\end{tabular}
\eqnum{5.6}
\end{equation}
and 
\begin{equation}
\begin{tabular}{l}
$R_3(x_1,x_2;y_1,y_2)_{\alpha \beta \rho \sigma }=-\delta ^4(x_1-y_1)(\Gamma
^{a\mu })_{\alpha \rho }(\overline{\Gamma }^{a\nu })_{\beta \lambda
}[\Lambda _{\mu \nu \kappa }^{{\bf c}abc}(x_1,x_2;y_2\mid x_2,y_2)_{\lambda
\delta }(\overline{\Gamma }^{c\kappa })_{\delta \sigma }$ \\ 
$-\int d^4v_2\Lambda _{\mu \nu }^{{\bf c}ab}(x_1,x_2\mid x_2,v_2)_{\lambda
\delta }\Sigma ^c(v_2,y_2)_{\delta \sigma }]$ \\ 
$-\delta ^4(x_2-y_2)(\overline{\Gamma }^{b\nu })_{\beta \sigma }((\Gamma
^{a\mu })_{\alpha \gamma }[\Lambda _{\mu \nu \kappa }^{abc}(x_1,x_2;y_1\mid
x_1,y_1)_{\gamma \tau }(\Gamma ^{c\kappa })_{\tau \rho }$ \\ 
$-\int d^4v_1\Lambda _{\mu \nu }^{ab}(x_1,x_2\mid x_1,v_1)_{\gamma \tau
}\Sigma (v_1,y_1)_{\tau \rho }]$%
\end{tabular}
\eqnum{5.7}
\end{equation}
which are respectively written from Eqs. (4.15), (4.19) and (4.36) by
dropping the terms related to the $q\overline{q}$ annihilation, and 
\begin{equation}
M(x_1,x_2;y_1,y_2)_{\alpha \beta \rho \sigma
}=\sum\limits_{i=1}^3M_i(x_1,x_2;y_1,y_2)_{\alpha \beta \rho \sigma } 
\eqnum{5.8}
\end{equation}
here

\begin{equation}
\begin{tabular}{l}
$M_1(x_1,x_2;y_1,y_2)_{\alpha \beta \rho \sigma }$ \\ 
$=-(\Gamma ^{a\mu })_{\alpha \gamma }\int d^4u_2\Sigma ^c(x_2,u_2)_{\beta
\lambda }[G_{\mu \nu \kappa }^{abc}(x_1;y_1,y_2\mid x_1,u_2;y_1,y_2)_{\gamma
\lambda \tau \delta }(\Gamma ^{b\nu })_{\tau \rho }(\overline{\Gamma }%
^{c\kappa })_{\delta \sigma }$ \\ 
$-\int d^4v_2G_{\mu \nu }^{ab}(x_1;y_1\mid x_{1,}u_2;y_1,v_2)_{\gamma
\lambda \tau \delta }\Sigma ^c(v_2,y_2)_{\delta \sigma }(\Gamma ^{b\nu
})_{\tau \rho }$ \\ 
$-\int d^4v_1G_{\mu \nu }^{ab}(x_1;y_2\mid x_{1,}u_2;v_1,y_2)_{\gamma
\lambda \tau \delta }\Sigma (v_1,y_1)_{\tau \rho }(\overline{\Gamma }^{b\nu
})_{\delta \sigma }],$%
\end{tabular}
\eqnum{5.9}
\end{equation}

\begin{equation}
\begin{tabular}{l}
$M_2(x_1,x_2;y_1,y_2)_{\alpha \beta \rho \sigma }$ \\ 
$=-(\overline{\Gamma }^{a\mu })_{\beta \lambda }\int d^4u_1\Sigma
(x_1,u_1)_{\alpha \gamma }[G_{\mu \nu \kappa }^{abc}(x_2;y_1,y_2\mid
u_1,x_2;y_1,y_2)_{\gamma \lambda \tau \delta }(\Gamma ^{b\nu })_{\tau \rho }(%
\overline{\Gamma }^{c\kappa })_{\delta \sigma }$ \\ 
$-\int d^4v_2G_{\mu \nu }^{ab}(x_2;y_1\mid u_{1,}x_2;y_1,v_2)_{\gamma
\lambda \tau \delta }\Sigma ^c(v_2,y_2)_{\delta \sigma }(\Gamma ^{b\nu
})_{\tau \rho }$ \\ 
$-\int d^4v_1G_{\mu \nu }^{ab}(x_2;y_2\mid u_{1,}x_2;v_1,y_2)_{\gamma
\lambda \tau \delta }\Sigma (v_1,y_1)_{\tau \rho }(\overline{\Gamma }^{b\nu
})_{\delta \sigma }]$%
\end{tabular}
\eqnum{5.10}
\end{equation}
and

\begin{equation}
\begin{tabular}{l}
$M_3(x_1,x_2;y_1,y_2)_{\alpha \beta \rho \sigma }$ \\ 
$=-(\Gamma ^{a\mu })_{\alpha \gamma }(\overline{\Gamma }^{a\nu })_{\beta
\lambda }[\int d^4v_2G_{\mu \nu \kappa }^{abc}(x_1,x_2;y_1\mid
x_1,x_2;y_1,v_2)_{\gamma \lambda \tau \delta }\Sigma ^c(v_2,y_2)_{\delta
\sigma }(\Gamma ^{c\kappa })_{\tau \rho }$ \\ 
$+\int d^4v_1G_{\mu \nu \kappa }^{abc}(x_1,x_2;y_2\mid
x_1,x_2;v_1,y_2)_{\gamma \lambda \tau \delta }\Sigma (v_1,y_1)_{\tau \rho }(%
\overline{\Gamma }^{c\kappa })_{\delta \sigma }]$%
\end{tabular}
\eqnum{5.11}
\end{equation}
which are written respectively from Eqs. (4.16), (4.20) and (4.37) with the
Green functions being replaced by the conventional ones.

The function $S(x_1,x_2;y_1,y_2)$ is defined by

\begin{equation}
\begin{tabular}{l}
$S(x_1,x_2;y_1,y_2)_{\alpha \beta \rho \sigma }$ \\ 
$=\int d^4u_1d^4u_2d^4v_1d^4v_2H_1(x_1,x_2;u_1,u_2)_{\alpha \beta \gamma
\lambda }G^{-1}(u_1,u_2;v_1,v_2)_{\gamma \lambda \tau \delta
}H_2(v_1,v_2;y_1,y_2)_{\tau \delta \rho \sigma }$%
\end{tabular}
\eqnum{5.12}
\end{equation}
where 
\begin{equation}
H_j(x_1,x_2;u_1,u_2)_{\alpha \beta \gamma \lambda
}=\sum\limits_{i=1}^3H_j^{(i)}(x_1,x_2;y_1,y_2)_{\alpha \beta \rho \sigma } 
\eqnum{5.13}
\end{equation}
here $j=1,2$,

\begin{equation}
H_1^{(1)}(x_1,x_2;y_1,y_2)_{\alpha \beta \rho \sigma }=-(\Gamma ^{a\mu
})_{\alpha \gamma }\int d^4u_2\Sigma ^c(x_2,u_2)_{\beta \lambda }G_\mu
^a(x_1\mid x_1,u_2;y_1,y_2)_{\gamma \lambda \rho \sigma },  \eqnum{5.14}
\end{equation}

\begin{equation}
H_1^{(2)}(x_1,x_2;y_1,y_2)_{\alpha \beta \rho \sigma }=-(\overline{\Gamma }%
^{b\nu })_{\beta \lambda }\int d^4u_1\Sigma (x_1,z_1)_{\alpha \gamma }G_\nu
^b(x_2\mid u_1,x_2;y_1,y_2)_{\gamma \lambda \rho \sigma },  \eqnum{5.15}
\end{equation}
\begin{equation}
H_1^{(3)}(x_1,x_2;y_1,y_2)_{\alpha \beta \rho \sigma }=(\Gamma ^{a\mu
})_{\alpha \gamma }(\overline{\Gamma }^{b\nu })_{\beta \lambda }G_{\mu \nu
}^{ab}(x_1,x_2\mid x_1,x_2;y_1,y_2)_{\gamma \lambda \rho \sigma }, 
\eqnum{5.16}
\end{equation}

\begin{equation}
H_2^{(1)}(x_1,x_2;y_1,y_2)_{\alpha \beta \rho \sigma }=-\int d^4v_2G_\mu
^a(y_1\mid x_1,x_2;y_1,v_2)_{\alpha \beta \tau \delta }\Sigma
^c(v_2,y_2)_{\delta \sigma }(\Gamma ^{a\mu })_{\tau \rho },  \eqnum{5.17}
\end{equation}
\begin{equation}
H_2^{(2)}(x_1,x_2;y_1,y_2)_{\alpha \beta \rho \sigma }=-\int d^4v_1G_\nu
^b(y_2\mid x_1,x_2;v_1,y_2)_{\alpha \beta \tau \delta }\Sigma
(v_1,y_1)_{\tau \rho }(\overline{\Gamma }^{b\nu })_{\delta \sigma } 
\eqnum{5.18}
\end{equation}
and 
\begin{equation}
H_2^{(3)}(x_1,x_2;y_1,y_2)_{\alpha \beta \rho \sigma }=G_{\mu \nu
}^{ab}(y_1,y_2\mid x_1,x_2;y_1,y_2)_{\alpha \beta \tau \delta }(\Gamma
^{a\mu })_{\tau \rho }(\overline{\Gamma }^{b\nu })_{\delta \sigma } 
\eqnum{5.19}
\end{equation}
which are written from Eqs. (2.36)-(2.38) and (3.21)-(3.23) with replacing
all the Green functions by the conventional ones. In the above, ${\cal H}%
_j^{(4)}(x_1,x_2;y_1,y_2)=0$ has been considered.

Now, some discussions on the kernel derived in this paper are in order. For
simplicity, we concentrate our attention on the expressions shown in Eqs.
(5.1)-(5.19). First, we note that the terms in the function $%
M(x_1,x_2;y_1,y_2)$ defined in Eqs. (5.8)-(5.11) are all related to the
quark and antiquark self-energies $\Sigma (x_1,y_1)$ and $\Sigma ^c(x_2,y_2)$%
. These terms are responsible for eliminating the corresponding terms
contained in the function $\overline{K}(x_1,x_2;y_1,y_2)$ defined in Eq.
(5.3). This point would be clearly seen from the one and two-particle
irreducible decompositions or the perturbative expansions of the Green
functions occurring in the functions $\overline{K}(x_1,x_2;y_1,y_2)$ and $%
M(x_1,x_2;y_1,y_2)$ ( detailed discussions of this point will be presented
later). By the irreducible decomposition of the Green function involved in
the function $\overline{K}(x_1,x_2;y_1,y_2)$, it would be found that the
irreducible diagrams of the function $\overline{K}(x_1,x_2;y_1,y_2)$ can be
divided into two classes: the first class consists of the diagrams with the
self-energies of the so-called external quark and antiquark (which mean the
quark and antiquark occurring in the B-S amplitude); the second class
contains the diagrams without such self-energies. The first class of
diagrams are precisely cancelled out by all the diagrams of the functions $%
M(x_1,x_2;y_1,y_2).$ Similarly, the self-energy-related terms in the
function $R(x_1,x_2;y_1,y_2)$ defined in Eqs. (5.4)-(5.7) and (4.48) are
responsible for cancelling the corresponding terms contained in the two
self-energy-irrelevant terms in the function $R_3(x_1,x_2;y_1,y_2)$ shown in
Eq. (5.7). From this cancellation, we see that the B-S kernel is, actually,
irrelevant to the self-energy corrections of the external fermion lines. It
only contains the terms consisting of vertices and internal propagators. In
particular, the function $R(x_1,x_2;y_1,y_2)$ gives all the higher order
perturbative corrections to the bare vertices included in the kernel $%
K_t^{(0)}(x_1,x_2;y_1,y_2)$ so as to make the vertices to be exact.
Therefore, the first two terms on the RHS of Eq. (5.2) actually give the
exact one-gluon exchange kernel. The self-energies of the external fermions
are included in the full propagators $S_F(x_1-z_1)$ and $S_F^c(x_2-z_2)$ in
Eq. (2.45). With these full propagators, the B-S equation has an advantage
that it can conveniently be renormalized. This advantage interprets why in
the derivation of the B-S equation, we operate on the Green functions with
the operators $(i{\bf \partial }_{x_1}-m_1+\Sigma )$ and $(i{\bf \partial }%
_{x_2}-m_2+\Sigma ^c)$ other than with the operators $(i{\bf \partial }%
_{x_1}-m_1)$ and $(i{\bf \partial }_{x_2}-m_2).$ Certainly, the B-S equation
can be set up by using the operators $(i{\bf \partial }_{x_1}-m_1)$ and $(i%
{\bf \partial }_{x_2}-m_2)$ to operate on the Green function $%
G(x_1,x_2;y_1,y_2)$. In this case, the fermion propagators in Eq. (2.45)
become free ones and, meanwhile, the B-S kernel, in the case that the quark
and antiquark are of different flavors, has an expression which may be
written down from Eqs. (5.1)-(5.19) by deleting out the self-energy-related
terms, as shown in the following 
\begin{equation}
\begin{tabular}{l}
$K(x_1,x_2;y_1,y_2)=K_t^{(0)}(x_1,x_2;y_1,y_2)+\overline{R}(x_1,x_2;y_1,y_2)$
\\ 
$+G(x_1,x_2;y_1,y_2\mid x_1,x_2;y_1,y_2)-\overline{S}(x_1,x_2;y_1,y_2)$%
\end{tabular}
\eqnum{5.20}
\end{equation}
where $K_t^{(0)}(x_1,x_2;y_1,y_2)$ is still represented in Eq. (4.34), 
\begin{equation}
\overline{R}(x_1,x_2;y_1,y_2)=-\delta (x_2-y_2)\Lambda (x_1,x_2,y_1\mid
x_1,y_1)-\delta (x_1-y_1)\Lambda ^c(x_1,x_2;y_2\mid x_2,y_2)  \eqnum{5.21}
\end{equation}
here 
\begin{equation}
\Lambda (x_1,x_2;y_1\mid x_1,y_1)=\left\langle 0^{+}\left| T\{{\bf J}(x_1)%
\overline{{\bf A}}(x_2)\overline{{\bf J}}(y_1)\}\right| 0^{-}\right\rangle 
\eqnum{5.22}
\end{equation}
\begin{equation}
\Lambda ^c(x_1,x_2;y_2\mid x_2,y_2)=\left\langle 0^{+}\left| T\{{\bf A}(x_1)%
{\bf J}^c(x_2)\overline{{\bf J}}^c(y_2)\}\right| 0^{-}\right\rangle 
\eqnum{5.23}
\end{equation}
in which 
\begin{equation}
\begin{tabular}{l}
${\bf A}(x)=\Gamma ^{a\mu }{\bf A}_\mu ^a(x),$ $\overline{{\bf A}}(x)=%
\overline{\Gamma }^{b\nu }{\bf A}_\nu ^b(x),$ ${\bf J}(x)={\bf A}(x){\bf %
\psi (}x),$ \\ 
${\bf J}^c(x)=\overline{{\bf A}}(x){\bf \psi }^c(x),\overline{\text{ }{\bf J}%
}(x)=\overline{{\bf \psi }}(x){\bf A(}x{\bf )},$ $\overline{{\bf J}}^c(x)=%
\overline{{\bf \psi }}^c(x)\overline{{\bf A}}(x),$%
\end{tabular}
\eqnum{5.24}
\end{equation}
\begin{equation}
\begin{tabular}{l}
$G(x_1,x_2;y_1,y_2\mid x_1,x_2;y_1,y_2)$ \\ 
$=\left\langle 0^{+}\left| T\{{\bf J}(x_1){\bf J}^c(x_2)\overline{{\bf J}}%
(y_1)\overline{{\bf J}}^c(y_2)\}\right| 0^{-}\right\rangle $%
\end{tabular}
\eqnum{5.25}
\end{equation}
and 
\begin{equation}
\begin{tabular}{l}
$\overline{S}(x_1,x_2;y_1,y_2)$ \\ 
$=\int d^4u_1d^4u_2d^4v_1d^4v_2G(x_1,x_2\mid
x_1,x_2;u_1,u_2)G^{-1}(u_1,u_2;v_1,v_2)G(y_1,y_2\mid v_1,v_2;y_1,y_2)$%
\end{tabular}
\eqnum{5.26}
\end{equation}
here 
\begin{equation}
G(x_1,x_2\mid x_1,x_2;u_1,u_2)=\left\langle 0^{+}\left| T\{{\bf J}(x_1){\bf J%
}^c(x_2)\overline{{\bf \psi }}(u_1)\overline{{\bf \psi }}^c(u_2)\}\right|
0^{-}\right\rangle ,  \eqnum{5.27}
\end{equation}
\begin{equation}
G(y_1,y_2\mid v_1,v_2;y_1,y_2)=\left\langle 0^{+}\left| T\{{\bf \psi }(v_1)%
{\bf \psi }^c(v_2)\overline{{\bf J}}(y_1)\overline{{\bf J}}^c(y_2)\}\right|
0^{-}\right\rangle .  \eqnum{5.28}
\end{equation}
The kernel shown in Eqs. (5.20)-(5.28) formally is simple though, its
content actually is more complicated than the kernel written in Eqs.
(5.1)-(5.19) because it contains the diagrams related to the self-energy
corrections of the external fermions.

It is well known that the B-S equation is invariant with respect to
renormalization. In other words, equation (2.45) keeps the same form before
and after renormalization. Therefore, the renormalized B-S kernel, for
example, in the case that the flavors of quark and antiquark are different,
is still represented by Eqs. (5.1)-(5.19) provided that all the quantities
such as the quark and antiquark masses, the coupling constant, the
self-energies and the Green functions are replaced by the renormalized ones.
In such a kernel, the first two terms in Eq. (5.2) give the exact
renormalized one-gluon exchange kernel. Especially, at the renormalization
point, as demonstrated in the renormalization theory [5, 38], the one gluon
exchange kernel is reduced to the form as given in the lowest order
approximation of perturbation and the function $M(x_1,x_2;y_1,y_2)$ vanishes
because at the renormalization, according to the renormalization boundary
conditions, the propagators become the free ones due to vanishing of the
self-energies and the vertices become the bare ones. Only in this case, the
B-S kernel has a simple form as written in Eq. (5.20) with the function $%
\overline{R}(x_1,x_2;y_1,y_2)$ defined in Eqs. (5.21)-(5.23) being absent.

Next, we would like to mention the role played by the last term in Eq.
(5.20). As pointed out in Sect.2, the Green function $G(x_1,x_2\mid
x_1,x_2;u_1,u_2)$ with two-gluon fields at the positions $x_1$ and $x_2$ is
B-S reducible and can be represented as 
\begin{equation}
G(x_1,x_2\mid x_1,x_2;u_1,u_2)=\int d^4z_1d^4z_2\widetilde{K}%
(x_1,x_2;z_1,z_2)G(z_1,z_2;u_1,u_2)  \eqnum{5.29}
\end{equation}
where $\widetilde{K}(x_1,x_2;z_1,z_2)$ is just the B-S kernel which is
generated from the Green function $G(x_1,x_2\mid x_1,x_2;u_1,u_2)$.
Similarly, the Green function $G(y_1,y_2\mid v_1,v_2;y_1,y_2)$ with two
gluon fields at $y_1$ and $y_2$ is also B-S reducible and can be expressed
in the form 
\begin{equation}
G(y_1,y_2\mid v_1,v_2;y_1,y_2)=\int d^4z_1^{\prime }d^4z_2^{\prime
}G(v_1,v_2;z_1^{\prime },z_2^{\prime })\widetilde{K}(z_1^{\prime
},z_2^{\prime };y_1,y_2).  \eqnum{5.30}
\end{equation}
Substituting Eqs. (5.29) and (5.30) into Eq. (5.26) and using the identity
denoted in Eq. (3.27), it is found that 
\begin{equation}
\overline{S}(x_1,x_2;y_1,y_2)=\int d^4u_1d^4u_2d^4v_1d^4v_2\widetilde{K}%
(x_1,x_2;u_1,u_2)G(u_1,u_2;v_1,v_2)\widetilde{K}(v_1,v_2;y_1,y_2). 
\eqnum{5.31}
\end{equation}
This expression shows the typical structure of the two-particle reducible
part of the B-S kernel. On the other hand, the Green function $%
G(x_1,x_2;y_1,y_2\mid x_1,x_2;y_1,y_2)$ for which at every position a gluon
field is nested, in general, can be split into a B-S irreducible part $%
G_{IR}(x_1,x_2;y_1,y_2\mid x_1,x_2;y_1,y_2)$ and a B-S reducible part $%
G_{RE}(x_1,x_2;y_1,y_2\mid x_1,x_2;y_1,y_2)$. The B-S reducible part just
equals to the function $\overline{S}(x_1,x_2;y_1,y_2)$. Therefore, both of
the functions $\overline{S}(x_1,x_2;y_1,y_2)$ and $G_{RE}(x_1,x_2;y_1,y_2%
\mid x_1,x_2;y_1,y_2)$ are cancelled with each other in Eq. (5.20). Thus, we
have 
\begin{equation}
K(x_1,x_2;y_1,y_2)=K_t(x_1,x_2;y_1,y_2)+G_{IR}(x_1,x_2,y_1,y_2\mid
x_1,x_2;y_1,y_2)  \eqnum{5.32}
\end{equation}
in which the B-S irreducible part of the Green function represents two and
more than two-gluon exchange interactions in the sense of perturbation
theory. Similar cancellations take place in Eqs. (4.55) and (5.1) where the
function ${\bf S}(x_1,x_2;y_1,y_2)$ and $S(x_1,x_2;y_1,y_2)$ just play the
role of eliminating the B-S reducible parts contained in the remaining terms
in the kernels.

At last, we would like to stress that unlike the Dyson-Schwinger (D-S)
equation [39,40] which contains an infinite set of equations, the B-S
equation\ is of a closed form. In particular, the interaction kernel in the
equation represents all the interactions taking place in the bound states.
As shown in the preceding section and this section, the closed expression of
the B-S kernel contains only a few types of Green functions. These Green
functions can easily be calculated by the perturbation method. In this kind
of calculations, we only need the perturbative expansions of the Green
functions without concerning the calculation of other more-point Green
functions as it is necessary to be done for the D-S equation. Especially,
the Green functions are possible to be evaluated nonperturbatively as
suggested by the lattice gauge theory. Therefore, the expression of the
kernel given in this paper provides a new formalism for exploring the QCD
nonperturbative effect existing in the bound state and the quark
confinement. For this purpose, it is appropriate to start from the kernel
represented in this section, particularly, as the first stage, from the
kernel in Eq. (5.32) in which the last term is given by the difference
between the $G(x_1,x_2;y_1,y_2\mid x_1,x_2;y_1,y_2)$ in Eq. (5.25) and the $%
\overline{S}(x_1,x_2;y_1,y_2)$ in Eq. (5.26). In the ordinary quark
potential model, the second term in Eq. (5.32) is usually simulated by a
linear potential kernel or some other modified confinement one. Obviously,
this simulation is oversimplified. To search for a sophisticated confining
potential including not only its spatial form, but also its spin and color
structures, it is necessary to call for a nonperturbative calculation of the
the B-S kernel given in this paper. The spin and color structures are
determined not only by the matrices $\Gamma ^{a\mu }$ and $\overline{\Gamma }%
^{b\nu }$ in Eq. (5.25), but also by the Green functions in the kernel. For
example, the color operator for the quark-antiquark system can generally be
decomposed into a combination of the two particle color operators $1\otimes $
$\overline{1}$, $T^a\otimes \overline{1}$, $1\otimes \overline{T}^a$ and $%
T^a\otimes \overline{T}^a$. The coefficients of the combination can be
determined in a certain perturbative calculation, but can not be concretely
given nonperturbativly before a nonperturbative calculation of the Green
functions in the kernel is implemented. Since the B-S amplitude in Eq.
(2.44) is color singlet, containing a color singlet wave function in it, to
solve the B-S equation for bound states, one needs to evaluate the
expectation value of the color operator between the color singlet. Clearly,
once the B-S kernel could be calculated nonperturbatively, the B-S equation
can be solved to give not only the exact masses of $q\overline{q}$ bound
states, but also the B-S amplitudes which may be used to investigate the
meson decay processes. This just is the advantage of the formalism of B-S
equation. At last, it is mentioned that the procedure of deriving the $q%
\overline{q}$ bound state B-S kernel can straightforwardly be applied to
derive the B-S kernels for other multi-quark and/or multi-gluon bound
systems as well as other fermion and/or boson systems either at
zero-temperature or at finite temperature.

\section{\bf Acknowledgment}

We would like to thank Mr. Su Feng for drawing the figures. This project was
supposed by National Natural Science Foundation of China.

\section{\bf Appendix: Equations of motion satisfied by Green functions}

In the previous sections, the B-S equations used to derive the B-S kernel
were derived based on the equations of motion satisfied by some kinds of
Green functions. In this appendix, we show how the equations of motion which
describe variations of the Green functions with the coordinates $x_1,x_2,y_1$
and $y_2$ are derived from the QCD generating functional.

\subsection{{\sf Equations of motion with respect to }${\sf x}_1$}

First, we write down the QCD generating functional [5] 
\begin{equation}
Z[J,\overline{\eta },\eta ,\overline{\xi },\xi ]=\frac 1N\int {\cal D(}A,%
\overline{\psi },\psi ,\overline{C},C)e^{iI}  \eqnum{A1}
\end{equation}
where 
\begin{equation}
I=\int d^4x[{\cal L}+J^{a\mu }A_\mu ^a+\overline{\eta }\psi +\overline{\psi }%
\eta +\overline{\xi }C+\overline{C}\xi ]  \eqnum{A2}
\end{equation}
in which ${\cal L}$ is the effective Lagrangian of QCD 
\begin{equation}
{\cal L}=\overline{\psi }(i{\bf \partial -}m+g{\bf A)}\psi -\frac 14F^{a\mu
\nu }F_{\mu \nu }^a-\frac 1{2\alpha }(\partial ^\mu A_\mu ^a)^2+\overline{C}%
^a\partial ^\mu (D_\mu ^{ab}C^b)  \eqnum{A3}
\end{equation}
here ${\bf A=}\gamma ^\mu T^aA_\mu ^a$ with A$_\mu ^a$ being the vector
potentials of gluon fields, 
\begin{equation}
F_{\mu \nu }^a=\partial _\mu A_\nu ^a-\partial _\nu A_\mu ^a+gf^{abc}A_\mu
^bA_\nu ^c  \eqnum{A4}
\end{equation}
are the strength tensors of the field, 
\begin{equation}
D_\mu ^{ab}=\delta ^{ab}\partial _\mu +gf^{abc}A_\mu ^bC^c  \eqnum{A5}
\end{equation}
are the covariant derivatives, $\overline{C}^a,C^b$ are the ghost fields,
and $J^{a\mu },\overline{\eta },\eta ,\overline{\xi }$ and $\xi $ denote the
external sources for the gluon, quark and ghost fields respectively. By the
charge conjugation transformations shown in Eq. (2.2) for the quark fields
and in the following for the external sources 
\begin{equation}
\eta ^c=C\overline{\eta }^T,\overline{\eta }^c=-\eta ^TC^{-1},  \eqnum{A6}
\end{equation}
it is easy to prove the following relation 
\begin{equation}
\overline{\psi }(i{\bf \partial -}m+g{\bf A)}\psi +\overline{\eta }\psi +%
\overline{\psi }\eta =\overline{\psi }^c(i{\bf \partial -}m+g\overline{{\bf A%
}}{\bf )}\psi ^c+\overline{\eta }^c\psi ^c+\overline{\psi }^c\eta ^c 
\eqnum{A7}
\end{equation}
where $\overline{{\bf A}}=\gamma ^\mu \overline{T}^aA_\mu ^a$ with $%
\overline{T}^a=-\lambda ^{a*}/2$ being the antiquark color matrix.

Now let us proceed to derive the equations of motion obeyed by the Green
functions. Taking the functional derivative of the generating functional in
Eq. (A1) with respect to the field function $\overline{\psi }_\alpha (x_1)$
and considering 
\begin{equation}
\frac{\delta Z}{\delta \overline{\psi }_\alpha (x_1)}=0  \eqnum{A8}
\end{equation}
and 
\begin{equation}
\frac{\delta I}{\delta \overline{\psi }_\alpha (x_1)}=\eta _\alpha (x_1)+[(i%
{\bf \partial }_{x_1}-m_1)_{\alpha \gamma }+g{\bf A(}x_1)_{\alpha \gamma
}]\psi _\gamma (x_1),  \eqnum{A9}
\end{equation}
we have 
\begin{equation}
\{\eta _\alpha (x_1)+[(i{\bf \partial }_{x_1}-m_1)_{\alpha \gamma }+(\Gamma
^{a\mu })_{\alpha \gamma }\frac \delta {i\delta J^{a\mu }(x_1)}]\frac \delta
{i\delta \overline{\eta }_\gamma (x_1)}\}Z=0  \eqnum{A10}
\end{equation}
where the fields $A_\mu ^a(x_1)$ and $\psi _\gamma (x_1)$ have been replaced
by the derivatives of the generating functional with respect to the sources $%
J^{a\mu }(x_1)$ and $\overline{\eta }_\gamma (x_1)$ and each of the
subscripts $\alpha ,\beta $ and $\gamma $ marks the components of color,
flavor and spinor. Differentiating Eq. (A10) with respect to the source $%
\eta _\rho (y_1)$ and then setting $\overline{\eta }=\eta =\overline{\xi }%
=\xi =0$, we obtain [5] 
\begin{equation}
\lbrack (i{\bf \partial }_{x_1}-m_1)_{\alpha \gamma }+(\Gamma ^{a\mu
})_{\alpha \gamma }\frac \delta {i\delta J^{a\mu }(x_1)}]S_F(x_1-y_1)_{%
\gamma \rho }^J=\delta _{\alpha \rho }\delta ^4(x_1-y_1)Z[J]  \eqnum{A11}
\end{equation}
where 
\begin{equation}
S_F(x_1-y_1)_{\gamma \rho }^J=\frac{\delta ^2Z[J,\overline{\eta },\eta ,%
\overline{\xi },\xi ]}{i\delta \overline{\eta }_\gamma (x_1)\delta \eta
_\rho (y_1)}\mid _{\overline{\eta }=\eta =\overline{\xi }=\xi =0} 
\eqnum{A12}
\end{equation}
is the quark propagator in the presence of source $J$. When the source $J$
is set to be zero, noticing 
\begin{equation}
\Lambda _\mu ^a(x_i\mid x_1,y_1)_{\gamma \rho }=\frac \delta {i\delta
J^{a\mu }(x_i)}S_F(x_1-y_1)_{\gamma \rho }^J\mid _{J=0}  \eqnum{A13}
\end{equation}
here $i=1,2$, Eq. (A11) straightforwardly goes over to the equation written
in Eq. (2.10).

Here, we think, it is appropriate to write down general definitions of some
kinds of Green functions in the presence of source $J$. These Green
functions will be encountered in later derivations and no longer defined
repeatedly later on for saving pages. First, we write the remaining fermion
propagators 
\begin{equation}
S_F^c(x_2-y_2)_{\beta \sigma }^J=\frac{\delta ^2Z[J,\overline{\eta },\eta ,%
\overline{\xi },\xi ]}{i\delta \overline{\eta }_\beta ^c(x_2)\delta \eta
_\sigma ^c(y_2)}\mid _{\overline{\eta }=\eta =\overline{\xi }=\xi =0}, 
\eqnum{A14}
\end{equation}
\begin{equation}
S_F^{*}(x_1-x_2)_{\alpha \beta }^J==\frac{\delta ^2Z[J,\overline{\eta },\eta
,\overline{\xi },\xi ]}{i^3\delta \overline{\eta }_\alpha (x_1)\delta 
\overline{\eta }_\beta ^c(x_2)}\mid _{\overline{\eta }=\eta =\overline{\xi }%
=\xi =0},  \eqnum{A15}
\end{equation}
and 
\begin{equation}
\overline{S}_F^{*}(y_1-y_2)_{\rho \sigma }^J=\frac{\delta ^2Z[J,\overline{%
\eta },\eta ,\overline{\xi },\xi ]}{i^3\delta \eta _\rho (y_1)\delta \eta
_\sigma ^c(y_2)}\mid _{\overline{\eta }=\eta =\overline{\xi }=\xi =0}. 
\eqnum{A16}
\end{equation}
On differentiating the above propagators with respect to the sources $%
J^{a\mu }(x_i),...$ and $J^{b\nu }(y_j),...$, we get the following Green
functions: 
\begin{equation}
\Lambda _{\mu ...\nu ...}^{a...b...}(x_i,...,y_{j,...}\mid x_1,y_1)_{\alpha
\rho }^J=\frac \delta {i\delta J^{a\mu }(x_i)}...\frac \delta {i\delta
J^{b\nu }(y_j)}...S_F(x_1-y_1)_{\alpha \rho }^J,  \eqnum{A17}
\end{equation}
\begin{equation}
\Lambda _{\mu ...\nu ...}^{{\bf c}a...b...}(x_i,...,y_{j,...}\mid
x_2,y_2)_{\beta \sigma }^J=\frac \delta {i\delta J^{a\mu }(x_i)}...\frac 
\delta {i\delta J^{b\nu }(y_j)}...S_F^c(x_2-y_2)_{\beta \sigma }^J, 
\eqnum{A18}
\end{equation}
\begin{equation}
\Lambda _{\mu ...\nu ...}^{*a...b...}(x_i,...,y_{j,...}\mid x_1,x_2)_{\alpha
\beta }^J=\frac \delta {i\delta J^{a\mu }(x_i)}...\frac \delta {i\delta
J^{b\nu }(y_j)}...S_F^{*}(x_1-x_2)_{\alpha \beta }^J  \eqnum{A19}
\end{equation}
and 
\begin{equation}
\overline{\Lambda }_{\mu ...\nu ...}^{*a...b...}(x_i,...,y_{j,...}\mid
y_1,y_2)_{\rho \sigma }^J=\frac \delta {i\delta J^{a\mu }(x_i)}...\frac 
\delta {i\delta J^{b\nu }(y_j)}...\overline{S}_F^{*}(y_1-y_2)_{\rho \sigma
}^J  \eqnum{A20}
\end{equation}
where $i,j=1,2$. The $q\overline{q}$ four-point Green function is defined,
in the presence of source $J$, as 
\begin{equation}
G(x_{1,}x_2;y_1,y_2)_{\alpha \beta \rho \sigma }^J=\frac{\delta ^4Z[J,%
\overline{\eta },\eta ,\overline{\xi },\xi ]}{\delta \overline{\eta }_\alpha
(x_1)\delta \overline{\eta }_\beta ^c(x_2)\delta \eta _\rho (y_1)\delta \eta
_\sigma ^c(y_2)}\mid _{\overline{\eta }=\eta =\overline{\xi }=\xi =0}. 
\eqnum{A21}
\end{equation}
By successively differentiating the above Green function with respect to the
sources $J^{a\mu }(x_i),...$ and $J^{b\nu }(y_j),...$, we obtain the Green
functions like this 
\begin{equation}
G_{\mu ...\nu ...}^{a...b...}(x_i,...,y_{j,...}\mid
x_{1,}x_2;y_1,y_2)_{\alpha \beta \rho \sigma }^J=\frac \delta {i\delta
J^{a\mu }(x_i)}...\frac \delta {i\delta J^{b\nu }(y_j)}%
...G(x_{1,}x_2;y_1,y_2)_{\alpha \beta \rho \sigma }^J.  \eqnum{A22}
\end{equation}

Taking differentiations of Eq. (A10) with respect to the sources $\overline{%
\eta }_\beta ^c(x_2),\eta _\rho (y_1)$ and $\eta _\sigma ^c(y_2)$ and
noticing the equality in Eq. (A7) and the following nonvanishing derivatives

\begin{equation}
\begin{tabular}{l}
$\frac{\delta \eta _\alpha ^c(x)}{\delta \overline{\eta }_\beta (y)}%
=C_{\alpha \beta }\delta ^4(x-y),\frac{\delta \overline{\eta }_\alpha ^c(x)}{%
\delta \eta _\beta (y)}=(C^{-1})_{\alpha \beta }\delta ^4(x-y),$ \\ 
$\frac{\delta \overline{\eta }_\alpha (x)}{\delta \eta _\beta ^c(y)}%
=(C^{-1})_{\alpha \beta }\delta ^4(x-y),\frac{\delta \eta _\alpha (x)}{%
\delta \overline{\eta }_\beta ^c(y)}=C_{\alpha \beta }\delta ^4(x-y),$%
\end{tabular}
\eqnum{A23}
\end{equation}
it can be found that

\begin{equation}
\begin{tabular}{l}
$(i{\bf \partial }_{x_1}-m_1)_{\alpha \gamma }G(x_{1,}x_2;y_1,y_2)_{\gamma
\beta \rho \sigma }^J=\delta _{\alpha \rho }\delta
^4(x_1-y_1)S_F^c(x_2-y_2)_{\beta \sigma }^J$ \\ 
$+C_{\alpha \beta }\delta ^4(x_1-x_2)\overline{S}_F^{*}(y_1-y_2)_{\rho
\sigma }^J-(\Gamma ^{a\mu })_{\alpha \gamma }G_\mu ^a(x_1\mid
x_1,x_2;y_1,y_2)_{\gamma \beta \rho \sigma }^J$%
\end{tabular}
\eqnum{A24}
\end{equation}
where we have set $\overline{\eta }=\eta =\overline{\xi }=\xi =0$ and the
Green functions involved have been defined in Eqs. (A14), (A16), (A21) and
(A22). When the source $J$ is turned off, Eq. (A.24) directly gives rise to
the equation of motion in Eq. (2.24).

A further differentiation of Eq. (A24) with respect to $J_\nu ^b(x_2)$ gives

\begin{equation}
\begin{tabular}{l}
$(i{\bf \partial }_{x_1}-m_1)_{\alpha \gamma }G_\nu ^b(x_2\mid
x_{1,}x_2;y_1,y_2)_{\gamma \beta \rho \sigma }^J=\delta _{\alpha \rho
}\delta ^4(x_1-y_1)\Lambda _\nu ^{{\bf c}b}(x_2\mid x_2,y_2)_{\beta \sigma
}^J$ \\ 
$+C_{\alpha \beta }\delta ^4(x_1-x_2)\overline{\Lambda }_\nu ^{*b}(x_2\mid
y_1,y_2)_{\rho \sigma }^J-(\Gamma ^{a\mu })_{\alpha \gamma }G_{\mu \nu
}^{ab}(x_1,x_2\mid x_1,x_2;y_1,y_2)_{\gamma \beta \rho \sigma }^J$%
\end{tabular}
\eqnum{A25}
\end{equation}
where the Green functions have been defined in Eqs. (A18), (A20) and (A22).
On letting the source $J$ to vanish, Eq. (A25) will lead to the equation of
motion in Eq. (2.30).

Now, let us differentiate Eq. (A10) with respect to $\overline{\eta }_\beta
^c(x_2).$ Once all the sources but $J$ are set to be zero, we obtain 
\begin{equation}
\lbrack (i{\bf \partial }_{x_1}-m_1)_{\alpha \gamma }+(\Gamma ^{a\mu
})_{\alpha \gamma }\frac \delta {i\delta J^{a\mu }(x_1)}%
]S_F^{*}(x_{1,}x_2)_{\gamma \beta }^J=(C^{-1})_{\alpha \beta }\delta
^4(x_1-x_2)Z[J]  \eqnum{A26}
\end{equation}
where the propagator was defined in Eq. (A15).

After differentiating Eq. (A26) with respect to $J^{b\nu }(x_2)$ and
noticing the definition given in Eq. (A19), one gets

\begin{equation}
\begin{tabular}{l}
$(i{\bf \partial }_{x_1}-m_1)_{\alpha \gamma }\Lambda _\nu ^{*b}(x_2\mid
x_{1,}x_2)_{\gamma \beta }^J=(C^{-1})_{\alpha \beta }\delta ^4(x_1-x_2)\frac{%
\delta Z[J]}{i\delta J^{b\nu }(x_2)}$ \\ 
$-(\Gamma ^{a\mu })_{\alpha \gamma }\Lambda _{\mu \nu }^{*ab}(x_1,x_2\mid
x_1,x_2)_{\gamma \beta }^J.$%
\end{tabular}
\eqnum{A27}
\end{equation}
When we set $J=0$ and considering $\frac{\delta Z[J]}{i\delta J^{b\nu }(x_2)}%
\mid _{J=0}=0$, Eq. (A27) simply gives the equation in Eq. (2.32).

\subsection{{\sf Equations of motion with respect to x}$_2$}

To derive the equations of motion with respect to $x_2$ for the Green
functions , Let us take a derivative of the generating functional in Eq.
(A1) with respect to the field $\overline{\psi }_\beta ^c(x_2).$ By the same
procedure as described in Eqs. (A8)-(A10), it can be found that 
\begin{equation}
\{\eta _\beta ^c(x_2)+[(i{\bf \partial }_{x_2}-m_2)_{\beta \lambda }+(%
\overline{\Gamma }^{b\nu })_{\beta \lambda }\frac \delta {i\delta J^{b\nu
}(x_2)}]\frac \delta {i\delta \overline{\eta }_\lambda ^c(x_2)}\}Z=0 
\eqnum{B1}
\end{equation}
In the above derivation, the relation in Eq. (A7) has been used.
Differentiating Eq. (B1) with respect to $\eta _\sigma ^c(y_2)$ and setting
all the sources except for $J$ to vanish, we have 
\begin{equation}
\lbrack (i{\bf \partial }_{x_2}-m_2)_{\beta \lambda }+(\overline{\Gamma }%
^{a\mu })_{\beta \lambda }\frac \delta {i\delta J^{b\nu }(x_2)}%
]S_F^c(x_2-y_2)_{\lambda \sigma }^J=\delta _{\beta \sigma }\delta
^4(x_2-y_2)Z[J]  \eqnum{B2}
\end{equation}
where $S_F^c(x_2-y_2)_{\lambda \sigma }^J$ was defined in Eq. (A14). When we
set $J=0$ and noticing 
\begin{equation}
\Lambda _\nu ^{cb}(x_2\mid x_2,y_2)_{\lambda \sigma }=\frac \delta {i\delta
J^{b\nu }(x_2)}S_F^c(x_2-y_2)_{\lambda \sigma }^J\mid _{J=0}  \eqnum{B3}
\end{equation}
and the relation in Eq. (2.15), Eq. (B2) will become the equation in Eq.
(2.14).

Upon differentiating Eq. (B1) with respect to the sources $\overline{\eta }%
_\alpha (x_1),\eta _\rho (y_1)$ and $\eta _\sigma ^c(y_2)$ and setting all
the sources but $J$ to be zero subsequently, it can be found that

\begin{equation}
\begin{tabular}{l}
$(i{\bf \partial }_{x_2}-m_2)_{\beta \lambda }G(x_{1,}x_2;y_1,y_2)_{\alpha
\lambda \rho \sigma }^J=\delta _{\beta \sigma }\delta
^4(x_2-y_2)S_F(x_1-y_1)_{\alpha \rho }^J$ \\ 
$+C_{\alpha \beta }\delta ^4(x_1-x_2)\overline{S}_F^{*}(y_1-y_2)_{\rho
\sigma }^J-(\overline{\Gamma }^{b\nu })_{\beta \lambda }G_\nu ^b(x_2\mid
x_1,x_2;y_1,y_2)_{\alpha \lambda \rho \sigma }^J$%
\end{tabular}
\eqnum{B4}
\end{equation}
where $G_\nu ^b(x_2\mid x_{1,}x_2;y_1,y_2)_{\alpha \lambda \rho \sigma }^J$
was defined in Eq. (A22). When the source $J$ is turned off, Eq. (B4) just
leads to the equation in Eq. (2.25).

\subsection{{\sf Equations of motion with respect to y}$_1$}

In order to find the equations with respect to $y_1$ for the Green
functions, it is necessary to differentiate the generating functional in Eq.
(A1) with respect to the field $\psi _\rho (y_1)$. Following the procedure
stated in Eqs. (A8)-(A10), it is easy to get 
\begin{equation}
\{\overline{\eta }_\rho (y_1)+\frac \delta {i\delta \eta _\tau (y_1)}[(i%
\overleftarrow{{\bf \partial }}_{y_1}+m_1)_{\tau \rho }-(\Gamma ^{a\mu
})_{\tau \rho }\frac \delta {i\delta J^{a\mu }(y_1)}]\}Z=0  \eqnum{C1}
\end{equation}
On differentiating the above equation further with respect to $\overline{%
\eta }_\alpha (x_1)$ and then keeping the source $J$ to be nonvanishing
only, we find 
\begin{equation}
S_F(x_1-y_1)_{\alpha \tau }^J[(i\overleftarrow{{\bf \partial }}%
_{y_1}+m_1)_{\tau \rho }-\frac{\overleftarrow{\delta }}{i\delta J^{a\mu
}(y_1)}(\Gamma ^{a\mu })_{\tau \rho }]=-\delta _{\alpha \rho }\delta
^4(x_1-y_1)Z[J]  \eqnum{C2}
\end{equation}
When the source $J$ is turned off , noticing 
\begin{equation}
\Lambda _\mu ^a(y_i\mid x_1,y_1)_{\alpha \tau }=\frac \delta {i\delta
J^{a\mu }(y_i)}S_F(x_1-y_1)_{\alpha \tau }^J\mid _{J=0}  \eqnum{C3}
\end{equation}
which is defined in Eq. (A17) and the relation 
\begin{equation}
\Lambda _\mu ^a(y_1\mid x_1,y_1)_{\alpha \tau }(\Gamma ^{a\mu })_{\tau \rho
}=\int d^4z_1S_F(x_1-z_1)_{\alpha \tau }\Sigma (z_1,y_1)_{\tau \rho } 
\eqnum{C4}
\end{equation}
which is easy to be proved, Eq. (C2) just gives rise to the equation shown
in Eq. (3.3).

On differentiating Eq. (C1) with respect to $\overline{\eta }_\alpha (x_1),%
\overline{\eta }_\beta ^c(x_2)$ and $\eta _\sigma ^c(y_2)$ and then setting $%
\overline{\eta }=\eta =\overline{\xi }=\xi =0$ , we have

\begin{equation}
\begin{tabular}{l}
$G(x_{1,}x_2;y_1,y_2)_{\alpha \beta \tau \sigma }^J(i\overleftarrow{{\bf %
\partial }}_{y_1}+m_1)_{\tau \rho }=-\delta _{\alpha \rho }\delta
^4(x_1-y_1)S_F^c(x_2-y_2)_{\beta \sigma }^J$ \\ 
$-C_{\rho \sigma }\delta ^4(y_1-y_2)S_F^{*}(x_1-x_2)_{\alpha \beta }^J+G_\mu
^a(y_1\mid x_1,x_2;y_1,y_2)_{\alpha \beta \tau \sigma }^J(\Gamma ^{a\mu
})_{\tau \rho }$%
\end{tabular}
\eqnum{C5}
\end{equation}
where the propagators were defined in Eqs. (A14) and (A15), while the Green
function $G_\mu ^a(y_1\mid x_1,x_2;y_1,y_2)^J$ was given in Eq. (A.22).
Clearly, the above equation directly leads to the equation in Eq. (3.5) when 
$J$ is set to vanish.

Taking the differentiation of Eq. (C5) with respect to $J^{a\mu }(x_1)$, we
get

\begin{equation}
\begin{tabular}{l}
$G_\mu ^a(x_1\mid x_{1,}x_2;y_1,y_2)_{\alpha \beta \tau \sigma }^J(i%
\overleftarrow{{\bf \partial }}_{y_1}+m_1)_{\tau \rho }=-\delta _{\alpha
\rho }\delta ^4(x_1-y_1)\Lambda _\mu ^{{\bf c}a}(x_1\mid x_2,y_2)_{\beta
\sigma }^J$ \\ 
$-C_{\rho \sigma }\delta ^4(y_1-y_2)\Lambda _\mu ^{*a}(x_1\mid
x_1,x_2)_{\alpha \beta }^J+G_{\mu \nu }^{ab}(x_1;y_1\mid
x_1,x_2;y_1,y_2)_{\alpha \beta \tau \sigma }^J(\Gamma ^{b\nu })_{\tau \rho }$%
\end{tabular}
\eqnum{C6}
\end{equation}
where the Green functions on the RHS of the above equation have already
defined in Eqs. (A18), (A19) and (A22). In the case of the source being
absent, we get from Eq. (C6) the equation of motion in Eq. (4.4). While,
differentiating Eq. (C6) with respect to $J^{b\nu }(x_2)$ further, one can
get

\begin{equation}
\begin{tabular}{l}
$G_{\mu \nu }^{ab}(x_1,x_2\mid x_{1,}x_2;y_1,y_2)_{\alpha \beta \tau \sigma
}^J(i\overleftarrow{{\bf \partial }}_{y_1}+m_1)_{\tau \rho }=-\delta
_{\alpha \rho }\delta ^4(x_1-y_1)\Lambda _{\mu \nu }^{{\bf c}ab}(x_1,x_2\mid
x_2,y_2)_{\beta \sigma }^J$ \\ 
$-C_{\rho \sigma }\delta ^4(y_1-y_2)\Lambda _{\mu \nu }^{*ab}(x_1,x_2\mid
x_1,x_2)_{\alpha \beta }^J+G_{\mu \nu \kappa }^{abc}(x_1,x_2;y_1\mid
x_1,x_2;y_1,y_2)_{\alpha \beta \tau \sigma }^J(\Gamma ^{c\kappa })_{\tau
\rho }$%
\end{tabular}
\eqnum{C7}
\end{equation}
where all the Green functions were already defined in Eqs. (A18), (A19) and
(A22). The equation given above straightforwardly gives rise to equation
(4.22) when the source is set to be zero.

Now we turn to derive an equation of motion obeyed by the propagator $%
\overline{S}_F^{*}(y_1-y_2).$ By differentiating Eq. (C1) with respect to $%
\eta _\sigma ^c(y_2)$ and then setting $\overline{\eta }=\eta =\overline{\xi 
}=\xi =0$, it will be found 
\begin{equation}
\overline{S}_F^{*}(y_1-y_2)_{\tau \sigma }^J(i\overleftarrow{{\bf \partial }}%
_{y_1}+m_1)_{\tau \rho }=C_{\rho \sigma }\delta ^4(y_1-y_2)Z[J]+\overline{%
\Lambda }_\nu ^{*b}(y_1\mid y_1,y_2)_{\tau \sigma }^J(\Gamma ^{b\nu })_{\tau
\rho }  \eqnum{C8}
\end{equation}
where $\overline{S}_F^{*}(y_1-y_2)_{\tau \sigma }^J$ and $\overline{\Lambda }%
_\nu ^{*b}(y_1\mid y_1,y_2)_{\tau \sigma }^J$ were defined in Eqs. (A16) and
(A20). Considering 
\begin{equation}
\overline{\Lambda }_\nu ^{*b}(y_i\mid y_1,y_2)_{\tau \sigma }=\frac \delta {%
i\delta J^{b\nu }(y_i)}\overline{S}_F^{*}(y_1-y_2)_{\tau \sigma }^J\mid
_{J=0}  \eqnum{C9}
\end{equation}
and the relation 
\begin{equation}
\overline{\Lambda }_\nu ^{*b}(y_1\mid y_1,y_2)_{\tau \sigma }(\Gamma ^{b\nu
})_{\tau \rho }=\int d^4z_1\overline{S}_F^{*}(z_1-y_2)_{\tau \sigma }\Sigma
(z_1,y_1)_{\tau \rho }  \eqnum{C10}
\end{equation}
which is easy to be proved by using Eqs. (2.7) and (2.11), we see, Eq. (C8)
isl immediately converted to thw equation in Eq. (3.7).

Taking a differentiation of Eq. (C8) with respect to $J^{a\mu }(x_2)$, we
obtain

\begin{equation}
\begin{tabular}{l}
$\overline{\Lambda }_\mu ^{*a}(x_2\mid y_1,y_2)_{\tau \sigma }^J(i%
\overleftarrow{{\bf \partial }}_{y_1}+m_1)_{\tau \rho }=C_{\rho \sigma
}\delta ^4(y_1-y_2)\frac{\delta Z[J]}{i\delta J^{a\mu }(x_2)}$ \\ 
$+\overline{\Lambda }_{\mu \nu }^{*ab}(x_2,y_1\mid y_1,y_2)_{\tau \sigma
}^J(\Gamma ^{b\nu })_{\tau \rho }$%
\end{tabular}
\eqnum{C11}
\end{equation}
where the Green functions were defined in Eq. (A20). On setting $J=0,$ the
above equation goes over to equation (4.39).

\subsection{{\sf Equations of motion with respect to y}$_2$}

For deriving the equations of motion with respect to y$_2$ for the Green
functions, we need to differentiate the generating functional in Eq. (A1)
with respect to $\psi _\sigma ^c(y_2).$ By the procedure shown in Eq.
(A8)-(A10), it can be found 
\begin{equation}
\{\overline{\eta }_\sigma ^c(y_2)+\frac \delta {i\delta \eta _\delta ^c(y_2)}%
[(i\overleftarrow{{\bf \partial }}_{y_2}+m_2)_{\delta \sigma }-(\overline{%
\Gamma }^{b\nu })_{\delta \sigma }\frac \delta {i\delta J^{b\nu }(y_2)}%
]\}Z=0.  \eqnum{D1}
\end{equation}
Then, we differentiate the above equation with respect to $\overline{\eta }%
_\beta ^c(x_2)$ and subsequently set all the sources, except for $J$, to be
zero. As a result , we get 
\begin{equation}
S_F^c(x_2-y_2)_{\beta \delta }^J[(\overleftarrow{{\bf \partial }}%
_{y_2}+m_2)_{\delta \sigma }-\frac{\overleftarrow{\delta }}{i\delta J^{b\nu
}(y_2)}(\overline{\Gamma }^{b\nu })_{\delta \sigma }]=-\delta _{\beta \sigma
}\delta ^4(x_2-y_2)Z[J].  \eqnum{D2}
\end{equation}
On setting $J=0$ and noticing 
\begin{equation}
\Lambda _\nu ^{cb}(y_2\mid x_2,y_2)_{\beta \delta }=\frac \delta {i\delta
J^{b\nu }(y_2)}S_F^c(x_2-y_2)_{\beta \delta }^J\mid _{J=0}  \eqnum{D3}
\end{equation}
and 
\begin{equation}
\Lambda _\nu ^{cb}(y_2\mid x_2,y_2)_{\beta \delta }(\overline{\Gamma }^{b\nu
})_{\delta \sigma }=\int d^4z_2S_F^c(x_2-z_2)_{\beta \delta }\Sigma
^c(z_2-y_2)_{\delta \sigma },  \eqnum{D4}
\end{equation}
we obtain the equation in Eq. (3.4) from Eq. (D2).

To derive Eq. (4.8), we need to differentiate Eq. (D2) with respect to $%
J^{a\mu }(x_1)$. The result is 
\begin{equation}
\begin{tabular}{l}
$\Lambda _\mu ^{{\bf c}a}(x_1\mid x_2,y_2)_{\beta \lambda }^J(\overleftarrow{%
{\bf \partial }}_{y_2}+m_2)_{\lambda \sigma }=-\delta _{\beta \sigma }\delta
^4(x_2-y_2)\frac{\delta Z[J]}{i\delta J^{a\mu }(x_1)}$ \\ 
$+\Lambda _{\mu \nu }^{{\bf c}ab}(x_1;y_2\mid x_2,y_2)_{\beta \lambda }^J(%
\overline{\Gamma }^{b\nu })_{\lambda \sigma }$%
\end{tabular}
\eqnum{D5}
\end{equation}
where the Green functions were represented in Eq. (A18). On setting $J=0$,
the above equation just leads to the equation in Eq. (4.8).

Upon differentiating Eq. (D5) with respect to $J^{b\nu }(x_2)$, it is found
that 
\begin{equation}
\begin{tabular}{l}
$\Lambda _{\mu \nu }^{{\bf c}ab}(x_1,x_2\mid x_2,y_2)_{\beta \lambda }^J(%
\overleftarrow{{\bf \partial }}_{y_2}+m_2)_{\lambda \sigma }=-\delta _{\beta
\sigma }\delta ^4(x_2-y_2)\frac{\delta ^2Z[J]}{i\delta J^{a\mu }(x_1)i\delta
J^{b\nu }(x_2)}$ \\ 
$+\Lambda _{\mu \nu \kappa }^{{\bf c}abc}(x_1,x_2;y_2\mid x_2,y_2)_{\beta
\lambda }^J(\overline{\Gamma }^{c\kappa })_{\lambda \sigma }$%
\end{tabular}
\eqnum{D6}
\end{equation}
where the Green's functions were also represented in Eq. (A18). In the case
of $J=0$, considering that 
\begin{equation}
\frac{\delta ^2Z[J]}{i\delta J^{a\mu }(x_1)i\delta J^{b\nu }(x_2)}\mid
_{J=0}=i\Delta _{\mu \nu }^{ab}(x_1-x_2)  \eqnum{D7}
\end{equation}
which is the gluon propagator, the above equation straightforwardly gives
rise to equation (4.26).

Now we turn to derive the equation of motion for the four-point Green
function. After completing the differentiations of Eq. (D1) with respect to $%
\overline{\eta }_\alpha (x_1),\overline{\eta }_\beta ^c(x_2)$ and $\eta
_\rho (y_1)$ and then letting $\overline{\eta }=\eta =\overline{\xi }=\xi =0$%
, we have 
\begin{equation}
\begin{tabular}{l}
$G(x_{1,}x_2;y_1,y_2)_{\alpha \beta \rho \delta }^J(i\overleftarrow{{\bf %
\partial }}_{y_2}+m_2)_{\delta \sigma }=-\delta _{\beta \sigma }\delta
^4(x_2-y_2)S_F(x_1-y_1)_{\alpha \rho }^J$ \\ 
$-C_{\rho \sigma }\delta ^4(y_1-y_2)S_F^{*}(x_1-x_2)_{\alpha \beta }^J+G_\nu
^b(y_2\mid x_1,x_2;y_1,y_2)_{\alpha \beta \rho \delta }^J(\overline{\Gamma }%
^{b\nu })_{\delta \sigma }$%
\end{tabular}
\eqnum{D8}
\end{equation}
where $G_\nu ^b(y_2\mid x_1,x_2;y_1,y_2)_{\alpha \beta \rho \delta }^J$ was
defined in Eq. (A22). It is clear that when $J=0$, the above equation leads
to the equation written in Eq. (3.10).

Going on further, it is seen that the differentiation of Eq. (D8) with
respect to $J^{a\mu }(y_1)$ gives 
\begin{equation}
\begin{tabular}{l}
$G_\mu ^a(y_1\mid x_{1,}x_2;y_1,y_2)_{\alpha \beta \rho \delta }^J(i%
\overleftarrow{{\bf \partial }}_{y_2}+m_2)_{\delta \sigma }=-\delta _{\beta
\sigma }\delta ^4(x_2-y_2)\Lambda _\mu ^a(y_1\mid x_1,y_1)_{\alpha \rho }^J$
\\ 
$-C_{\rho \sigma }\delta ^4(y_1-y_2)\Lambda _\mu ^{*a}(y_1\mid
x_1,x_2)_{\alpha \beta }^J+G_{\mu \nu }^{ab}(y_1,y_2\mid
x_1,x_2;y_1,y_2)_{\alpha \beta \rho \delta }^J(\overline{\Gamma }^{b\nu
})_{\delta \sigma }$%
\end{tabular}
\eqnum{D9}
\end{equation}
where the Green functions were defined in Eqs. (A17), (A19) and (A22).
Obviously, when the source $J$ is turned off, the equation shown above goes
over to equation (3.14).

Differentiating Eq. (D8) with respect to $J^{a\mu }(x_1)$, we obtain an
equation like this 
\begin{equation}
\begin{tabular}{l}
$G_\mu ^a(x_1\mid x_{1,}x_2;y_1,y_2)_{\alpha \beta \rho \delta }^J(i%
\overleftarrow{{\bf \partial }}_{y_2}+m_2)_{\delta \sigma }=-\delta _{\beta
\sigma }\delta ^4(x_2-y_2)\Lambda _\mu ^a(x_1\mid x_1,y_1)_{\alpha \rho }^J$
\\ 
$+(C^{-1})_{\rho \sigma }\delta ^4(y_1-y_2)\Lambda _\mu ^{*a}(x_1\mid
x_1,x_2)_{\alpha \beta }^J+G_{\mu \nu }^{ab}(x_1;y_2\mid
x_1,x_2;y_1,y_2)_{\alpha \beta \rho \delta }^J(\overline{\Gamma }^{b\nu
})_{\delta \sigma }$%
\end{tabular}
\eqnum{D10}
\end{equation}
where the Green functions can also be found in Eqs. (A17), (A19) and (A22).
Letting the source to be zero, the equation shown above will lead to the
equation in Eq. (4.9).

Taking a further differentiation of Eq. (D10) with respect to $J^{b\nu
}(y_1) $, it is seen that 
\begin{equation}
\begin{tabular}{l}
$G_{\mu \nu }^{ab}(x_1;y_1\mid x_{1,}x_2;y_1,y_2)_{\alpha \beta \rho \delta
}^J(i\overleftarrow{{\bf \partial }}_{y_2}+m_2)_{\delta \sigma }=-\delta
_{\beta \sigma }\delta ^4(x_2-y_2)\Lambda _{\mu \nu }^{ab}(x_1;y_1\mid
x_1,y_1)_{\alpha \rho }^J$ \\ 
$+(C^{-1})_{\rho \sigma }\delta ^4(y_1-y_2)\Lambda _{\mu \nu
}^{*ab}(x_1;y_1\mid x_1,x_2)_{\alpha \beta }^J+G_{\mu \nu \kappa
}^{abc}(x_1;y_1,y_2\mid x_1,x_2;y_1,y_2)_{\alpha \beta \rho \delta }^J(%
\overline{\Gamma }^{c\kappa })_{\delta \sigma }.$%
\end{tabular}
\eqnum{D11}
\end{equation}
When the source is turned off, the equation above just gives the equation in
Eq. (4.11).

Now, let us take a differentiation of Eq. (D10) with respect to $J^{b\nu
}(x_2).$ The result is 
\begin{equation}
\begin{tabular}{l}
$G_{\mu \nu }^{ab}(x_1,x_2\mid x_{1,}x_2;y_1,y_2)_{\alpha \beta \rho \delta
}^J(i\overleftarrow{{\bf \partial }}_{y_2}+m_2)_{\delta \sigma }=-\delta
_{\beta \sigma }\delta ^4(x_2-y_2)\Lambda _{\mu \nu }^{ab}(x_1,x_2\mid
x_1,y_1)_{\alpha \rho }^J$ \\ 
$+(C^{-1})_{\rho \sigma }\delta ^4(y_1-y_2)\Lambda _{\mu \nu
}^{*ab}(x_1,x_2\mid x_1,x_2)_{\alpha \beta }^J+G_{\mu \nu \kappa
}^{abc}(x_1,x_2;y_2\mid x_1,x_2;y_1,y_2)_{\alpha \beta \rho \delta }^J(%
\overline{\Gamma }^{c\kappa })_{\delta \sigma }$%
\end{tabular}
\eqnum{D12}
\end{equation}
where the Green functions can be read from Eqs. (A17), (A19) and (A22). When
the source $J$ is absent, the above equation will be converted to the
equation in Eq. (4.28). A further differentiation of the above equation with
respect to $J^{c\kappa }$($y_1)$ leads to 
\begin{equation}
\begin{tabular}{l}
$G_{\mu \nu \kappa }^{abc}(x_1,x_2;y_1\mid x_{1,}x_2;y_1,y_2)_{\alpha \beta
\rho \delta }^J(i\overleftarrow{{\bf \partial }}_{y_2}+m_2)_{\delta \sigma }$
\\ 
$=-\delta _{\beta \sigma }\delta ^4(x_2-y_2)\Lambda _{\mu \nu \kappa
}^{abc}(x_1,x_2;y_1\mid x_1,y_1)_{\alpha \rho }^J$ \\ 
$+(C^{-1})_{\rho \sigma }\delta ^4(y_1-y_2)\Lambda _{\mu \nu \kappa
}^{*abc}(x_1,x_2,y_1\mid x_1,x_2)_{\alpha \beta }^J$ \\ 
$+G_{\mu \nu \kappa \theta }^{abcd}(x_1,x_2;y_1,y_2\mid
x_1,x_2;y_1,y_2)_{\alpha \beta \rho \delta }^J(\overline{\Gamma }^{c\lambda
})_{\delta \sigma }$%
\end{tabular}
\eqnum{D13}
\end{equation}
where the Green functions can also be read off from Eqs. (A17), (A19) and
(A22). When the source is turned off, Eq. (D13) will go over to the equation
in Eq. (4.30).

To derive the equations in Eqs. (3.11) and (3.17), it is necessary to
differentiate Eq. (D1) with respect to $\eta _\rho (y_1).$ On completing
this differentiation, we find 
\begin{equation}
\overline{S}_F^{*}(y_1-y_2)_{\rho \delta }^J[(\overleftarrow{{\bf \partial }}%
_{y_2}+m_2)_{\delta \sigma }-\frac{\overleftarrow{\delta }}{i\delta J^{b\nu
}(y_2)}(\overline{\Gamma }^{b\nu })_{\delta \sigma }]=C_{\rho \sigma }\delta
^4(y_1-y_2)Z[J]  \eqnum{D14}
\end{equation}
where we have set $\overline{\eta }=\eta =\overline{\xi }=\xi =0.$ Noticing 
\begin{equation}
\overline{\Lambda }_\nu ^{*b}(y_2\mid y_1,y_2)_{\rho \delta }=\frac \delta {%
i\delta J^{b\nu }(y_2)}\overline{S}_F^{*}(y_1-y_2)_{\rho \delta }^J{}\mid
_{J=0}  \eqnum{D15}
\end{equation}
and 
\begin{equation}
\overline{\Lambda }_\nu ^{*b}(y_2\mid y_1,y_2)_{\rho \delta }(\overline{%
\Gamma }^{b\nu })_{\delta \sigma }=\int d^4z_2\overline{S}%
_F^{*}(y_1-z_2)_{\rho \delta }^J\Sigma ^c(z_2-y_2)_{\delta \sigma }, 
\eqnum{D16}
\end{equation}
equation (D14) will be brought to the equation in Eq. (3.11).

By differentiating Eq. (D14) with respect to $J^{a\mu }(y_1)$ and noticing
the definition given in Eq. (A20), we obtain 
\begin{equation}
\begin{tabular}{l}
$\overline{\Lambda }_\mu ^{*a}(y_1\mid y_1,y_2)_{\rho \delta }^J(%
\overleftarrow{{\bf \partial }}_{y_2}+m_2)_{\delta \sigma }=C_{\rho \sigma
}\delta ^4(y_1-y_2)\frac{\delta Z[J]}{i\delta J^{a\mu }(y_1)}$ \\ 
$+\overline{\Lambda }_{\mu \nu }^{*ab}(y_1,y_2\mid y_1,y_2)_{\rho \delta }^J(%
\overline{\Gamma }^{b\nu })_{\delta \sigma }.$%
\end{tabular}
\eqnum{D17}
\end{equation}
Equation (3.17) directly follows from the above equation when we set $J=0$.
On taking the derivative of Eq. (D14) with respect to $J^{a\mu }(x_2)$ , it
is clearly seen that 
\begin{equation}
\begin{tabular}{l}
$\overline{\Lambda }_\mu ^{*a}(x_2\mid y_1,y_2)_{\rho \delta }^J(%
\overleftarrow{{\bf \partial }}_{y_2}+m_2)_{\delta \sigma }=C_{\rho \sigma
}\delta ^4(y_1-y_2)\frac{\delta Z[J]}{i\delta J^{a\mu }(x_2)}$ \\ 
$+\overline{\Lambda }_{\mu \nu }^{*ab}(x_2;y_2\mid y_1,y_2)_{\rho \delta }^J(%
\overline{\Gamma }^{b\nu })_{\delta \sigma }$%
\end{tabular}
\eqnum{D18}
\end{equation}
where the Green functions were all shown in Eq. (A20). After setting $J=0$,
the equation above becomes the equation in Eq. (4.41). Differentiating the
above equation further with respect to $J^{b\nu }(y_1)$, we get 
\begin{equation}
\begin{tabular}{l}
$\overline{\Lambda }_{\mu \nu }^{*ab}(x_2;y_1\mid y_1,y_2)_{\rho \delta }^J(%
\overleftarrow{{\bf \partial }}_{y_2}+m_2)_{\delta \sigma }=C_{\rho \sigma
}\delta ^4(y_1-y_2)\frac{\delta ^2Z[J]}{i\delta J^{a\mu }(x_2)i\delta
J^{b\nu }(y_1)}$ \\ 
$+\overline{\Lambda }_{\mu \nu \kappa }^{*abc}(x_2;y_1,y_2\mid
y_1,y_2)_{\rho \delta }^J(\overline{\Gamma }^{c\kappa })_{\delta \sigma }$%
\end{tabular}
\eqnum{D19}
\end{equation}
where the Green functions were represented in Eq. (A20). When the source is
set to vanish and noticing the definition in Eq. (D7) for the gluon
propagator, the above equation just gives the equation in Eq. (4.42).

.

\section{\bf REFERENCES}

[1] E. E. Salpeter and H. A. Bethe, Phys. Rev. 84 (1951) 1232.

[2] M. Gell-Mann and F. E. Low, Phys. Rev. 84 (1951) 350.

[3] N. Nakanishi, Prog. Theor. Phys. Suppl. 42 (1959) 1 ; Prog. Theor. Phys.
Suppl. 95 (1988) 1,\ A great deal of references are cited therein.

[4] N. Seto, Prog. Theor. Phys. Suppl. 95 (1988) 25.

[5] C. Itzykson and J. B. Zuber, Quantum Field Theory, McGraw-Hill, New
York, 1980.

[6] E. E. Salpeter, Phys. Rev. 87 (1952) 328 .

[7] M. B\"ohm, Nucl. Phys. B91 (1975) 494.

[8] H. Pagels, Phys. Rev. D15 (1977) 2991.

[9] A. N. Mitra, Z. Phys. C8 (1981) 25 .

[10] A. A. Logunov and A. N. Tavkhhelidze, Nuovo Cimento 29 (1963) 380.

[11] Y. A. Alessandrini and R. L. Omnes, Phys. Rev. 139 (1965) 167.

[12] R. Blankenbecler and R. Sugar, Phys. Rev. 142 (1966) 1051.

[13] F. Gross, Phys. Rev. 186 (1969) 1448.

[14] I. T. Todorov, Phys. Rev. D3 (1971) 2351.

[15] T. N. Ruan, H. Q. Zhu, T. X. Ho, C. R. Qing and W. Q. Chao, Phys.
Energ. Fortis.\ Phys. Nucl. 5 (1981) 393, 537.

[16] A. Bial and P. Schuck, Phys. Rev. D31 (1985) 2045.

[17] R. S. Bhalerao and C. S. Warke, Phys. Rev. C34 (1986) 1920 .

[18] J. Bijtebier and J. Broekaert, J. Phys. G: Nucl. Part. Phys. 22 (1996)
559 .

[19] S. S. Wu, J. Phys. G: Nucl. Part. Phys. 16 (1990) 1447.

[20] J. C. Su and D. Z. Mu, Commun. Theor. Phys. 15 (1991) 437.

[21] J. C. Su J. Phys.G: Nucl. Part. Phys. 30 (2004) 1309.

[22] H. A. Bethe and E. E. Salpeter, Quantum Mechanics of One and
Two-Electron Atoms,\ Springer-Verlag/Academic, New York, 1957.

[23] M. A. Stroscio, Phys. Rep. 22 (1975) 215.

[24] G. T. Bodwin, D. R. Yennie and M. A. Gregorio, Rev. Mod. Phys. 57
(1985) 723.

[25] T. Murota, Prog. Theor. Phys. Suppl. 95 (1988) 46.

[26] J. H. Connell, Phys.Rev. D43 (1991) 1393.

[27] W. Lucha, F. F. Sch\"oberl and D. Gromes, Phys. Rep. 200 (1991) 127,
many references concerning the $q\overline{q}$ bound state B-S equation can
be found therein.

[28] D. Gromes, Z. Phys. C11 (1981) 147.

[29] J. C. Su. Y. B. Dong and S. S Wu, J. Phys. G: Nucl. Part. Phys. 18
(1992) 1347.

[30] K. Wilson, Phys. Rev. D10 (1974) 2445.

[31] L. S. Brown and W. I. Weisberger, Phys. Rev. D20 (1979) 3239.

[32] E. Eichten and F. Feinberg, Phys. Rev. D23 (1981) 2724.

[33] I. Montvay and G. M\"unster, Quantum Fields on A Lattice, Cambridge
University Press, 1994 ( A large number of references are quoted therein).

[34] K. Erkelenz, Phys. Rep. 13 (1974) 191.

[35] K. Nishijima, Phys. Rev. 111 (1958) 995 ; R. Haag, Phys. Rev. 112
(1958) 669.

[36] E. S. Abers and B. W. Lee, Phys. Rep. C9 (1973) 1.

[37] J. C. Su, Z. Q. Chen, S. S. Wu, Nucl. Phys. A254 (1991) 615;\ J. X.
Chen, Y. H. Cao and J. C. Su, Phys. Rev.\ C64 (2001) 065201; H. J. Wang, H.
Yang and J. C. Su, Phys. Rev. C 68 (2003) 076002

[38] J. C. Su, X. X. Yi and Y.H.Cao, J. Phys. G: Nucl. Part. Phys. 25 (1999)
665; J. C. Su and H. J. Wang, Phys. Rev. C 70 (2004) 044003.

[39] F. J. Dyson, Phys. Rev. 75 (1949) 1736.

[40] J. S. Schwinger, Proc. Nat. Acad. Sc. 37 (1951) 452, 455; Phys. Rev.
125 (1962) 397; 128 (1962) 2425.

\end{document}